\documentclass[a4paper,11pt]{article}
\pdfoutput=1 % if your are submitting a pdflatex (i.e. if you have
\usepackage{jheppub} % for details on the use of the package, please
\usepackage[T1]{fontenc} % if needed
\DeclareMathOperator{\Tr}{Tr} 
\usepackage{slashed}
\usepackage{booktabs}
\usepackage{feynmp}

\usepackage{graphicx}
\allowdisplaybreaks
\usepackage{longtable}

\renewcommand{\imath}{\mathrm{i}}
 
\title{\boldmath Higgs boson mass and electroweak observables in the MRSSM}

\author[a,1]{Philip Diessner,\note{Corresponding author.}}
\author[b]{Jan Kalinowski,}
\author[b]{Wojciech Kotlarski}
\author[a]{and Dominik St\"ockinger}

\affiliation[a]{Institut f\"ur Kern- und Teilchenphysik, TU Dresden\\
 01069 Dresden, Germany}
\affiliation[b]{Faculty of Physics, University of Warsaw,\\ Pasteura 5, 02093 Warsaw, Poland}

\emailAdd{philip.diessner@mailbox.tu-dresden.de}
\emailAdd{jan.kalinowski@fuw.edu.pl}
\emailAdd{wojciech.kotlarski@fuw.edu.pl}
\emailAdd{dominik.stoeckinger@tu-dresden.de}

\abstract{R-symmetry is a fundamental symmetry which can solve the
   SUSY flavor problem and relax the search limits on SUSY
  masses. Here we provide a complete next-to-leading order computation
  and discussion of the lightest Higgs boson mass, the W boson mass
  and muon decay in 
the minimal R-symmetric SUSY model (MRSSM). This model contains non-MSSM
particles including a Higgs triplet, Dirac gauginos and higgsinos,
and leads to significant
new tree-level and one-loop contributions to these observables. We
show that the model can accommodate the measured values of the
observables for interesting regions of parameter space with stop masses of order 1 TeV in spite of the absence of stop mixing.  We
characterize these regions and provide typical benchmark points, which
are also checked against further experimental constraints. A detailed exposition of the model, its mass matrices and its 
Feynman rules relevant for computations in this paper is also provided.}

\begin{document} 
\maketitle
\flushbottom

\newpage
\section{Introduction}
Supersymmetry is one of the most attractive concepts for physics
beyond the Standard Model (SM). The Haag--\L{}opusza\'nski--Sohnius theorem
states that supersymmetry is the only possible non-trivial extension
of the Poincar\'e algebra in a relativistic quantum field
theory \cite{HLS}. However, this theorem allows for a
continuous R-symmetry  \cite{Salam:1974xa,Fayet:1974pd}, an internal symmetry which 
does not commute
with supersymmetry, but such a continuous R-symmetry is not present in
models such as the minimal supersymmetric standard model (MSSM). A
continuous R-symmetry forbids Majorana gaugino masses, so it is
phenomenologically only viable provided the gaugino masses are Dirac
masses, which in turn implies that the gauge/gaugino sector is the one
of an $N=2$ supersymmetric theory \cite{Fayet:1984wm,Fayet:1985kt}, i.e.\ every 
gauge boson has a Dirac
gaugino and a scalar superpartner. Properties of Dirac gauginos have
already been extensively investigated independently of R-symmetry in the
literature \cite{Fox:2002bu,Abel:2011dc,Benakli:2011vb,Kribs:2013oda}, in particular
the required theory of supersymmetry breaking and $\beta$ functions
have been developed \cite{Jack:1999ud,Goodsell:2012fm,{Benakli:2008pg},{Benakli:2010gi}}. 

Thus supersymmetry with R-symmetry is
motivated and distinct from the MSSM since it contains a continuous
R-symmetry, Dirac gauginos and an $N=2$ sector. A minimal
viable model for this, the MRSSM, has been constructed in
ref.\ \cite{Kribs:2007ac}; for models with different R-charges see 
refs.\ \cite{Davies:2011mp,Frugiuele:2011mh,Frugiuele:2012kp,Riva:2012hz,Chakraborty:2013gea}. Excitingly,
the MRSSM alleviates some of the most important experimental
constraints on the MSSM: the R-symmetry forbids large contributions to
flavor-violating observables, so the MRSSM is generically in
agreement with flavor data even for an anarchic flavor structure in
the sfermion sector and for sfermion masses below the TeV
scale   \cite{Kribs:2007ac,Fok:2010vk}. Similarly, heavy Dirac gluinos
suppress the production cross section for squarks, making squarks
below the TeV scale generically compatible with LHC
data  \cite{Kribs:2012gx}. Furthermore, models with R-symmetry and/or
Dirac gauginos contain promising dark matter
candidates \cite{Buckley:2013sca,Chun:2009zx,Belanger:2009wf}, and the collider
physics of the extra, non-MSSM-like states has been studied 
\cite{Plehn:2008ae,Choi:2009jc,Choi:2009ue,Choi:2010gc,Choi:2010an,Kalinowski:2010ii,Kalinowski:2011zzc,Kotlarski:2011zz,Kotlarski:2013lja,Benakli:2012cy}.

In this paper we investigate a question of obvious importance after the Higgs
boson discovery at the LHC: Is the MRSSM compatible with the measured
Higgs boson mass of 125~GeV, and what are the implications of this
measurement on the MRSSM parameter space? The answer is not
immediately obvious since at tree level the lightest Higgs boson mass
is typically reduced compared to the MSSM, due to mixing with
additional scalar states. Furthermore, an important enhancement mechanism
in the MSSM --- large stop mixing --- is not available due to
R-symmetry. Therefore, additional loop effects
have to increase the Higgs boson mass. 

The parameters entering the Higgs
boson mass prediction also affect electroweak precision observables,
in particular the W boson mass $m_W$. In addition, R-symmetry  necessarily introduces 
an SU(2) scalar triplet, which can increase $m_W$ already at tree level.   Hence we also study
constraints on the model parameters and the Higgs boson mass from
$m_W$. A similar set of questions has been studied recently in
ref.\ \cite{Gregoire}, with the affirmative answer that the MRSSM is
in agreement with the measured Higgs boson mass. Our study
extends the one of that reference. We take into account the full
one-loop corrections to both observables and perform extensive
investigations of the parameter space. We define benchmark points
illustrating different viable parameter regions and also verify that
further theoretical and experimental constraints from Higgs, collider
and low-energy physics are satisfied.

The following section presents the model in detail.  All relevant mass
and mixing matrices are provided, and the differences to the MSSM are
explained. A complementary Appendix A presents useful Feynman
rules. Section \ref{ch:Mh} describes the computation of the Higgs
boson mass and presents a comparison of the full one-loop calculation with an approximate
result for the leading MRSSM-specific contributions. Even though stops do not mix in 
the MRSSM  a 125 GeV Higgs mass can be obtained with stop masses 
below 1 TeV. Section
\ref{ch:mW} provides a detailed discussion of the W boson mass and the
leading contributions to the $T$-parameter; the details of the calculations are relegated 
to Appendix B. Our main results are given
in section \ref{ch:results}. The parameter dependence of both
observables is discussed, viable parameter regions are identified and
the detailed predictions for a set of benchmark points are given.

\section{Model definition}
\label{ch:ModDef}
\subsection{Field content, superpotential and soft masses}
\label{ch:basics}
If we assign R-charges to the MSSM superfields in such a way that all Standard Model particles have R=0 
(in analogy to discrete R-parity), Majorana gaugino, the $\mu$-term and higgsino mass terms are forbidden.  
Therefore, imposing R-symmetry on the MSSM requires an
enlarged field content to account for non-vanishing gaugino and higgsino masses.  
The minimal R-symmetric  extension of the supersymmetric model (MRSSM),
 which we will analyze, consists of the standard
MSSM matter, Higgs and gauge superfields augmented by adjoint chiral
superfields $\hat{\cal O},\hat{T},\hat{S}$ with R-charge 0 for each gauge sector, and two Higgs iso-doublet superfields
${\hat{R}}_{d,u}$ with R-charge 2. 
The R-charges of the MRSSM superfields and their component
fields are as listed in table~\ref{tab:Rcharges}, if by convention the fermionic coordinates  $\theta,  \, \bar{\theta}$ have R-charge $+1,\,-1$. 
%%%%%%%%%%%%%%%%%%%%%%%%%%%%%%%%%%%%%%%%%%%%%%%%%%%%%%%%%%%%%%%%%%%%%%%%%%%%%
\begin{table}[th]
\begin{center}
\begin{tabular}{c|l|l||l|l|l|l}
%\hline
\multicolumn{1}{c}{Field} & \multicolumn{2}{c}{Superfield} &
                              \multicolumn{2}{c}{Boson} &
                              \multicolumn{2}{c}{Fermion} \\
\hline 
 \phantom{\rule{0cm}{5mm}}Gauge Vector    &\, $\hat{g},\hat{W},\hat{B}$        \,& \, $\;\,$ 0 \,
          &\, $g,W,B$                 \,& \, $\;\,$ 0 \,
          &\, $\tilde{g},\tilde{W}\tilde{B}$             \,& \, +1 \,  \\
Matter   &\, $\hat{l}, \hat{e}$                    \,& \,\;+1 \,
          &\, $\tilde{l},\tilde{e}^*_R$                 \,& \, +1 \,
          &\, $l,e^*_R$                                 \,& $\;\;\,$\,\;0 \,    \\
          &\, $\hat{q},{\hat{d}},{\hat{u}}$       \,& \,\;+1 \,
          &\, $\tilde{q},{\tilde{d}}^*_R,{\tilde{u}}^*_R$ \,& \, +1 \,
          &\, $q,d^*_R,u^*_R$                             \,& $\;\;\,$\,\;0 \,    \\
 $H$-Higgs    &\, ${\hat{H}}_{d,u}$   \,& $\;\;\,$\, 0 \,
          &\, $H_{d,u}$               \,& $\;\;\,$\, 0 \,
          &\, ${\tilde{H}}_{d,u}$     \,& \, $-$1 \, \\ \hline
\phantom{\rule{0cm}{5mm}} R-Higgs    &\, ${\hat{R}}_{d,u}$   \,& \, +2 \,
          &\, $R_{d,u}$               \,& \, +2 \,
          &\, ${\tilde{R}}_{d,u}$     \,& \, +1 \, \\
  Adjoint Chiral  &\, $\hat{\cal O},\hat{T},\hat{S}$     \,& \, $\;\,$ 0 \,
          &\, $O,T,S$                \,& \, $\;\,$ 0 \,
          &\, $\tilde{O},\tilde{T},\tilde{S}$          \,& \, $-$1 \,  \\
%\hline
\end{tabular}
\end{center}
\caption{The R-charges of the superfields and the corresponding bosonic and
             fermionic components.
        }
\label{tab:Rcharges}
\end{table}
%%%%%%%%%%%%%%%%%%%%%%%%%%%%%%%%%%%%%%%%%%%%%%%%%%%%%%%%%%%%%%%%%%%%%%%%%%%%%%

With the above assignments the gauge, matter-gauge and Higgs-gauge actions
of the MSSM are R-invariant, as well as the action associated with the
trilinear Yukawa terms $Y_d \,\hat{d}\,\hat{q}\,\hat{H}_d$ etc
in the superpotential. In contrast, the standard $\mu$ term is forbidden by the R-symmetry. 
However, the presence of  two iso-doublet superfields $\hat{R}_u$ and $\hat{R}_d$  with R-charge 2  
together with the standard iso-doublets $\hat{H}_u$ and $\hat{H}_d $ allows to generate R-symmetric $\mu$-type terms and the corresponding higgsino masses,
\begin{align}\label{eq:muterm}
W \;\ni \mu_d\,   \hat{R}_d \cdot \hat{H}_d         +\mu_u\,  \hat{R}_u\cdot \hat{H}_u.
\end{align}
The appearance of the  supersymmetric Higgs-higgsino mass term in eq.~(\ref{eq:muterm}) can be thought of as arising from the R-symmetric operator involving the hidden sector F-type spurion $X=\theta^2 F$
\begin{equation}
\int d^4 \theta\; \left(\frac{X^\dagger}{M} \hat{R}_d \cdot \hat{H}_d   +   \frac{X^\dagger}{M}  \hat{R}_u\cdot \hat{H}_u \right)	\ni \mu_d\,   \hat{R}_d\cdot \hat{H}_d         +\mu_u\,  \hat{R}_u\cdot \hat{H}_u.
\end{equation}
Similarly, the trilinear couplings of the electroweak $\hat{T},\,\hat{S}$ adjoint chiral superfields to the Higgs doublets can be generated,
\begin{equation}
W\ni\;  \Lambda_d \hat{R}_d\cdot\hat{T}\,\hat{H}_d\,+\lambda_d\,\hat{S}\,\hat{R}_d\cdot\hat{H}_d \;\; +(d\to u)\,,
\end{equation}
where the dot denotes $\epsilon$ contraction with $\epsilon_{12} = +1$ and where the triplet is defined as
\begin{equation}
\hat{T}=
\begin{pmatrix}
\hat{T}_0/\sqrt{2}&\hat{T}_+\\
\hat{T}_-&-\hat{T}_0/\sqrt{2}\\
\end{pmatrix}
\,,
\end{equation} to have canonical kinetic terms. 
As we will see, these trilinear couplings will play an important role in  obtaining a Higgs boson mass close to 125 GeV.

 Thus the MRSSM superpotential takes the form of
\begin{align}
\nonumber W = & \mu_d\,\hat{R}_d \cdot \hat{H}_d\,+\mu_u\,\hat{R}_u\cdot\hat{H}_u\,+\Lambda_d\,\hat{R}_d\cdot \hat{T}\,\hat{H}_d\,+\Lambda_u\,\hat{R}_u\cdot\hat{T}\,\hat{H}_u\,\\ 
 & +\lambda_d\,\hat{S}\,\hat{R}_d\cdot\hat{H}_d\,+\lambda_u\,\hat{S}\,\hat{R}_u\cdot\hat{H}_u\,
 - Y_d \,\hat{d}\,\hat{q}\cdot\hat{H}_d\,- Y_e \,\hat{e}\,\hat{l}\cdot\hat{H}_d\, +Y_u\,\hat{u}\,\hat{q}\cdot\hat{H}_u\, .
\label{eq:superpot}
 \end{align} 

Note that R-symmetry also forbids all baryon- and lepton-number violating terms in the superpotential 
(as well as dimension-five operators mediating proton decay) \cite{Weinberg:1979sa,Sakai:1981pk} 
and that the R-symmetry is being preserved when the electroweak symmetry is broken.  

Turning to soft breaking, the usual soft mass terms of the MSSM scalar fields are allowed just like in the MSSM. The MSSM A-terms, however, are zero due to R-symmetry.
Importantly, soft Majorana gaugino mass terms are also forbidden by R-symmetry, but the fermionic components of the chiral adjoints, $\hat{\Phi}_i=\hat{\cal O},\,\hat{T}, \,\hat{S}$ 
for each standard model gauge group $i=SU(3)$, $SU(2)$, $U(1)$ respectively, can be paired with standard gauginos $\tilde{g},\tilde{W},\tilde{B}$ 
to build Dirac fermions and the corresponding mass terms. The Dirac masses for the gauginos can be considered as elements of a general supersymmetric 
theory broken softly by hidden sector spurions $W'_\alpha=\theta_\alpha D$ from a hidden sector $U(1)'$ that acquires a D-term.  
The Dirac gaugino mass is generated from the R-symmetric operator involving the D-type spurion \cite{Fox:2002bu},
\begin{equation}
\label{eq:mdirac}
\int d^2\theta\frac{\hat{W'_\alpha}}{M} W_i^\alpha \hat{\Phi}_i  \ni M_i^D\, \tilde{g}_i \tilde{g}'_i,
\end{equation}
where M is the mediation scale of SUSY breaking from the hidden sector to the visible sector, $W_i^\alpha$ represents the gauge superfield strength tensors, $\tilde{g}_i=\tilde{g},\tilde{W},\tilde{B}$ is the gaugino, and $\tilde{g}'_i=\tilde{O},\tilde{T},\tilde{S}$  is the corresponding Dirac partner with opposite R-charge.  
Note that integrating out the spurion in eq.~\eqref{eq:mdirac} generates additional terms with the D-fields and the scalar components, which leads to the appearance of Dirac masses in the scalar sector as well,
\begin{align}
V_D= \, & M_B^D (\tilde{B}\,\tilde{S}-\sqrt{2} \mathcal{D}_B\, S)+
M_W^D(\tilde{W}^a\tilde{T}^a-\sqrt{2}\mathcal{D}_W^a T^a)+
M_g^D(\tilde{g}^a\tilde{O}^a-\sqrt{2}\mathcal{D}_g^a O^a)
 %-\sqrt{2} (M_B^D \mathcal{D}_B S + M_W^D \mathcal{D}_W^a T^a+ M_g^D \mathcal{D}_B^a O^a)
+ \mbox{h.c.}\,.
\label{eq:potdirac}
\end{align}

In a similar fashion, the soft  SUSY breaking $B_\mu$, the Higgs and R-Higgs scalar masses can be generated  by R-symmetric interactions with the hidden sector spurion $X$ 
\begin{align}\label{eq:Bmu}
&\int d^4 \theta \;\frac{X^\dagger \hat X}{M^2} \hat{H}_d\cdot \hat{H}_u \; \ni \; B_\mu H_d \cdot H_u, \\
&\int d^4 \theta \;\frac{X^\dagger \hat X}{M^2} \hat{R}_d^\dagger \hat{R}_d \; \ni \; m^2_{R_d}
( |R^+_d|^2+|R^0_d|^2),  \label{eq:Rmass}
\;\; \text{and }d\rightarrow u\, ,
\end{align}
as well as 
the SUSY breaking adjoint scalar mass terms can be generated ($\Phi_i=O,T,S$) 
\begin{align}
\int d^4 \theta \;\frac{X^\dagger \hat X}{M^2} \hat{\Phi}_i^\dagger \hat{\Phi}_i \; \ni \; m^2_i
|{\Phi}_i|^2.  \label{eq:nonholo}
\end{align}
suggesting that $\mu_{d,u}$, $B_\mu$ and soft scalar masses could be of the same order, $i.e.$  of the order of SUSY breaking scale.
Note that the bilinear coupling of the R-Higgs
fields, like in eq.~(\ref{eq:Bmu}), has  R-charge  4 and therefore is forbidden in the
R-symmetric theory. Thus only the $B_\mu$ term destroys the exchange
symmetry between the $H$ and R-Higgs fields. 

For completeness, it should be noticed that bilinear couplings of the
 $\hat{X}$ fields with the adjoint chiral fields can also
generate contributions 
\begin{align}
\int d^4 \theta \;\frac{X^\dagger \hat X}{M^2} Tr[\hat{\Phi}_i \hat{\Phi}_i] \; \ni \; B_i
{\Phi}_i {\Phi}_i.  \label{eq:holo}
\end{align}
 to the (holomorphic) soft masses of the adjoint scalars
in addition to the above soft terms. Such terms split scalar and pseudoscalar masses squared by equal amounts. If the pseudoscalar gets a negative contribution to its mass squared, 
it may generate color- or charge-breaking minima.  Although there is no symmetry that allows non-holomorphic (\ref{eq:nonholo}), 
but forbids holomorphic (\ref{eq:holo}) mass terms, these terms are technically independent \cite{Fox:2002bu}. In the following
we will not study theoretical points such as SUSY breaking mechanisms, but
explore only the phenomenology of the R-symmetric low-energy theory.
Therefore, for simplicity we will neglect holomorphic mass terms for $S$ and $T$ adjoint scalars, as well as their 
trilinear couplings among themselves and to the Higgs bosons since their presence does not influence our results significantly (however, see also \cite{Gregoire}). 
Thus the soft-breaking scalar mass terms read
\begin{align}
V^{EW}_{SB}= \, &  B_{\mu}(H_d^- H_u^+- H_d^0 H_u^0 ) + \mbox{h.c.} \nonumber
\\%[2mm] 
\, & +m_{H_d}^2 (|H_d^0|^2 + |H_d^-|^2) +m_{H_u}^2 (|H_u^0|^2 + |H_u^+|^2) 
\nonumber  \\ &
+m_{R_d}^2 (|R_d^0|^2 + |R_d^+|^2)   +m_{R_u}^2 |R_u^0|^2+m_{R_u}^2 |R_d^-|^2 \nonumber
\\ &
+m_S^2 |S|^2 +m_T^2 |T^0|^2 +m_T^2 |T^-|^2 +m_T^2 |T^+|^2 +m_O^2 |O|^2\nonumber 
\\
  &+\tilde{d}^*_{L,{i }}  m_{q,{i j}}^{2} \tilde{d}_{L,{j }} 
 +\tilde{d}^*_{R,{i }}  m_{d,{i j}}^{2} \tilde{d}_{R,{j }}  +\tilde{u}^*_{L,{i }} m_{q,{i j}}^{2} \tilde{u}_{L,{j }} +\tilde{u}^*_{R,{i }}  m_{u,{i j}}^{2} \tilde{u}_{R,{j }}
 \nonumber \\ 
 & +\tilde{e}^*_{L,{i}} m_{l,{i j}}^{2} \tilde{e}_{L,{j}} +\tilde{e}^*_{R,{i}} m_{e,{i j}}^{2} \tilde{e}_{R,{j}} +\tilde{\nu}^*_{L,{i}} m_{l,{i j}}^{2} \tilde{\nu}_{L,{j}} \,.
\label{eq:othersoftpot}
\end{align}
Note that R-symmetry forbids all trilinear scalar couplings involving Higgs bosons to squarks and sleptons, which in the MSSM are notoriously unwanted sources of flavor violation. Since in this paper we do not discuss lepton flavor violating phenomena, for simplicity we assume that slepton mass matrices are diagonal. 

\subsection{Tree-level Higgs boson masses}
\label{ch:treehiggs}
The electroweak part of the tree-level scalar potential for the neutral fields takes the form of
\begin{align}%multline} 
\label{eq:scalarpot} %1
 & V^{EW}=   \left(m_{H_d}^2+\mu_d^2 \right) \left|H_d^{0} \right|^2  
+ \left(m_{H_u}^2+\mu_u^2 \right) \left|H_u^{0} \right|^2  -B_\mu \left(H_d^{0} H_u^{0}+h.c. \right) 
%\notag
\\ %2
& + \left(m_{R_d}^2 + \mu_d^2 \right) \left|R_d^{0} \right|^2
+\left(m_{R_u}^2 + \mu_u^2 \right) \left|R_u^{0} \right|^2 + m_T^2 \left|T^{0} \right|^2 +     m_S^2 \left|S \right|^2 \notag
\\ 
& +  \left(2 M^D_W\Re \left(T^0 \right)\right)^2  - \left(2 M_B^D \Re(S) \right)^2 
+ \textstyle{\frac{1}{8}}(g_1^2  +  g_2^2)\left(  \left |H_d^{0}\right|^2- \left |H_u^{0}\right|^2 
-\left |R_d^0\right|^2+\left |R_u^0\right|^2\right)^2
\notag 
\\ %4
& - \left[\left({g_1} M^D_B   -  \sqrt{2}\lambda_d \mu_d \right)\sqrt{2}\Re(S)  - 
 \left({g_2}M^D_W + {\Lambda_d \mu_d}\right)\sqrt{2}\Re \left( T^0 \right) 
 -\left|\lambda_d S+\textstyle{\frac{1}{\sqrt{2}}}{\Lambda_d} T^0 \right|^2\right] 
 \left|H_d^{0} \right|^2  
\notag 
\\  %5
& +  \left[\left({g_1} M^D_B   - \sqrt{2} \lambda_u \mu_u \right)\sqrt{2}\Re(S)
 -\left({g_2} M^D_W  + \Lambda_u \mu_u \right)\sqrt{2}\Re \left(T^{0} \right) 
 +\left|\lambda_u S-\textstyle{\frac{1}{\sqrt{2}}}\Lambda_u T^0 \right|^2 \right]
 \left|H_u^{0} \right|^2  
 \notag 
 \\  %6
& + \left[
  \left(g_1 M^D_B +  \sqrt{2}   \lambda_d \mu_d \right) \sqrt{2} \Re(S)  
 - \left(g_2 M^D_W   -\Lambda_d \mu_d \right) \sqrt{2}\Re \left(  T^{0} \right)
+\left|\lambda_d S+\textstyle{\frac{1}{\sqrt{2}}}\Lambda_d T^0 \right|^2\right] \left|R_d^{0} \right|^2 
\notag
\\  %7
 & - \left[ 
  \left(g_1 M^D_B   - \sqrt{2}  \lambda_u \mu_u \right) \sqrt{2} \Re( S) 
 - \left(g_2 M^D_W  - \Lambda_u \mu_u \right)\sqrt{2}\Re \left( T^{0} \right)
  - \left|\lambda_u S-\textstyle{\frac{1}{\sqrt{2}}}\Lambda_u T^0 \right|^2
  \right] \left|R_u^{0} \right|^2 
\notag
\\  %8
& + \left( \lambda_d^2 + \textstyle{\frac{1}{2}}\Lambda_d \right) \left|H_d^{0} R_d^{0}\right|^2 
 + \left( \lambda_u^2 + \textstyle{\frac{1}{2}}\Lambda_u^2  \right) \left|H_u^{0} 
 R_u^{0} \right|^2  
 \notag
   -  \left(\lambda_d \lambda_u -  \textstyle{\frac{1}{2}}\Lambda_d \Lambda_u \right) 
    \left(H_d^{0} R_d^{0} H_u^{0\dagger} R_u^{0\dagger}+ \text{h.c.} \right) 
  \end{align}%multline}

\newcommand{\muu}[1]{\mu_u^{\text{eff,}#1}}
\newcommand{\mud}[1]{\mu_d^{\text{eff,}#1}}

Since R-Higgs bosons carry R-charge 2, they do not develop vacuum expectation values. Therefore, for electroweak symmetry breaking (EWSB)
we write the $R=0$ neutral EW scalars as
\begin{align*} 
H_d^0=& \, \frac{1}{\sqrt{2}} (v_d + \phi_{d}+i  \sigma_{d}) \;,& 
H_u^0=& \, \frac{1}{\sqrt{2}} (v_u + \phi_{u}+i  \sigma_{u}) \;,\\ 
T^0  =& \, \frac{1}{\sqrt{2}} (v_T + \phi_T +i  \sigma_T)   \;,&
S   = & \, \frac{1}{\sqrt{2}} (v_S + \phi_S +i  \sigma_S)   \;. 
\end{align*} 
After EWSB the singlet and triplet vacuum expectation values
effectively modify the $\mu$-parameters of the model, and it is useful
to define the abbreviations
\begin{align}
\mu_i^{\text{eff,}\pm}&
=\mu_i+\frac{\lambda_iv_S}{\sqrt2}
\pm\frac{\Lambda_iv_T}{2}
,
&\mu_i^{\text{eff,}0}&
=\mu_i+\frac{\lambda_iv_S}{\sqrt2}
,&i=u,d.
\end{align}

The minimization conditions for the scalar potential, or tadpole equations, then read
\begin{equation}\label{eq:tadpoles}
0= t_{d} = t_{u} = t_{T} = t_{S},
\end{equation}
where the tree-level tadpoles $t_i\equiv\frac{\partial
V^{EW}}{\partial \phi_i}$ are
\begin{align} 
{t_{d}} =&v_d [\textstyle{\frac{1}{8}}\left(g_1^2+g_2^2\right) \left(v_d^2-v_u^2\right)- g_1 M_B^D v_S + g_2 M_W^D v_T+m_{H_d}^2+ (\mud{+})^2]- v_u B_\mu\;, \notag\\
{t_{u}}=&v_u [\textstyle{\frac{1}{8}}\left(g_1^2+g_2^2\right) \left(v_u^2-v_d^2\right)
+g_1 M_B^D v_S- g_2 M_W^D v_T+m_{H_u}^2+ (\muu{-})^2]- v_d B_\mu\;, \notag\\
{t_{T}}=& \textstyle{\frac{1}{2}} g_2
M_W^D \left(v_d^2-v_u^2\right)+
\textstyle{\frac{1}{2}}\Lambda_d v_d^2  \mud{+} -\Lambda_uv_u^2 \muu{-} 
+4 (M_W^D)^2 v_T+ m_T^2 v_T\;, \notag\\
{t_{S}}=& \textstyle{\frac{1}{2}}g_1
M_B^D \left(v_u^2-v_d^2\right)+
\textstyle{\frac{1}{\sqrt{2}}} \lambda_d v_d^2 \mud{+}
+ \lambda_uv_u^2 \muu{-} +4 (M_B^D)^2 v_S+ m_S^2 v_S\;.
\label{eq:tadp}
\end{align}

The tadpole eqs.~\eqref{eq:tadpoles} are used to substitute $m_{H_d}^2$ and $m_{H_u}^2$ 
as usual with $v_d$ and $v_u$, 
introducing $v=\sqrt{v_u^2+v_d^2}$ and $\tan\beta=v_u/v_d$  as in the MSSM.
The other two equations are solved for $v_T$ and $v_S$, allowing us to use $m_S^2$ and $m_T^2$ as input parameters. 
This can be done analytically, but the
expressions have no simple form  and will 
not be written here. 
Because of this, we will use those equations in the following
analytical analyses only in cases where they simplify the expressions
significantly,  as they do in the case of the pseudo-scalar Higgs mass matrix, 
and only to some extent in the case of the scalar Higgs one. 
In other cases, the appearing $v_T$ and $v_S$
are always functions of the other model parameters and this is taken into account in numerical calculations.

Masses for the gauge bosons arise in the usual form for the Z boson, but the W boson has a
contribution from the triplet vev:
\begin{equation}
m_Z^2 = \frac{g^2_1+g^2_2}{4} v^2\;,\qquad m_W^2= \frac{g^2_2}{4} v^2+g^2_2 v_T^2\;,\qquad 
\hat\rho_{\text{tree}} = 1 + \frac{4 v_T^2}{v^2}\;.
\label{eq:bosonmasses}
\end{equation}
This will shift the $\rho$ parameter away from one already at the tree level and gives an upper limit on
$\left | v_T \right|\lesssim 4$ GeV, which is taken into account when studying parameter points. 
Further consequences at higher orders will be investigated in section \ref{ch:mW}.

Using the tadpole eqs.~\eqref{eq:tadp}, the pseudo-scalar Higgs boson mass matrix in the basis $(\sigma_d,\sigma_u,\sigma_S,\sigma_T)$ 
has a simple form (in Landau gauge)
\begin{equation}
\mathcal{M}_A=
\begin{pmatrix}
 B_\mu \frac{v_u}{v_d} & B_\mu & 0 & 0 \\
 B_\mu &  B_\mu \frac{v_d}{v_u} & 0 & 0 \\
 0 & 0 & m_S^2+\frac{\lambda_d^2 v_d^2+\lambda_u^2 v_u^2 }{2} & \frac{\lambda_d\Lambda_d v_d^2-\lambda_u\Lambda_u v_u^2}{2 \sqrt{2}} \\
 0 & 0 & \frac{\lambda_d\Lambda_d v_d^2-\lambda_u\Lambda_u v_u^2}{2 \sqrt{2}}& m_T^2+ \frac{\Lambda_d^2 v_d^2+\Lambda_u^2 v_u^2 }{4}\\
\end{pmatrix}
\;.
\end{equation}
Due to the absence of mixing terms, the MSSM-like upper left and the singlet-triplet bottom right
sub matrices are independent of each other. Therefore, the neutral Goldstone boson and
one of the pseudo-scalar Higgs bosons with $m_A^2=2B_\mu/\sin2\beta$  will be as in the MSSM while the other two
are singlet- and triplet-like.
To ensure that EWSB occurs correctly, meaning no tachyons arise when diagonalizing the mass matrices, we restrict the values of $m_S^2$ and $m_T^2$ 
to always being positive.

The scalar Higgs boson mass matrix in the weak basis $(\phi_d,\phi_u,\phi_S,\phi_T)$  is given by
\begin{equation}\label{eq:scalarmassmatrix}
\mathcal{M}_{H^0}=
\begin{pmatrix}
\mathcal{M}_{\text{MSSM}}
&\mathcal{M}_{21}^T\\
\mathcal{M}_{21} &
\mathcal{M}_{22}
\\
\end{pmatrix}
\end{equation}
with the  sub-matrices ($c_\beta=\cos\beta$, $s_\beta=\sin\beta$, etc)
\begin{align}
\mathcal{M}_{\text{MSSM}}&= 
\begin{pmatrix}
 m_Z^2 c_\beta^2+m_A^2 s_\beta^2 \; & -(m_Z^2 + m_A^2)s_\beta c_\beta  \\
  -(m_Z^2 + m_A^2)s_\beta c_\beta \; &  m_Z^2 s_\beta^2+m_A^2 c_\beta^2\\
\end{pmatrix}
\;, \notag\\
\mathcal{M}_{22}&= 
\begin{pmatrix} 4 (M_B^D)^2+m_S^2+\frac{\lambda_d^2 v_d^2+\lambda_u^2 v_u^2}{2} \;
& \frac{\lambda_d \Lambda_d v_d^2-\lambda_u \Lambda_u v_u^2}{2 \sqrt{2}} \\
 \frac{\lambda_d \Lambda_d v_d^2-\lambda_u \Lambda_u v_u^2}{2 \sqrt{2}} \;
 & 4 (M_W^D)^2+m_T^2+\frac{\Lambda_d^2 v_d^2+\Lambda_u^2 v_u^2}{4}\\
\end{pmatrix}
\,, \notag\\
\mathcal{M}_{21}&= 
\begin{pmatrix}
 v_d ( \sqrt{2}\lambda_d \mud{+} -g_1 M_B^D )\; & \;
v_u (\sqrt{2} \lambda_u\muu{-} +g_1 M_B^D) \\
v_d ( \Lambda_d \mud{+} + g_2 M_W^D) \;& - 
 v_u (\Lambda_u  \muu{-} + g_2 M_W^D) \\
\end{pmatrix}
\;.\notag
\end{align}
       
Again, $\mathcal{M}_{\text{MSSM}}$ is MSSM-like and $\mathcal{M}_{22}$ singlet-triplet-like, but 
compared to the pseudo-scalar case, additional mixing exists through $\mathcal{M}_{21}$. 
Generally, this will lead to a reduction of the lightest Higgs boson mass at tree-level compared to the MSSM.
An approximate formula for the reduction can be given for the lightest Higgs boson mass when using the MSSM mixing angle $\alpha$ to diagonalize 
$\mathcal{M}_{\text{MSSM}}$ for large $m_A^2$ when $\alpha=\beta-\pi/2$, 
further assuming $\lambda=\lambda_u=-\lambda_d$, $\Lambda=\Lambda_u=\Lambda_d$, $\mu_u=\mu_d=\mu$ and $v_S\approx v_T\approx0$: 
\begin{equation}
m_{H_1,\text{approx}}^2 = m_Z^2 \cos^2 2\beta - v^2 \left(
\frac{\left(g_1 M^D_B+\sqrt{2}\lambda\mu\right)^2}{4(M^D_B)^2+ m_S^2}
+
\frac{\left(g_2 M^D_W+\Lambda\mu\right)^2}{4(M_W^D)^2+ m_T^2}
\right) \cos^2 2\beta\;.
\label{eq:approx_treehiggs}
\end{equation}
It can be seen that the upper limit of the MSSM at the tree-level is reduced by terms depending on
the new model parameters. Thus one-loop contributions are important, and a
detailed study will be presented in section \ref{ch:Mh}. 

The charged Higgs boson mass matrix in the weak basis $(H^{-*}_d,H^{+}_u,T^{-*},T^+)$  is given as
\begin{equation}
\mathcal{M}_{H^\pm}=
\begin{pmatrix}
\mathcal{M}_{\text{MSSM},\pm}
&\mathcal{M}_{21,\pm}^T\\
\mathcal{M}_{21,\pm} &
\mathcal{M}_{22,\pm}
\\
\end{pmatrix}
\end{equation}
with the  sub-matrices
\begin{align}
\mathcal{M}_{\text{MSSM},\pm}&= 
\begin{pmatrix}
 m_{H^\pm}^2 c_\beta^2-2 v_T (g_2 M^D_W+\Lambda_d\mud{0})\; &
  m_{H^\pm}^2s_\beta c_\beta  \\
  m_{H^\pm}^2s_\beta c_\beta \; &  m_{H^\pm}^2c_\beta^2+2 v_T (g_2 M^D_W+ \Lambda_u\muu{0})\\
\end{pmatrix}
\;, \notag\\
\mathcal{M}_{22,\pm}&=
\begin{pmatrix} 
  2 (M^D_W)^2+ m_T^2 \;
 & 2 (M^D_W)^2 \\
 2 (M^D_W)^2 \;
 & 2 (M^D_W)^2+ m_T^2\\
\end{pmatrix}
 \\
&+\begin{pmatrix} 
\frac{ g_2^2 v_T^2+\Lambda_d^2 v^2c_\beta^2}{2}
 -\frac{g_2^2v^2\cos 2 \beta }{4} \;
 &-\frac{g_2^2 v_T^2}{2} \\
 -\frac{g_2^2 v_T^2}{2}\; 
 &
\frac{ g_2^2 v_T^2+\Lambda_u^2 v^2s_\beta^2}{2}
 +\frac{g_2^2v^2 \cos 2 \beta}{4} \\
\end{pmatrix}
\;, \notag\\
\mathcal{M}_{21,\pm}&=
\begin{pmatrix}
\frac{ v_d}{2\sqrt{2}}
(2\Lambda_d \mud{-}+2{g_2 M^D_W }+{v_T g_2^2}) \; &
\frac{ v_u}{2\sqrt{2}}
(2\Lambda_u \muu{-}+2{ g_2 M^D_W}+{v_T g_2^2}) \\
\frac{ v_d}{2\sqrt{2}}
(2\Lambda_d \mud{+}+2{ g_2 M^D_W}-{v_T g_2^2}) \;  &
\frac{ v_u}{2\sqrt{2}}
(2\Lambda_u\muu{+}+2{ g_2 M^D_W}-v_T g_2^2 ) \\
\end{pmatrix},
\;\notag
\end{align}
where the charged Higgs mass parameter $m^2_{H^{\pm}}=m^2_A+\frac{g_2^2v^2}{4}$ reads as in the MSSM.

The mass matrix of neutral R-Higgs bosons is
\begin{equation}
\mathcal{M}_R = \left (
\begin{matrix}
  m^2_{R_d R_d} & \frac{1}{4} \left ( \Lambda_u \Lambda_d - 2 \lambda_u \lambda_d \right ) v_u v_d \\
  \frac{1}{4} \left ( \Lambda_u \Lambda_d - 2 \lambda_u \lambda_d \right ) v_u v_d & m^2_{R_u R_u}
\end{matrix}
\right )
\end{equation}
with
\begin{align*}
 &m^2_{R_d R_d}=
 m_{R_d}^2+ (\mud{+})^2+g_1 M^D_B v_S- g_2M^D_W v_T
+\textstyle{\frac{1}{8}} [(g_1^2+g_2^2) (v_u^2- v_d^2)+4 \lambda_d^2 v_d^2+2 \Lambda_d^2 v_d^2],\\
&m^2_{R_u R_u}=m_{R_u}^2+(\muu{-})^2- g_1 M^D_B v_S+ g_2 M^D_W v_T
+\textstyle{\frac{1}{8}} [(g_1^2 +g_2^2 )(v_d^2-v_u^2)+4 \lambda_u^2 v_u^2+2 \Lambda_u^2  v_u^2].\\
\end{align*}
It is diagonalized by an orthogonal matrix with mixing angle $\theta_R$.

Because of the R-symmetry the charged R-Higgs bosons do not mix  
and the mass eigenvalues are given as
\begin{align}
& m^2_{R^+_1}=
 m_{R_d}^2+  (\mud{-})^2+g_1 M^D_B v_S+ g_2M^D_W v_T
+\textstyle{\frac{1}{8}} [(g_1^2-g_2^2) (v_u^2- v_d^2)+4 \Lambda_d^2 v_d^2] \notag\\
&m^2_{R^+_2}= m_{R_u}^2 + (\muu{+})^2 - g_1 M^D_B v_S- g_2 M^D_W v_T
+\textstyle{\frac{1}{8}} [(g_1^2 -g_2^2 )(v_d^2-v_u^2)+4 \Lambda_u^2  v_u^2]\;.
\end{align}
\subsection{Tree-level chargino and neutralino masses}
For completeness and to fix the notation, we give below expressions for tree-level masses and mixing matrices of the charginos and neutralinos. 

%The mass matrix of neutralinos 
In the weak basis of eight neutral  electroweak two-component fermions: $ {\xi}_i=({\tilde{B}}, \tilde{W}^0, \tilde{R}_d^0, \tilde{R}_u^0)$,  $\zeta_i=(\tilde{S}, \tilde{T}^0, \tilde{H}_d^0, \tilde{H}_u^0) $ with R-charges $+1$, $-1$ respectively, the neutralino mass matrix takes the form of
\begin{equation} 
\label{eq:neut-massmatrix}
m_{{\chi}} = \left( 
\begin{array}{cccc}
M^{D}_B &0 &-\frac{1}{2} g_1 v_d  &\frac{1}{2} g_1 v_u \\ 
0 &M^{D}_W &\frac{1}{2} g_2 v_d  &-\frac{1}{2} g_2 v_u \\ 
- \frac{1}{\sqrt{2}} \lambda_d v_d  &-\frac{1}{2} \Lambda_d v_d  & - \mud{+} &0\\ 
\frac{1}{\sqrt{2}} \lambda_u v_u  &-\frac{1}{2} \Lambda_u v_u  &0 & \muu{-}
\end{array} 
\right) .
 \end{equation} 
The transformation to a diagonal mass matrix and mass eigenstates $\kappa_i$ and $\psi_i$ is performed by two unitary mixing matrices \(N^1\) and \(N^2\) as
\begin{align} \nonumber
N^{1,*} m_{{\chi}} N^{2,\dagger} &= m^{diag}_{{\chi}} \,,
&
{\xi}_i&=\sum_{j=1}^4 N^{1,*}_{ji}{\kappa}_j\,,
&
\zeta_i=\sum_{j=1}^4 N^{2,*}_{ij}{\psi}_j\,,
\end{align} 
and physical four-component Dirac neutralinos are constructed as
\begin{equation}
{\chi}_i=\left(\begin{array}{c}
\kappa_i\\
{\psi}^{*}_i\end{array}\right)\qquad i=1,2,3,4.
\end{equation}

The mass matrix of charginos in the weak basis of eight charged two-component
fermions breaks  into two $(2\times2)$ submatrices. 
The first, in the
basis \( (\tilde{T}^-, \tilde{H}_d^-), (\tilde{W}^+, \tilde{R}_d^+) \)
of spinors with R-charge equal to electric charge,
 takes the form of
\begin{equation} 
m_{{\chi}^+} = \left( 
\begin{array}{cc}
g_2 v_T  + M^{D}_W \; &\frac{1}{\sqrt{2}} \Lambda_d v_d \\ 
\frac{1}{\sqrt{2}} g_2 v_d \; &+ \mud{-}
\end{array} 
\right) 
\label{eq:cha1-massmatrix}
 \end{equation} 
The diagonalization and transformation to mass eigenstates $\lambda^\pm_i$ is
performed by two unitary matrices \(U^1\) and \(V^1\)  as
\begin{align} 
U^{1,*} m_{{\chi}^+} V^{1,\dagger} &= m^{diag}_{{\chi}^+} \,,
&\tilde{T}^- &= \sum_{j=1}^2U^{1,*}_{j 1}\lambda^-_{{j}}\,,& %\hspace{1cm} 
\tilde{H}_d^- &= \sum_{j=1}^2U^{1,*}_{j 2}\lambda^-_{{j}}\,,\\ 
&&
\tilde{W}^+ &= \sum_{j=1}^2V^{1,*}_{1 j}\lambda^+_{{j}}\,,&% \hspace{1cm} 
R_d^+ &= \sum_{j=1}^2V^{1,*}_{2 j}\lambda^+_{{j}}
\end{align} 
and the corresponding physical four-component charginos are built as 
\begin{equation}
{\chi}^+_i=\left(\begin{array}{c}
\lambda^+_i\\
\lambda^{-*}_i\end{array}\right)\qquad i=1,2.
\end{equation}
The second submatrix, in the basis $ (\tilde{W}^-, R_u^-)$, $(\tilde{T}^+, \tilde{H}_u^+) $
 of spinors with R-charge equal to minus electric charge, reads
\begin{equation} 
m_{{\rho}^-} = \left( 
\begin{array}{cc}
- g_2 v_T  + M^{D}_W \;&\frac{1}{\sqrt{2}} g_2 v_u \\ 
- \frac{1}{\sqrt{2}} \Lambda_u v_u \; &  - \muu{+} \end{array} 
\right) 
\label{eq:cha2-massmatrix}
 \end{equation} 
The diagonalization and transformation to mass eigenstates $\eta^\pm_i$  is
performed by \(U^2\) and \(V^2\) as 
\begin{align} 
U^{2,*} m_{{\rho}^-} V^{2,\dagger} & = m^{diag}_{{\rho}^-} \,,&
\tilde{W}^- & = \sum_{j=1}^2U^{2,*}_{j 1}\eta^-_{{j}}\,, &%\hspace{1cm} 
R_u^- & = \sum_{j=1}^2U^{2,*}_{j 2}\eta^-_{{j}}\\ 
&&\tilde{T}^+ & = \sum_{j=1}^2V^{2,*}_{1 j}\eta^+_{{j}}\,, &%\hspace{1cm} 
\tilde{H}_u^+ & = \sum_{j=1}^2V^{2,*}_{2 j}\eta^+_{{j}}
\end{align} 
and the other two physical four-component charginos are built as 
\begin{equation}
{\rho}^-_i=\left(\begin{array}{c}
\eta^-_i\\
\eta^{+*}_i\end{array}\right)\qquad i=1,2.
\end{equation}
%%%%%%%
%
\subsection{Calculation setup and benchmark points \label{benchmark-points}}
An analysis of the phases of fields and parameters in the superpotential eq.~\eqref{eq:superpot},
the soft-breaking terms, eqs.~\eqref{eq:potdirac} and \eqref{eq:othersoftpot}, and the tadpole eqs.~\eqref{eq:tadp}
leads to the conclusion that it is possible to choose $M_B^D$, $M_W^D$,$M_O^D$, $\mu_d$ and $\mu_u$  
positive using the free phases of $\hat{S}$, $\hat{T}$, $\hat{\cal O}$, $\hat{R}_d$, $\hat{R}_u$. 
It then must be allowed for $v_S$ and $v_T$ to become positive or negative, as  required  by the  tadpole equations. Also the sign of the couplings $\lambda$ and $\Lambda$ is not fixed.
$B_\mu$ can be chosen to be positive by the usual PQ-symmetry.
All other soft masses need to be positive to avoid unwanted charge or color symmetry breaking.
In table \ref{tab:BMP}\footnote{Parameters are defined following the SPA convention \cite{AguilarSaavedra:2005pw}.} we define  benchmark points which represent regions of the MRSSM parameter space with distinct characteristics.  It will be shown in the following
sections that they  are in agreement with experiment and they will serve as starting points for scans of the parameter space. The SM input parameters are \cite{PDG}
\begin{equation}
\begin{aligned}[c]
\hat{\alpha}^{\overline{\text{MS}}}(m_Z) &= 127.940^{-1}\\
G_\mu &= 1.1663787 \times 10^{-5} \text{ GeV}^{-2}\\
m_Z &= 91.1876 \text{ GeV}
\end{aligned}
\qquad
\begin{aligned}[c]
\hat{m}^{\overline{\text{MS}}}_b ( \hat{m}^{\overline{\text{MS}}}_b ) &= 4.18  \text{ GeV}\\
m_t &= 173.34 \text{ GeV}
\end{aligned}
\label{eq:SMparams}
\end{equation}
\begin{table}[t]
\begin{center}
\begin{tabular}{l|rrr}
%\toprule
&BMP1&BMP2&BMP3\\
\hline
$\tan\beta$  &  3  &  10  &  40\\
$B_\mu$      &  $500^2$  &  $300^2$  &  $200^2$\\
$\lambda_d$, $\lambda_u$&   $1.0,-0.8$ &  $1.1,-1.1$  &   $0.15,-0.15$\\
$\Lambda_d$, $\Lambda_u$&  $-1.0,-1.2$ &  $-1.0,-1.0$ & $-1.0,-1.15$\\
$M_B^D$&$600$&$1000$&$250$\\
$m_{R_u}^2$&$2000^2$&$1000^2$&$1000^2$\\
\midrule
$\mu_d$, $\mu_u$&\multicolumn{3}{c}{$400,400$}\\
$M_W^D$&\multicolumn{3}{c}{$500$}\\
$M_O^D$&\multicolumn{3}{c}{$1500$}\\
$m_T^2$, $m_S^2$, $m_O^2$&\multicolumn{3}{c}{$3000^2,2000^2,1000^2$}\\
$m_{Q;1,2}^2$, $m_{Q;3}^2$&\multicolumn{3}{c}{$2500^2,1000^2$}\\
$m_{D;1,2}^2$, $m_{D;3}^2$&\multicolumn{3}{c}{$2500^2,1000^2$}\\
$m_{U;1,2}^2$, $m_{U;3}^2$&\multicolumn{3}{c}{$2500^2,1000^2$}\\
$m_L^2$, $m_E^2$&\multicolumn{3}{c}{$1000^2$}\\
$m_{R_d}^2$&\multicolumn{3}{c}{$700^2$}\\
\midrule
$v_S$& $5.9$ & $1.3$ &$-0.14$\\
$v_T$&$-0.33$ &$-0.19$ &$-0.34$\\
$m_{H_d}^2$&$671^2$ &$761^2$ &$1158^2$ \\
$m_{H_u}^2$& $-532^2$ &$-544^2$ & $-543^2$ \\
%\bottomrule
\end{tabular}
\end{center}
\caption{Benchmark points. Dimensionful parameters are given in
GeV or GeV${}^2$, as appropriate. The first part gives input parameters, where the values are
specific for each point, the second, where the values are common for all points. 
The last part shows the derived parameters.}
\label{tab:BMP}
\end{table}
For the computational calculations in the MRSSM the \textit{Mathematica}~\cite{Mathematica} package
\texttt{SARAH} \cite{SA1,SA2,SA3,SA4,SA5} \textit{v4.3.5} is used.  A model file for 
the MRSSM  defined
there has been modified  to match with our model definitions as specified in section \ref{ch:basics}.
With \texttt{SARAH}, an MRSSM version of the spectrum generator 
\texttt{SPheno}~\cite{SP1,SP2}
has been created. This allows us to calculate the mass spectrum of the 
model at the full one-loop
level. With the recent
framework \texttt{FlexibleSUSY} \cite{FlexibleSUSY} \textit{v1.0.2}, a second spectrum
generator was generated which has similarities with \texttt{Softsusy} \cite{Allanach:2001kg,Allanach:2013kza}
and which was
used for comparison. The correctness of all calculations was checked at great length, and
especially details for the calculation of the Higgs and W boson mass will be presented in the following 
sections \ref{ch:Mh} and \ref{ch:mW}. In section \ref{ch:results} we will then show how
it is possible for MRSSM to accommodate both the lightest Higgs boson mass of $125$ GeV and the correct
W boson mass  and which regions of parameter space are still viable for the model.

\section{Higgs boson mass prediction at one loop}
\label{ch:Mh}
\newcommand{\ovq}[1][]{\frac{#1}{Q^2}}
%\subsection{General  remarks} % combine with other sections?
In section \ref{ch:ModDef} it was noted that 
mixing with the additional scalars can reduce the tree-level lightest Higgs boson mass significantly, see eq.~\eqref{eq:approx_treehiggs}. 
Therefore, loop corrections to the Higgs boson mass play even more significant role than in the MSSM 
\cite{Haber:1990aw,Ellis:1991zd,Chankowski:1991md}. New, non-MSSM loop corrections 
can balance the tree-level reduction and can be helpful in lowering the fine-tuning in the model.\footnote{For the discussion of fine-tuning in the MRSSM we refer to \cite{Gregoire}.}

To calculate the Higgs boson pole mass at the one-loop level, we renormalize the parameters of the model 
in the  $\overline{\text{DR}}$ scheme
and choose $v_d$, $v_u$, $v_S$ and $v_T$ to be given by the minimum of the 
loop-corrected effective potential.\footnote{%
This is the definition commonly used and for which the results of
ref.~\cite{Sperling:2013} apply; for a recent comparison with
alternatives, see ref.~\cite{Bednyakov:2013cpa}.}
Then the pole mass $m^2_{\text{pole}}$ of a field  is given by the pole
of the full propagator 
\begin{equation}
\label{eq:PropPole}
0 \overset{!}{=} \det\left[p^2 \delta_{ij}- \hat{m}_{ij}^2 + \Re(\hat{\Sigma}_{ij}(p^2))\right]_{p^2=m^2_{\text{pole}}}\;,
\end{equation}
where $p$ is the momentum, $\hat{m}^2$ the tree-level mass matrix and $\hat{\Sigma}(p^2)$ 
the finite part of the self-energy corrections. Here and in the
following sections, mass parameters with hats denote $\overline{\text{DR}}$-renormalized
tree-level quantities fulfilling the relations of section \ref{ch:ModDef}, while
mass parameters without hats denote loop-corrected pole masses.

\subsection{Full one-loop calculation}
The one-loop self energies have been computed exactly using 
\texttt{FeynArts}~\cite{FA}, \texttt{FormCalc} \cite{Hahn:1998yk,FC} and Feynman rules generated by \texttt{SARAH}.
They are also implemented in the spectrum generators generated
by \texttt{SARAH} and by \texttt{FlexibleSUSY}
as described in section~\ref{benchmark-points}. 
We have verified that all three implementations agree.
As a further check we computed the self energies in the 
effective potential approximation (see below)
and found agreement for the gauge-independent part.

Therefore, eq.~\eqref{eq:PropPole} can be analyzed at complete one-loop order.
Due to the momentum dependence of self-energies, no direct solution for $m^2_{\text{pole}}$ can be given,
but has to be found iteratively. 

\subsection{Effective potential approach}
It is beneficial to complement the full calculation with a compact approximation of the 
leading behavior.
Loop contributions to the Higgs boson masses can be approximated by neglecting the $p^2$ dependence within
the self-energy in eq.~\eqref{eq:PropPole}. Then the self energy $\hat{\Sigma}(0)$ can be obtained from 
the second derivatives of the effective potential \cite{Coleman:1973jx}, which is given by
\begin{equation}
V^{\text{1L}}_{\text{eff}}=\frac{1}{64 \pi^2} \sum_i (-1)^{2 S_i+1}(2 S_i+1)\Tr \left[M_i^4 ( 
\log\ovq[M_i^2]- \frac{3}{2} )\right]\;,
\label{eq:effpot}
\end{equation}
summing over all fields with Higgs field dependent mass matrix $M_i^2$ and spin $S_i$ in the $
\overline{\text{DR}}$ scheme and Landau gauge.  As no implicit dependence on the momentum remains, 
eq.~\eqref{eq:PropPole} can be solved directly
for the one-loop corrected mass matrix:
\begin{equation}
m^2_{\text{pole, approx};ij}= \hat{m}_{ij}^2 + \left. \frac{\partial^2 V^{\text{1L}}_{\text{eff}}}{\partial 
\phi_i\partial \phi_j} \right|_{\phi_i=0,\phi_j=0}
\label{eq:effpole}
\end{equation}
Eigenvalues of this matrix give then an approximation to the pole masses of corresponding mass 
eigenstates.

For illustration we use again the limit leading to eq. \eqref{eq:approx_treehiggs}, 
$\lambda=\lambda_u=-\lambda_d$, $\Lambda=\Lambda_u=\Lambda_d$ and $v_S\approx v_T\approx0$, 
and assume large $\tan\beta$. Then, the lightest Higgs state is given mainly by the $\phi_u$ 
component and only the $(\phi_u,\phi_u)$-component of the mass matrix in eq.~\eqref{eq:scalarmassmatrix} needs 
to be considered. Simple analytical expressions can then be derived by expanding 
$V^{\text{1L}}_{\text{eff}}$ in powers of $\phi_u$, where terms of higher order than $\mathcal{O}(\phi_u^4)$ 
will be suppressed  by denominators containing $m_S^2$ and $m_T^2$. 

The most important contributions are the ones governed by four powers of $\lambda/\Lambda$. 
The one-loop terms are given by 
\begin{equation}
\label{eq:effpothiggs}
\begin{split}
\Delta m_{H_1,\text{eff.pot},\lambda}^2 &=\frac{2 v^2}{16 \pi^2}\Bigg[\frac{\Lambda^2\lambda^2}{2}   +  \frac{4\lambda^4 + 4\lambda^2 \Lambda^2 + 5 \Lambda^4}{8} \log \ovq[m^2_{R_u}]\\
  &+\left(\frac{\lambda^4}{2}-\frac{\lambda^2\Lambda^2}{2} \frac{m^2_S}{m^2_T-m^2_S}\right)\log \ovq[m^2_S]\\
&+\left(\frac{5}{8}\Lambda^4+\frac{\lambda^2\Lambda^2}{2} \frac{m^2_T}{m^2_T-m^2_S}\right)\log\ovq[m^2_T]\\
&- \left(\frac{5}{4}\Lambda^4-\lambda^2\Lambda^2 \frac{(M^D_W)^2}{(M^D_B)^2-(M^D_W)^2}\right)\log \ovq[(M^D_W)^2]\\
& -\left(\lambda^4+\lambda^2\Lambda^2 \frac{(M^D_B)^2}{(M^D_B)^2-(M^D_W)^2} 
  \right)\log\ovq[(M^D_B)^2] \Bigg]\;.
\end{split}
\end{equation}
This result agrees with \textit{ver.~2} of  \cite{Gregoire}.  Note that the logarithmic terms have a similar form as the stop-top contributions
 \begin{equation}
 \Delta m_{H_1,\text{eff.pot},y_t}^2 =\frac{6 v^2}{16 \pi^2}\left[
  Y_t^4 \log\left(\frac{m_{\tilde{t}_1}m_{\tilde{t}_2}}{m^2_t}\right) \right]\;.
 \end{equation}
This, of course, can be explained with similarity of the Yukawa and $\lambda/\Lambda$ terms in the superpotential in combination with the $\hat{H}_u$ field.

For the numerical analysis no approximation and expansions are performed and the full dependencies of the effective potential are taken into account.

\subsection{Comparison of different  calculations and higher-order uncertainties}
\begin{figure}[th]
\centering
\includegraphics[width=\textwidth]{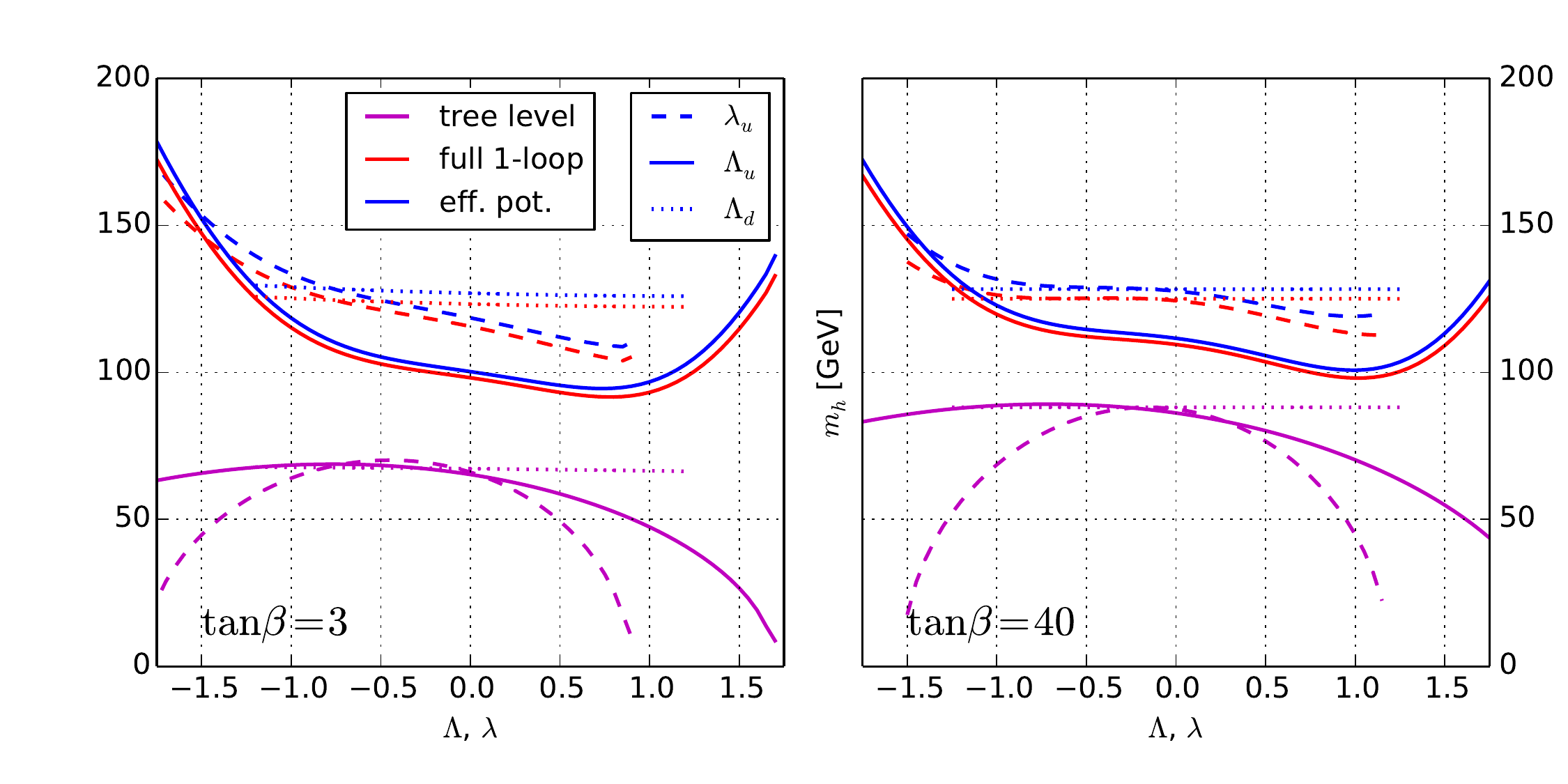}
\caption{Comparison of the lightest Higgs boson mass calculated using the effective potential
approach (blue)  and the full one-loop calculation (red), as well as the tree-level mass (magenta).  Results are shown as functions of one of the couplings: $\Lambda_u$ (solid), $\Lambda_d$ (dots), $\lambda_u$ (dashes), 
for benchmark point 1 (left, $\tan\beta=3$) and 3 (right, $\tan\beta=40$).}
\label{img:effpotvs1l}
\end{figure}

In the plots of figure~\ref{img:effpotvs1l} the different predictions
(tree-level, effective potential, full one-loop) for the 
lightest Higgs boson mass are shown as functions of one of the couplings 
$\Lambda_{u}$ (solid),  $\lambda_{u}$ (dashes), and $\Lambda_d$ (dots).  For every line, the corresponding parameter is varied and all the others are set to values
given by benchmark point 1 (left plot) and 3 (right plot) from table~\ref{tab:BMP}\footnote{%
For large couplings $\lambda/\Lambda$ the mixing with singlet and/or triplet states gets too strong and tachyonic states appear
at tree level. The lines in the plots end at 
those values of $\lambda/\Lambda$ beyond which this happens.}.
As described in section~\ref{ch:treehiggs}, the tree-level mass
(magenta) is reduced significantly below  $m_Z \cos\beta$  for 
large values of $\Lambda_{u}$ and $\lambda_{u}$ as the mixing of the doublets with the singlet
and triplet components gets enhanced.
The parameter $\Lambda_{d}$ generally has no strong influence on the prediction even for
a low $\tan\beta=3$.
The general asymmetry of the plots with regard to the sign of the $\Lambda$'s originates from the dependence of
the off-diagonal elements in the tree-level Higgs boson mass matrix.

The one-loop effective potential (blue) and full one-loop (red) predictions both show the large 
positive contribution from top Yukawa as well as the $\Lambda/\lambda$ enhancement. 
As described by eq.~\eqref{eq:approx_treehiggs}, the tree-level Higgs boson mass will
show a quadratic dependence on the $\Lambda/\lambda$, while the one-loop contribution
given by eq.~\eqref{eq:effpothiggs} has a quartic dependence, explaining the behavior of the sum.

The contribution from the top/stop Yukawa alone can be read 
from the solid 
lines at $\Lambda_u=0$, since then only stop/top contributions are significant.
It can be seen that in the MRSSM a stop mass of one TeV, as set in the benchmark points, is not 
enough to achieve the correct value of the Higgs boson mass, due to the absence of left-right mixing.
But the additional $\Lambda/\lambda$ contributions can push it to the experimental 
result for values 
of $\Lambda_{u}$ close to unity\footnote{Note that we define $\Lambda_{u/d}$ differing
 by a factor of $\sqrt{2}$ compared to the conventional form of a Yukawa term due to 
our normalization of the triplet field.}. This is clearly seen in table \ref{tab:higgs}, 
where contributions from different sectors of the MRSSM model to the 
lightest Higgs boson mass are shown. 
In particular, the low values of $\tan\beta$  require a larger value of $\Lambda$ 
and heavier R-Higgs bosons to compensate the significant reduction of the 
tree-level value. 

To get an estimate of the higher-order corrections, we follow the discussion in 
\cite{Allanach:2004rh}. 
There, two-loop corrections from the strongly interacting sector and their implementation in 
different spectrum generators for the MSSM and, among other things, different higher order contributions in the 
$\overline{\text{DR}}$ scheme are presented.
These two-loop corrections  are usually positive and contribute around $3-5$~GeV.
Similar effects from the strongly interacting sector should also exists in the MRSSM. Calculating in the $\overline{\text{DR}}$ scheme
we expect similar positive corrections of order $\mathcal{O}(\alpha_S\alpha_t)$.
Although quantitative differences  due to the Dirac nature of the gluino in the MRSSM  might appear, the main contribution from gluon diagrams should be identical.
Effects of order $\mathcal{O}(\alpha_t^2)$ should behave similar to the MSSM, with small and 
usually  negative contributions, see table~5 of \cite{Allanach:2004rh}.

As in the one-loop calculation, we expect that effects of
the order $\mathcal{O}(\alpha_\Lambda^2+\alpha_\lambda^2+\alpha_\Lambda\alpha_\lambda)$
will have a similar impact as these subleading effects, due to the already noted
similarity of the parameters in the superpotential. 
A quantitative analysis of all these two-loop effects is left for future work and
we estimate a theoretical uncertainty for the lightest Higgs boson mass of 6~GeV for our one-loop mass
calculation, based on  an expected two-loop contribution of up to 5~GeV and the two-loop
uncertainty of $1-3$~GeV given in \cite{Allanach:2004rh}.

\begin{table}
\centering 
\begin{tabular}{l|cccccc}
 & tree-level & \begin{tabular}{@{}c@{}}gauge \\ ghost \\ Higgs\end{tabular} & \begin{tabular}{@{}c@{}}gluino/sgluon \\ quark/squark\end{tabular} & \begin{tabular}{@{}c@{}}chargino\\ neutralino\end{tabular} & R-Higgs &all\\
\hline
BMP1 & 69.9 & 81.5 & 110.3 & 117.8 & 124.2 &125.3\\
BMP2 & 86.5 & 96.3 & 120.7 & 123.2 & 124.1 &125.1\\
BMP3 & 90.6 & 93.9 & 118.4 & 123.4 & 124.1 &125.1 
\end{tabular}
\caption{The lightest Higgs boson mass (in GeV) for the benchmark points: tree-level value and after adding one-loop contributions sector by sector of the MRSSM.    \label{tab:higgs}
 }
\end{table}

\section{W boson mass prediction at one loop}
\label{ch:mW}
The mass of the W boson is one of the most precisely known electroweak
precision observables and plays a crucial role in constraining
extensions of the SM. The SM and MSSM loop corrections to $m_W$ are
dominated by the top Yukawa coupling. We have seen that the new
superpotential couplings $\lambda_{u,d}$, $\Lambda_{u,d}$ of the MRSSM
influence the Higgs boson mass similarly to the top Yukawa coupling
--- hence it is to be expected that the W boson mass has a significant
dependence on these new couplings. Here we present the higher-order
computation of the W boson mass in the MRSSM, which can then be used
to constrain the MRSSM parameter space.

\subsection{Master formula for $m_W$} 

Like for the Higgs boson mass, an appropriate scheme is the 
$\overline{\text{DR}}$ scheme. The calculation can then be organized
as in ref.\ \cite{degrassi} for the SM; a major difference is the
appearance of the additional Higgs triplet with non-vanishing vev, which enters already  at
lowest order (for
computations of $m_W$ in models with Higgs triplets and singlets see refs.\ \cite{blank&hollik,chankowski} 
and \cite{Lopez-Val:2014jva}, respectively).

Beyond tree-level the W boson pole mass $m_W$ can be obtained from the precisely
known muon decay constant using the relation
\begin{equation}
  \frac{G_\mu}{\sqrt{2}} = \frac{\pi \hat{\alpha}}{2 \hat{s}^2_{W}
 m_{W}^2} \frac{1}{1 - \Delta \hat{r}_W }
 .\label{w-mass-master-formula}
\end{equation}
As in the previous section,
$\hat{\alpha},\hat{s}_W^2={\hat{g}_1^2}/({\hat{g}_1^2
+ \hat{g}_2^2})$ are $\overline{\text{DR}}$-renormalized running
parameters of the MRSSM. The denominator $(1-\Delta \hat{r}_W)$
contains quantum corrections from the W boson self-energy, process
dependent box- and vertex-type contributions and counterterms; it also
properly resums leading two-loop SM corrections \cite{degrassi}.

Expressing the weak mixing angle as $  \hat{s}_W^2
 = 1 - \frac{m_W^2}{m_Z^2 \hat{\rho}} $
 in terms of the parameter $\hat\rho$,
defined as 
\begin{equation}
\hat{\rho} =  \frac{m_W^2}{m_Z^2 \hat{c}_W^2} ,
\label{Definitionrho}
\end{equation}
we obtain the master formula for the W boson mass
\begin{equation}
m_W^2 = \frac{1}{2} m_Z^2 \hat{\rho} \left [ 1 + \sqrt{1
- \frac{4 \pi \hat{\alpha}}{\sqrt{2} G_\mu m_Z^2 \hat{\rho}
(1-\Delta \hat{r}_W)}}\;\right ] \label{w-mass-master-formula2}.
\end{equation}
Hence we need to compute the three quantities $\hat\alpha$,
$\hat\rho$, and $\Delta\hat{r}_W$, all of which depend on the entire
particle content of the model. Before discussing these computations it
is instructive to expand eq.~ \eqref{w-mass-master-formula2}  
in small shifts by writing each of the
quantities $q=m_W,\hat\alpha,\hat\rho,\Delta\hat{r}_W$ in the form of
$q=q^\text{ref}+\delta(q)$.
We
then obtain
\begin{equation}
% m_W \approx m_W^\text{ref} + \Big[56\,\delta({ \hat{\rho}}) -
% 16\, \delta({\Delta\hat{r}_W}) -16\, \delta({\hat\alpha})\Big]\text{GeV},
m_W \approx m_W^{\text{ref}} + \frac{m_Z\hat{c}_W}{2(\hat{c}_W^2-\hat{s}_W^2)}\Big[\hat{c}^2_W
\delta({\hat{\rho}}) - \hat{s}^2_W (\delta({\Delta\hat{r}_W}) + \delta({\hat\alpha}))\Big],
\label{linearapproximation}
\end{equation}
which shows the relative importance of these contributions. Note that
the form of eq.\ (\ref{w-mass-master-formula2}) resums leading
two-loop corrections, and that there are cancellations between the
terms on the right-hand side of eq.\ (\ref{linearapproximation}).

\subsection{Computational framework}

We now describe the computation of the three quantities
$\hat\alpha,\hat\rho,\Delta\hat{r}_W$. The $\overline{\text{DR}}$
running electromagnetic coupling $\hat\alpha$ in the MRSSM can be 
obtained from the known running SM coupling
$\hat{\alpha}^{\overline{\text{MS}},\text{SM}}(m_Z)$ by 
applying MRSSM threshold corrections and adding finite counterterm
which converts from
$\overline{\text{MS}}$ to $\overline{\text{DR}}$. These are
\begin{align}
\frac{2 \pi}{\alpha} \Delta \hat{\alpha}^{\overline{\text{DR}},\text{MRSSM}}(m_Z)
& =  \frac{1}{3}
- \sum_{i=1}^6 \left(\frac{1}{3} \log \frac{m_{\tilde{l}_i^\pm}}{m_Z} 
+
 \frac{1}{9} \log \frac{m_{\tilde{d}_i}}{m_Z} 
+ \frac{4}{9} \log \frac{m_{\tilde{u}_i}}{m_Z}
\right )
\notag\\
&- \sum_{i=1}^3\left( \frac{4}{3} \log \frac{m_{l^\pm_i}}{m_Z}
+
\frac{4}{9} \log \frac{m_{d_i}}{m_Z}
+  \frac{16}{9} \log \frac{m_{u_i}}{m_Z}
\right )
\notag\\
&- \sum_{i=1}^3 \frac{1}{3} \log \frac{m_{H_i^\pm}}{m_Z} - \sum_{i=1}^2 \frac{1}{3}  \log \frac{m_{R^\pm_i}}{m_Z} 
- \sum_{i=1}^2 
\frac{4}{3} \left (
 \log \frac{m_{\chi^\pm_i}}{m_Z}
+  \log \frac{m_{\rho^\pm_i}}{m_Z}
\right )
\end{align}
where  $\alpha$ is the electromagnetic coupling   in the Thomson limit. In our case
this expression is always negative, reducing the value of the running coupling
\begin{equation}
\hat{\alpha}(m_Z) = \frac{\hat{\alpha}^{\overline{\text{MS}},\text{SM}}(m_Z)}{1-\Delta \hat{\alpha}^{\overline{\text{DR}},\text{MRSSM}}(m_Z)} \leq \hat{\alpha}^{\overline{\text{MS}},\text{SM}}(m_Z)
\end{equation}
For the benchmark points defined in section \ref{benchmark-points} we obtain 
\begin{equation}
\hat{\alpha}^{-1}(m_Z) \approx 132 \, .
\end{equation}

Large corrections to the W boson mass originate in the $\hat\rho$
parameter defined in eq.\ (\ref{Definitionrho}). In the SM, the
dominant contributions arise from top/bottom loop; in the MRSSM there
are not only loop contributions but already a tree-level
contribution due to the presence of the Higgs triplet with a vev
$v_T$ as already pointed out in eq. \eqref{eq:bosonmasses}.
This is used to define the tree-level shift $\Delta\hat\rho_{\text{tree}}$ using
\begin{equation}
  \hat{\rho}_{\text{tree}} = \frac{\hat{m}^2_W}{\hat{m}^2_Z \hat{c}_W^2}\equiv1+\Delta\hat\rho_{\text{tree}}
 =  1 + \frac{4 v_T^2}{v^2}. \label{tree-rho}
\end{equation}
Here $\hat{m}_{V}$ ($V=W,Z$) are the tree-level $\overline{\text{DR}}$
masses given by eq.\ (\ref{eq:bosonmasses}) and related to the pole
masses $m_V$ by
\begin{equation}
\hat{m}_V^2 = m_V^2 + \Re(\hat{\Pi}_{VV}^T (m_V^2)),
\end{equation}
where $\hat{\Pi}_{VV}^T$ denotes the finite part of the respective
transverse vector boson self energy. 
The loop contributions to $\hat{\rho}$ are given by
\begin{eqnarray}
\frac{\hat\rho}{\hat{\rho}_\text{tree}}\equiv\frac{1}{1-\Delta\hat\rho}
 &=& \frac{1+ \frac{\Re(\Pi_{ZZ}^T
 (m_Z^2))}{m_Z^2}}{1+ \frac{\Re(\Pi_{WW}^T(m_W^2))}{m_W^2}},
\end{eqnarray}
and the full $\hat\rho$ can then be approximated by 
\begin{eqnarray}
\hat\rho&=& \frac{1}{1-\Delta\hat\rho_{\text{tree}}-\Delta\hat\rho},
\end{eqnarray}
neglecting products of the form $\Delta\hat\rho_{\text{tree}}\Delta\hat\rho$.

The remaining quantity $\Delta\hat{r}_W$ can then be written
as \cite{degrassi}
\begin{align}
\Delta\hat{r}_W &= \Delta\hat\rho(1-\Delta\hat{r})+\Delta\hat{r},
\\
\Delta\hat{r}
&= \Re\left(\frac{1}{1-\Delta\hat{\rho}}\frac{\hat\Pi_{WW}^T(0)}{m_W^2}
-\frac{\hat\Pi_{ZZ}^T(m_Z^2)}{m_Z^2}\right)
+\frac{1}{1-\Delta\hat{\rho}}\delta_{VB}.
\end{align}
The term $\delta_{VB}$ contains vertex and box diagram contributions
to muon decay; it is detailed in the Appendix \ref{sec:deltaVB}.
It is worth noting that in these equations only the loop contributions
to $\hat\rho$ appear. This way of writing the contributions and the
master formula (\ref{w-mass-master-formula}) automatically resums
leading reducible two-loop contributions; further leading irreducible
SM-like two-loop contributions can also be incorporated
easily \cite{degrassi,Fanchiotti:1992tu,Pierce:1996zz}. 
This calculation has been implemented numerically with the help
of \texttt{SARAH}, appropriately modified to take into account the
Higgs triplet contributions.

\subsection{Qualitative discussion}
\label{ch:mw-discussion}
In our numerical analysis and plots of the W boson mass we use
the full MRSSM one-loop results for
$\Delta\hat\alpha^{\overline{\text{DR}},\text{MRSSM}}$,
$\Delta\hat\rho$, $\Delta\hat{r}_W$ including leading two-loop
contributions as 
incorporated in eq.\ (\ref{w-mass-master-formula}). 

Here we discuss the leading behavior, particularly focusing on 
the dependence on the new parameters $\lambda_{u,d},\Lambda_{u,d}$. 
These enter in the sectors
involving charginos/neutralinos, Higgs bosons and R-Higgs bosons.
The pure one-loop approximation, eq.~\eqref{linearapproximation}, 
can be recast in a form using
the electroweak precision parameters $S$, $T$ and $U$ \cite{Peskin:1990zt,Marciano:1990dp,Peskin:1991sw,Kennedy:1990ib,Kennedy:1991sn,Altarelli:1990zd} as
\begin{equation}
m_W= m_W^{\text{ref}} +\frac{\hat{\alpha} m_Z \hat{c}_W}{2(\hat{c}_W^2-\hat{s}_W^2)}\left (-\frac{S}{2}+\hat{c}^2_WT+\frac{\hat{c}^2_W-\hat{s}^2_W}{4 \hat{s}_W^2}U \right)\;.
\label{eq:mW-STU}
\end{equation}
The main contribution to $m_W$ from the MRSSM can be described
using the $T$ parameter 
\begin{equation}
\hat{\alpha} T = \left.\left(\frac{\hat\Pi_{ZZ}^T(0)}{\hat{m}_Z^2}-
\frac{\hat\Pi_{WW}^T(0)}{\hat{m}_W^2}\right) \right|_{\text{New Physics}}
\end{equation}
for realistic scenarios,
i.e.\ for each sector the contribution to the bracket on the r.h.s.\ of eq.\
(\ref{linearapproximation}) is well approximated by $\hat{\alpha}\hat{c}_W^2T$.

Calculating $S$, $T$ and $U$ has the additional benefit that theses parameters can be used in the 
calculation of many electroweak precision observables. Therefore, strong bounds exist 
from fits to a large number of such observables \cite{PDG}.

In order to qualitatively understand the leading behavior and to
further illustrate the complexity and richness of the MRSSM, we will
consider all three sectors in turn and derive various limits of
phenomenological and conceptual interest. All limits assume large
$\tan\beta$ and neglect $v_{d,S,T}$.
\begin{itemize}
\item Charginos/neutralinos, interplay of $\Lambda_u$ and Dirac
masses: In the parameter regions considered in the present paper, the
chargino/neutralino sector provides the largest contributions to
$m_W$, and the most influential parameters are $\Lambda_u$,
$\lambda_u$ and the Dirac-type mass parameters $\mu_u$, $M_{W,B}^D$. 
For the simplified case 
$\lambda_u=g_1=0$ and $\mu_u=M^D_W$ we obtain
\begin{equation}
T=\frac{1}{16\pi \hat{s}_W^2 \hat{m}_W^2}\frac{v_u^4}{(M_W^{D})^2}\left[\frac{13 g_2^4 +2g_2^3\Lambda_u +18g_2^2\Lambda_u^2 +2g_2\Lambda_u^3 +13\Lambda_u^4}{96} \right]+\mathcal{O}\left(\frac{v_u^4}{(M_W^{D})^4}\right),
\label{chaneumainlimit}
\end{equation}
while for the similar case $\Lambda_u=g_2=0$ and $\mu_u=M^D_B$ we
obtain the same formula with an additional prefactor $1/5$ and the
replacement $\Lambda_u\to\sqrt{2}\lambda_u$, $M_W^D\to M_B^D$.
Hence the main dependence on the new superpotential couplings
$\Lambda_u$, $\lambda_u$ is the one of a fourth-order polynomial.
\newcommand{\xmu}{x}
\item Charginos/neutralinos as vector-like fermions: The dependence on
the Dirac-type mass parameters can also be understood in a different
way. If all couplings $g_{1,2}$, $\lambda_{u,d}$, $\Lambda_{u,d}$
vanish, the charginos and neutralinos simply correspond to new,
vector-like fermions (a singlet, a triplet, and two doublets) with
vanishing $T$. As an example with non-vanishing couplings and
mixing we consider the $N=2$ SUSY limit
$g_1=\sqrt{2} \lambda_u=\sqrt{2} \lambda_d$ and
$g_2=\Lambda_u=-\Lambda_d$. For the vector-like limit we then set $\Lambda_u=0$ 
in the neutralino/chargino tree-level mass matricies and the
$\tilde{S}$--$\tilde{B}$ singlet and the $\tilde{R}_u$--$\tilde{H}_u$
doublet mix just like the vector-like fermions considered in
ref.\ \cite{Cynolter:2008ea}. Keeping the masses $\mu_u$ and $M_B^D$
independent, with  $\xmu=\mu_u/M_B^D$, we obtain
%% \begin{align}
%% T=&\frac{ v_u^4 \lambda_u^4}{16\pi \hat{s}_W^2 \hat{m}_W^2}\left[\frac{(M_B^D)^3((M_B^D)^2-M_B^D\mu_u+4 \mu_u^2)}{(M_B^D-\mu_u)^5(M_B^D+\mu_u)^2}  \log\frac{(M_B^D)^2}{\mu^2_u}\right. \notag\\
%% &\left. +\frac{-13 (M_B^D)^3+(M_B^D)^2\mu_u -17 (M_B^D)\mu^2_u+5 \mu_u^3}{6(M_B^D-\mu_u)^4(M_B^D+\mu_u)}\right]+\mathcal{O}\left(\frac{v_u^4}{(M_B^D)^4},\frac{v_u^4}{\mu_u^4}\right)\;.
%% \label{vectorlikelimit}
%% \end{align}
\begin{align}
T=&\frac{
1}{16\pi \hat{s}_W^2 \hat{m}_W^2}\frac{v_u^4 \lambda_u^4}{(M_B^D)^2}
\left[\frac{(\xmu-1-4 \xmu^2)  \log{\xmu^2}}{(1-\xmu)^5(1+\xmu)^2}
-\frac{13 -\xmu
 +17 \xmu^2-5 \xmu^3}{6(1-\xmu)^4(1+\xmu)}\right]
+\mathcal{O}\left(\frac{v_u^4}{(M_B^D)^4}\right).
\label{vectorlikelimit}
\end{align}
Compared to eq.\ (\ref{chaneumainlimit}), the couplings are fixed but
the Dirac-type masses enter individually.
In the limit $\mu_u \rightarrow M_B^{D}$, or $\xmu\to1$, 
this equation reduces to
\begin{equation}
T=\frac{1}{16\pi\hat{s}_W^2 \hat{m}_W^2}\frac{2v_u^4 \lambda_u^4}{5(M_B^{D})^2}+\mathcal{O}\left(\frac{v_u^4}{(M_B^D)^4}\right)\,,
\label{eq:drhominimal}
\end{equation}
which can also be obtained from eq.\ (\ref{chaneumainlimit}),
applied in the appropriate limit.
\item Charginos/neutralinos, no Dirac masses: The above formulas apply
for sufficiently heavy Dirac masses. The opposite case is
$M_{B,W}^D=\mu_{u,d}=0$. In this limit we obtain
\begin{equation}
T=\frac{v_u^2 }{16\pi\hat{s}_W^2 \hat{m}_W^2}\left[
\frac{2\lambda_u^2-5\Lambda_u^2}{4}+\frac{\Lambda_u^2(2 \lambda_u^2+3\Lambda_u^2)}{2 \lambda_u^2
-\Lambda_u^2}\log\frac{2 \lambda_u^2+ \Lambda_u^2}{2  \Lambda_u^2}+\text{``}g_{1,2}\text{-terms''}\right].
\label{eq:drhom0}
\end{equation}
The generic behavior is $\propto\lambda^2v^2$ as opposed to
$\propto\lambda^4v^4/M_D^2$ as in the previous formulas. The first
term (corresponding to $\Lambda_u\to0$) is identical to the
SM top/bottom contribution up to the color factor and replacements
$Y_t\to\lambda_u$, $Y_b\to0$, because of the parallel structure of the
$\lambda_u$ and $Y_t$-terms in the superpotential. The gauge-coupling
terms in eq.\ (\ref{eq:drhom0}) are obtained by the replacements
$\sqrt2\lambda_u\to g_1$, $\Lambda_u\to g_2$, reflecting the structure
of the chargino/neutralino mass matrices and the $N=2$ SUSY limit.
\item Higgs sector:
The Higgs sector contributions depend on the same coupling constants,
but the Higgs boson masses are also governed by the soft parameters
$m_{S,T}^2$, which are large in all our examples.
The approximation when using the gauge-less limit and setting $\mu_u=\mu_d=M_B^D=M_W^D=\lambda_u=0$ 
to fulfill the tree-level tadpole equations is given as
\begin{equation}
T=\frac{ v_u^2 }{16\pi\hat{s}_W^2\hat{m}_W^2}
\frac{\Lambda_u^4 v_u^2}{12
m_T^2}+\mathcal{O}\left(\frac{v_u^4}{m_T^4}
\right).
\label{eq:drhohh}
\end{equation}
It has a qualitatively similar
coupling dependence, but the contribution is significantly smaller
than the ones of the chargino/neutralino sector as $m_T$ is much greater than $M_D^W$.
\item
R-Higgs sector:
The contributions from the R-Higgs sector have a structure similar to
the stop-sbottom contribution \cite{Drees:1990dx},
% Q = T3+ Y
% Rd = (1,0)
% Ru = (0, 1)
\begin{multline}
T \approx \frac{1}{16\pi\hat{s}_W^2 \hat{m}_W^2} \left \{ -\sin^2 2 \theta_R F_{0} \left (m_{R_1}^2, m_{R_2}^2 \right ) + 
\cos^2 \theta_R \left [ F_0 \left ( m_{R_1}^2, m_{R^+_1}^2 \right ) + (1 \rightarrow 2)
%F_0 \left ( m_{R_2}^2, m_{R^+_2}^2 \right ) 
\right ] 
\right . \\+ \left . \sin^2 \theta_R \left [ F_0 \left ( m_{R_1}^2, m_{R^+_2}^2 \right ) + %F_0 \left ( m_{R_2}^2, m_{R^+_1}^2 \right ) 
(1 \leftrightarrow 2)
\right ] \right \} ,
\label{eq:drhorh}
\end{multline}
where
\begin{equation}
F_0 (x, y) = x + y + \frac{2 x y}{x-y} \log \frac{y}{x} \, ,
\end{equation}
and we have neglected contributions proportional to $\left ( v_T/v \right )^2$.
This means that in our case, for large soft masses $m_{R_u}^2$ and
$m_{R_d}^2$, mixing between the R-Higgs bosons and the mass-splitting
between neutral and charged R-Higges is suppressed leading 
to negligible contributions to $T$.
\end{itemize}

\subsection{Comparison of full and approximate results}

\begin{figure}[th]
\centering
\includegraphics[width=0.45\textwidth]{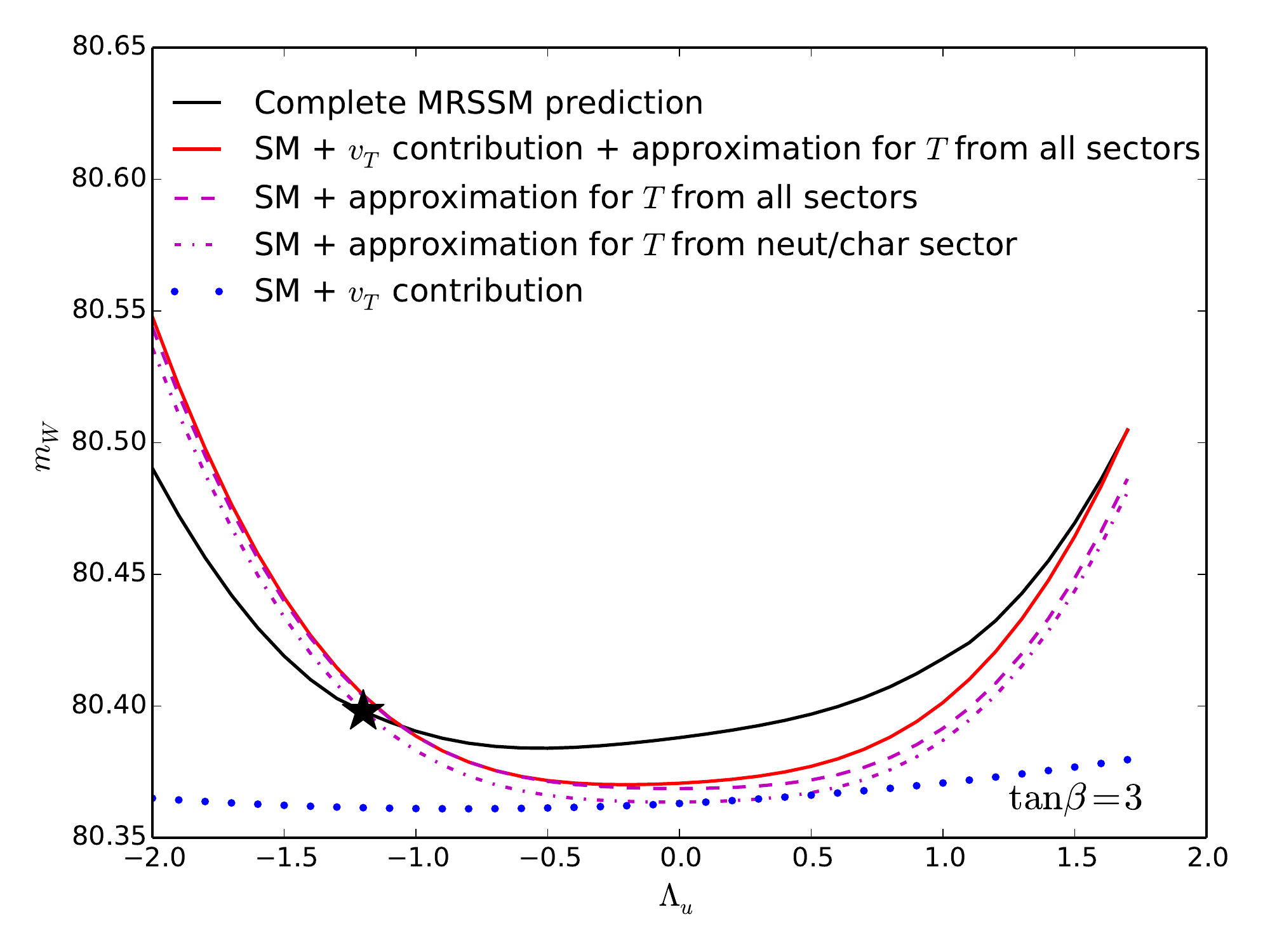}
\includegraphics[width=0.45\textwidth]{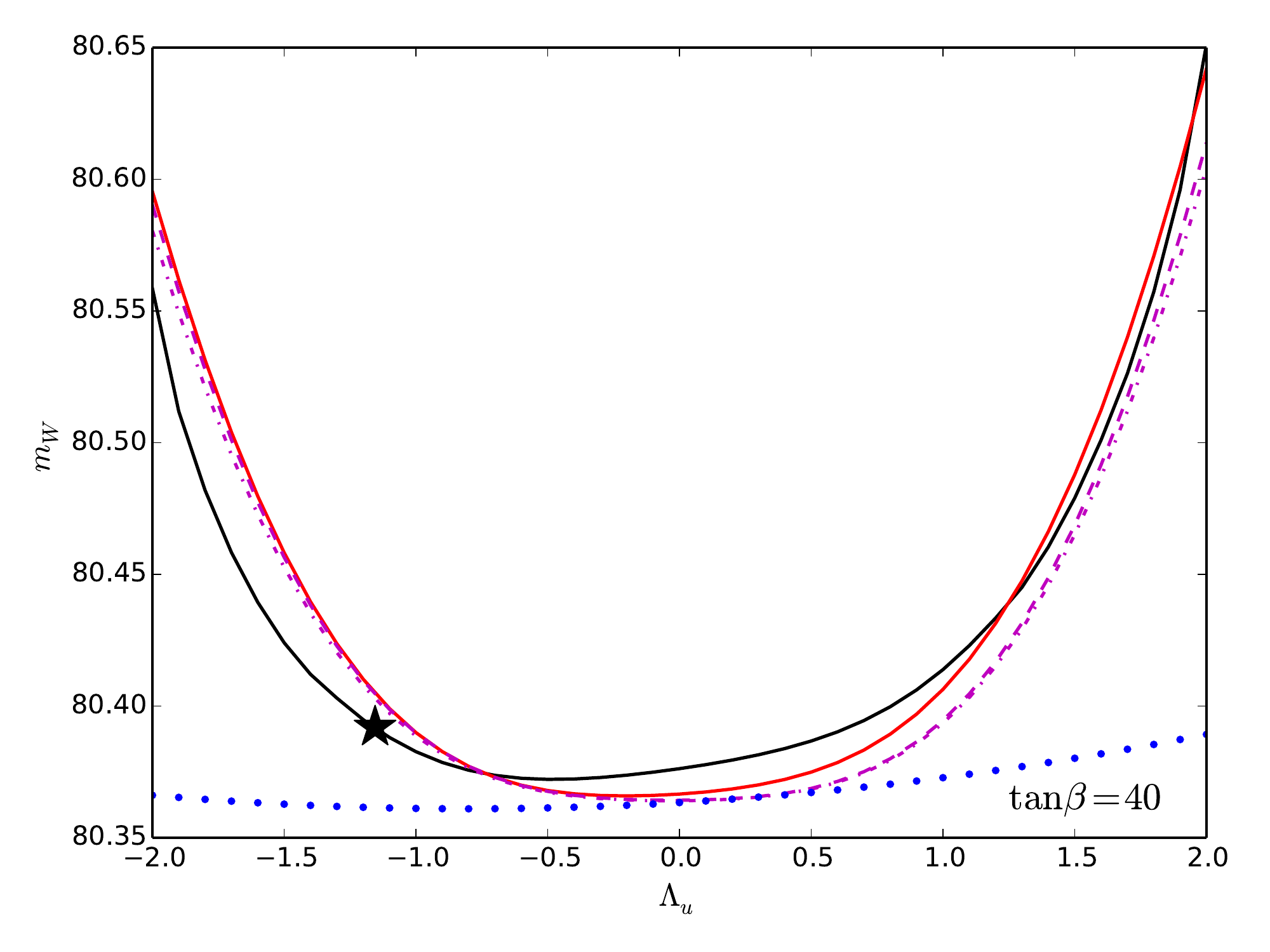}
\caption{Comparison of the mass of the W boson depending on $\Lambda_u$, calculated using full MRSSM contributions and different approximations
for the $T$-parameter (neutralino and chargino sector (neut/char) eq. \eqref{chaneumainlimit}, 
Higgs sector eq. \eqref{eq:drhohh} and R-Higgs sector eq. \eqref{eq:drhorh}) and the tree-level contribution from the triplet vev
for Benchmark point 1 (left, $\tan\beta=3$) and 3 (right, $\tan\beta=40$). The black stars mark the corresponding Benchmark points.}
\label{img:mwlam}
\end{figure}

The previous approximations, particularly eqs.\
(\ref{chaneumainlimit}, \ref{eq:drhorh}), provide useful qualitative
insight into the parameter dependence of $m_W$. Figure \ref{img:mwlam}
shows how well various approximations agree quantitatively with the
full calculation of $m_W$. All lines plotted in the figure contain the
full SM contribution, but the MRSSM contributions are
taken into account either completely, or only via the $T$-parameter in
various approximations, or from the tree-level triplet vev
contribution.

We see that the chargino/neutralino approximation
(\ref{chaneumainlimit}) already gives an excellent approximation to
the full $T$-parameter. The $T$-parameter, together with the
tree-level triplet vev contribution, provides a good approximation to
the full result. The remaining difference from non-$T$-parameter
oblique corrections, vertex and box contributions, and leading higher
loop contributions, is between $\pm20$~MeV, except for
$\Lambda_u$>1.5.

\section{Numerical predictions }
\label{ch:results}
In the following we present a thorough analysis of the viable parameter space of the MRSSM.  As was shown in sections \ref{ch:Mh} and \ref{ch:mW}, the benchmark points of table \ref{tab:BMP} can accommodate both the mass of the lightest 
Higgs boson as well as are in agreement with the measured W boson
mass. % see table \ref{tab:exp-constraints}. 
In the following we will present the mass spectra for the benchmark points and explore the parameter space around them.
\subsection{Mass spectra}
\begin{figure}
\centering
\includegraphics[scale=0.49]{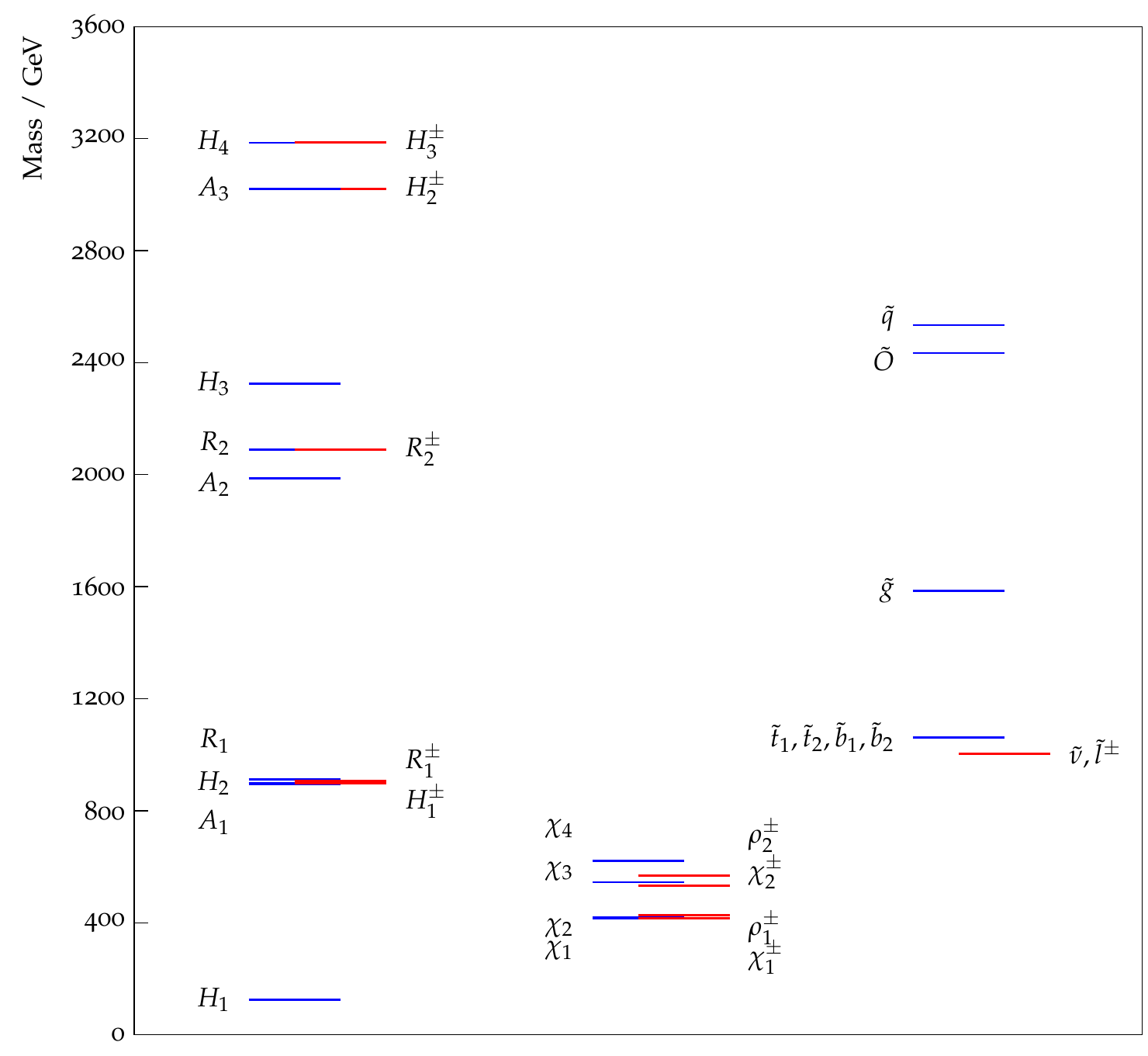}
\includegraphics[scale=0.49]{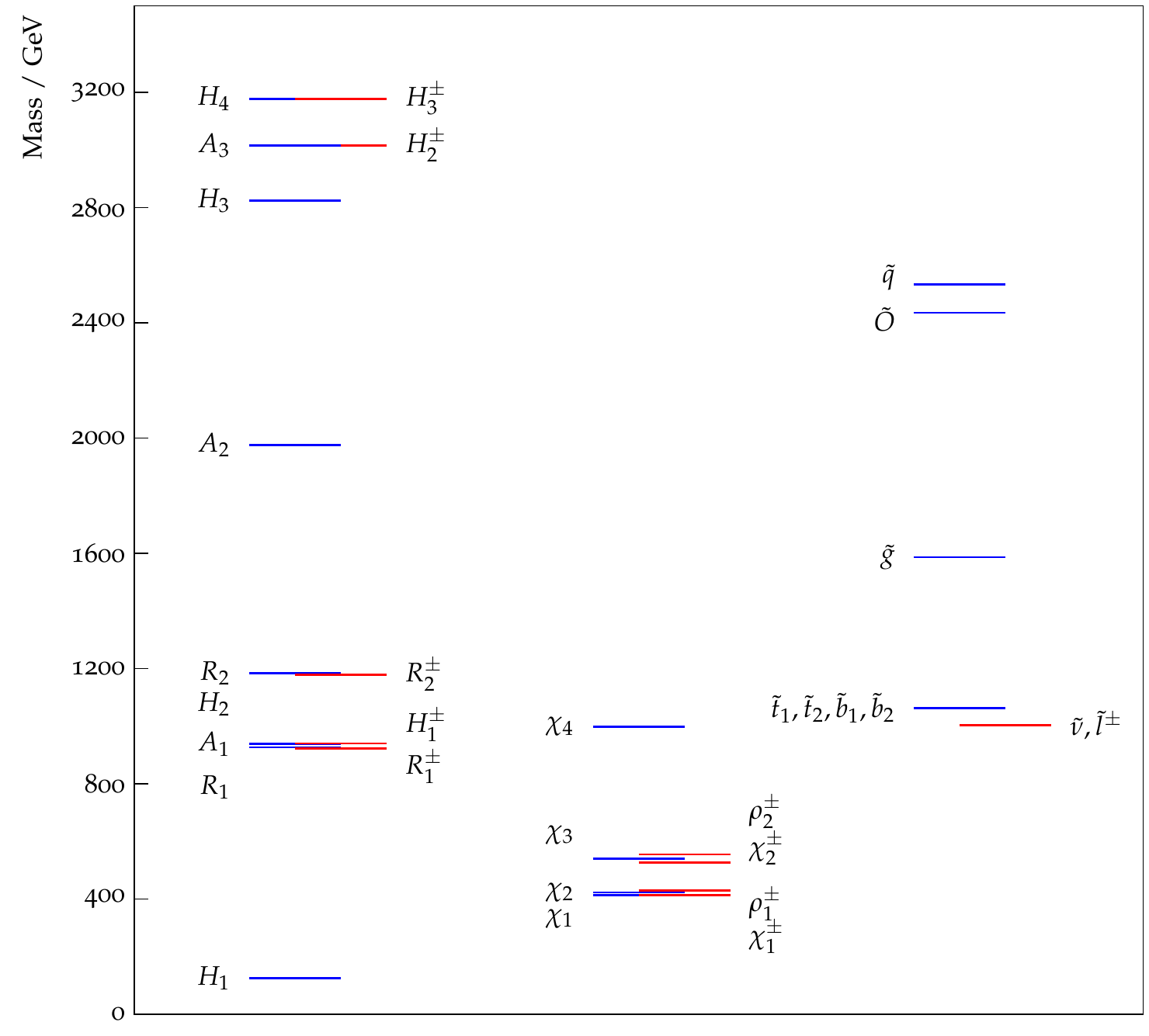}\\
\begin{center}
\includegraphics[scale=0.49]{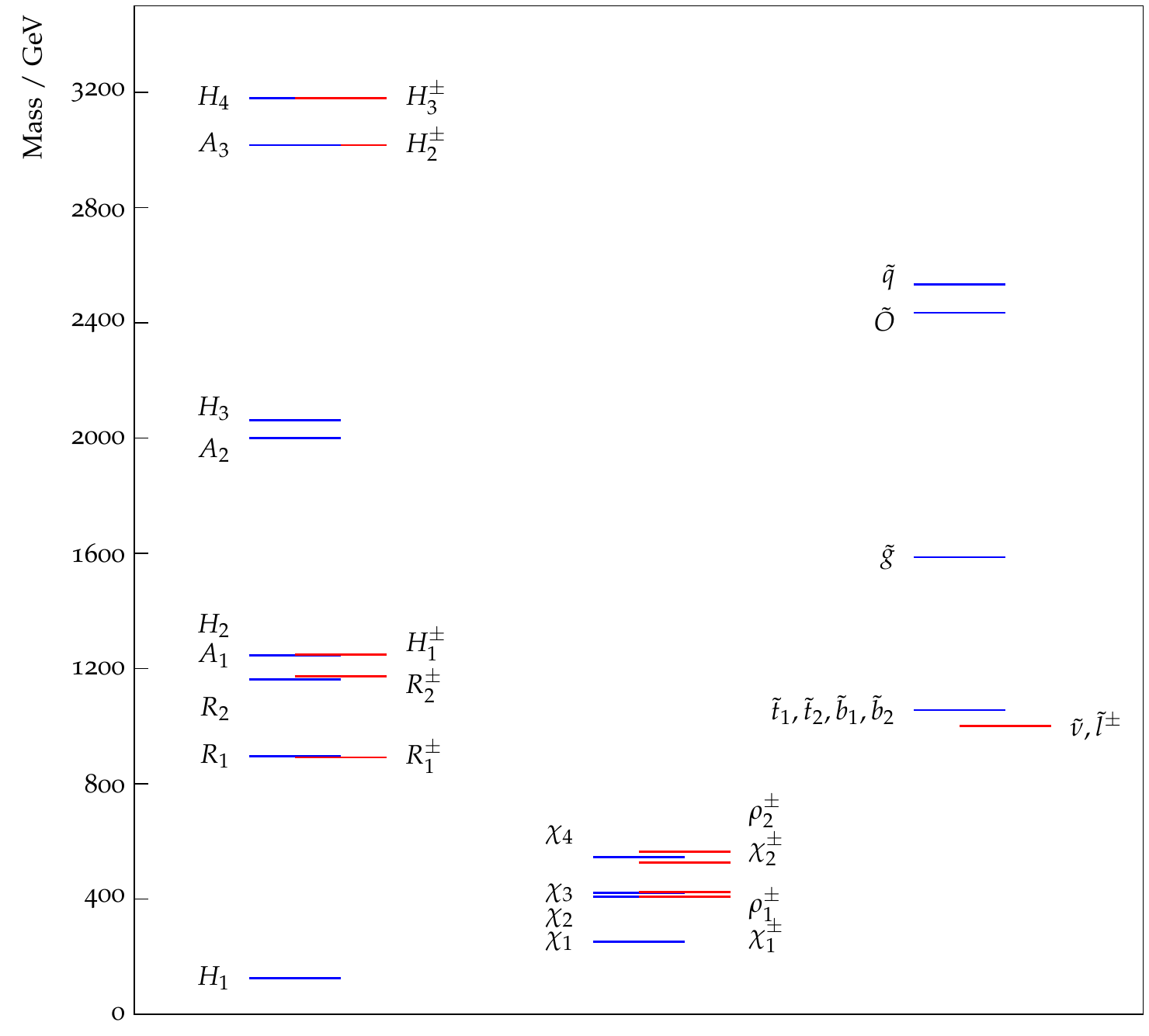}
\end{center}
\caption{Particle mass spectra for $\tan \beta = 3$ (top-left), $\tan \beta = 10$ (top-right) and $\tan \beta = 40$ (bottom). $\tilde{q}$ denotes all 1st and 2nd generation squarks. Plots done using \texttt{PySLHA}~\cite{Buckley:2013jua}. \label{fig:mass_spectrum}}
\end{figure}

\begin{table}
\centering
\begin{tabular}{l|ccccc|cccc|cc|c}
& $H_2$ & $A_1$ & $H_1^\pm$ & $R_1$ & $R_1^\pm$ & $\chi_1$ & $\chi_2$ & $\chi_1^\pm$ & $\rho_1^\pm$ & $\tilde{t}_1$ & $\tilde{b}_1$ & $\tilde{\nu}$ \\
\hline
BMP1 & 897 & 896  & 899  & 912 & 906 & 415 & 420 & 416 & 427 & 1059 & 1061 & 1002 \\
BMP2 & 937 & 937  & 940  & 926 & 921 & 413 & 423 & 413 & 429 & 1061 & 1062 & 1003 \\
BMP3 & 1245 & 1245 & 1248 & 896 & 891 & 251 & 408 & 408 & 424 & 1060 & 1056 & 1000
\end{tabular}
\caption{ Masses of selected particles (in GeV).  \label{tab:mass_spectrum} }
\end{table}

After fixing the critical parameters to account for the Higgs and $W$ boson masses, 
other parameters of BMP1-BMP3 have been chosen to produce 
sufficiently heavy  states as to not come 
into conflict with exclusion bounds from direct collider searches. 
Figure \ref{fig:mass_spectrum}  shows mass spectra for the benchmark points 
of table \ref{tab:BMP} and table \ref{tab:mass_spectrum} gives numerical values for lightest states in each sector.  The calculation of particle spectra have been performed including one-loop corrections using the  \texttt{SPheno} code generated by \texttt{SARAH}. The results  
have been  checked against the ones from \texttt{FlexibleSUSY},  and an agreement of the order of a few percent has been achieved.    

A number of comments is in order:
\begin{itemize}
\item Heavy charged and neutral  $\mathcal{CP}$-even and odd Higgs bosons are 
composed mainly from $S$ and $T$ fields. Their masses are driven by large 
$m_S$ and $m_T$ soft masses, which in turn are needed to make $v_S$ and 
$v_T$ small\footnote{The mass parameter $m_S$ could in principle be much 
smaller than $m_T$ as $v_S$ is not as constrained as $v_T$.}. This leads to the 
hierarchy between them visible in the first column of every plot in  
figure \ref{fig:mass_spectrum}.
\item Due to the lack of left-right sfermion mixing and choice of common value for 
soft masses  the small mass splitting between left-  and right-chiral sfermions 
is caused by the corresponding fermion masses and the Dirac mass parameters that enter 
via D-tems, see eq.~\eqref{eq:potdirac}. In all BMPs the third-generation squarks and 
all sleptons are roughly mass degenerate with masses of order 1 TeV
\item The $R_2, R^\pm_2$ states for the BMP1 are significantly heavier than in BMP2 
and BMP3 since the loop-corrections from $R$-Higgs bosons have to account for 
a lower tree-level mass of the lightest Higgs boson, c.f. table \ref{tab:higgs}.
\item In all benchmark points charginos and neutralinos are well below 1 TeV; in the 
BMP3 the lightest neutralino is as light as $\sim 250$ GeV. 
\item Table   \ref{tab:mass_spectrum} shows that many of the new states should be 
accessible kinematically at the LHC.  Note that the Dirac nature of gauginos, together with R-charge conservation, implies distinctly different signatures in comparison to 
the MSSM. The study of phenomenological implications of our BMPs for the 
LHC physics is left for future work.
\end{itemize}

\subsection{Exploring the parameter space}
\begin{figure}[th]
\centering
\begin{minipage}{0.3\textwidth}
\includegraphics[width=\textwidth]{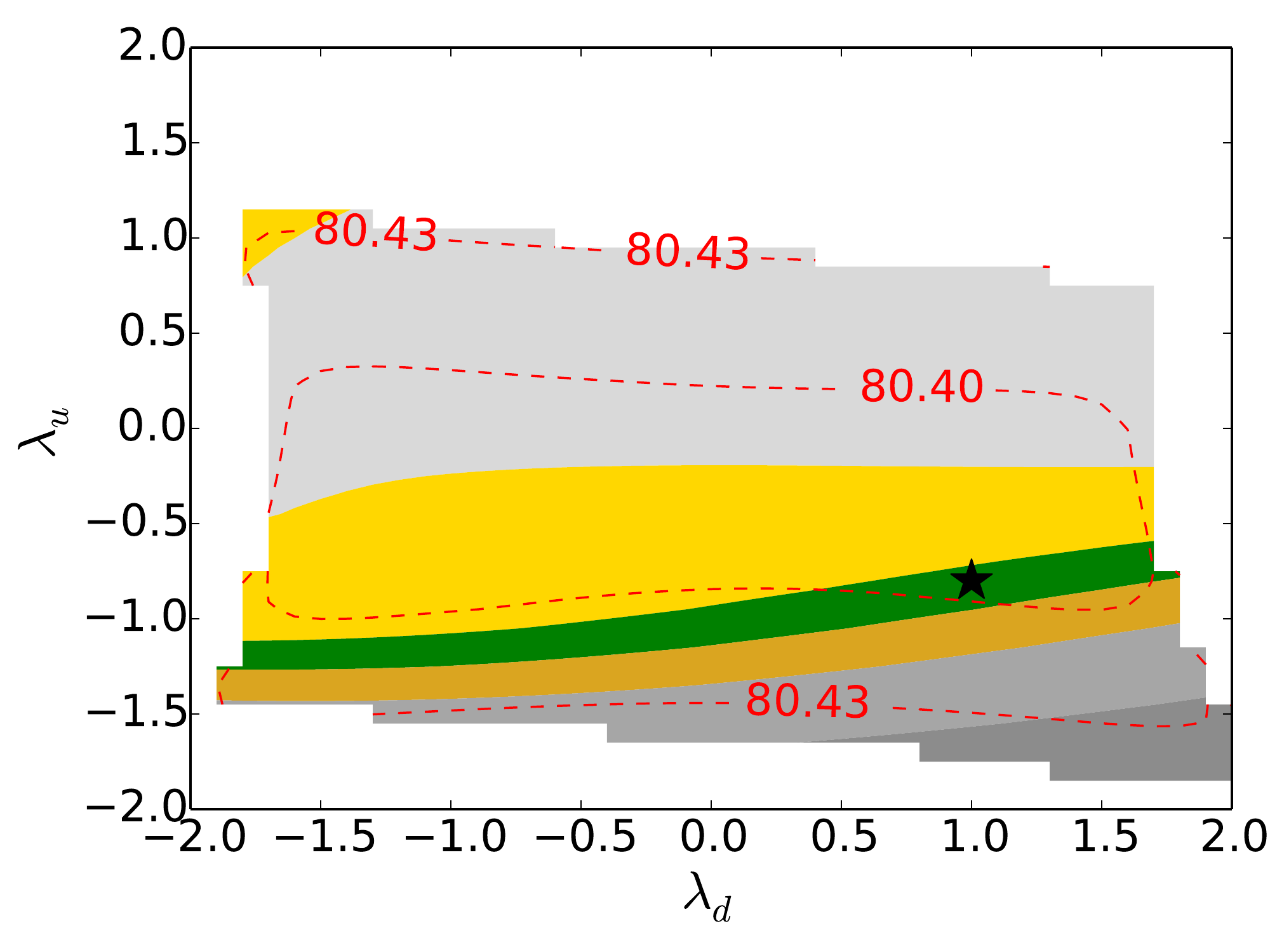}\\
\includegraphics[width=\textwidth]{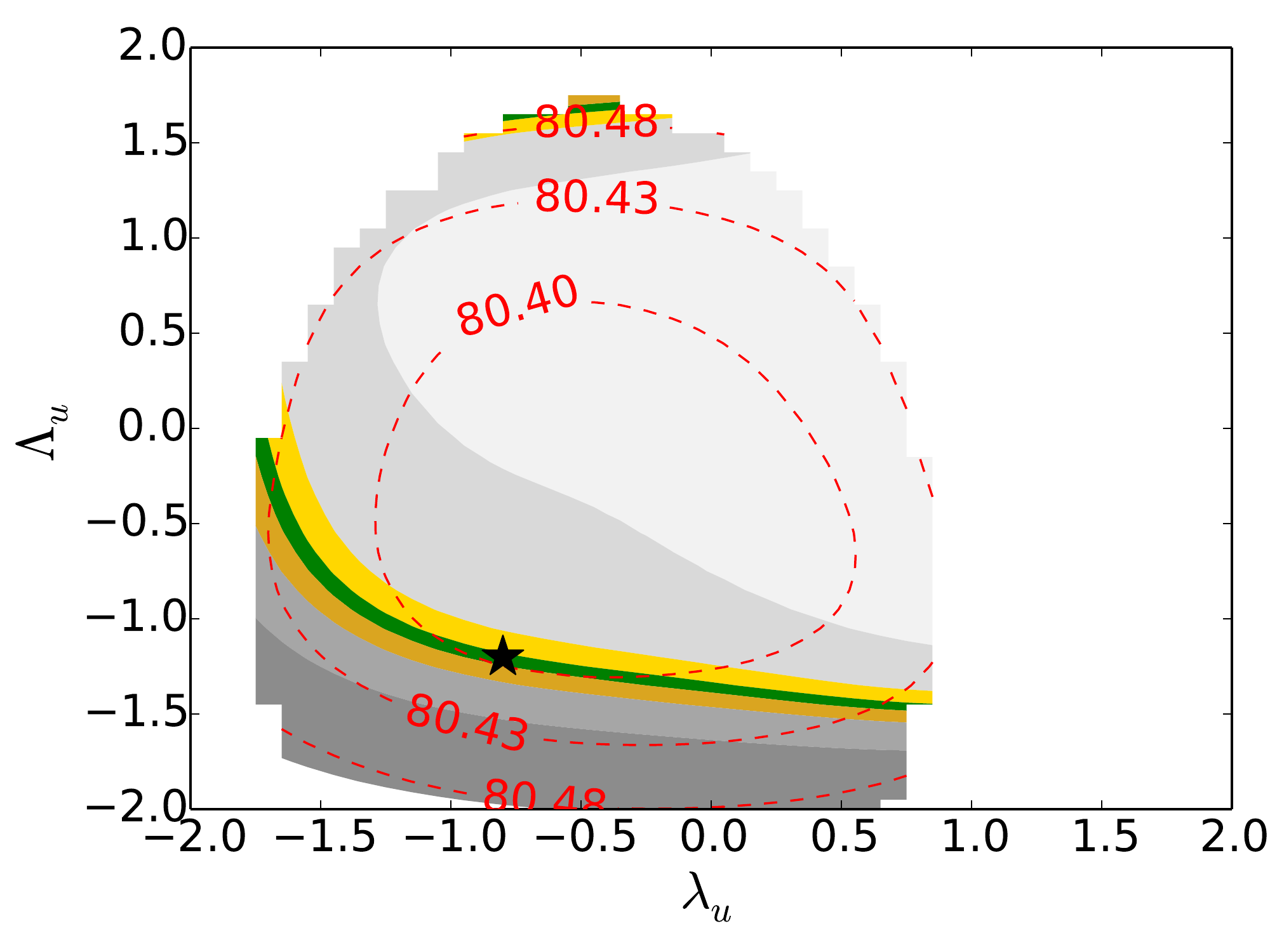}
\end{minipage}
\begin{minipage}{0.3\textwidth}
\includegraphics[width=\textwidth]{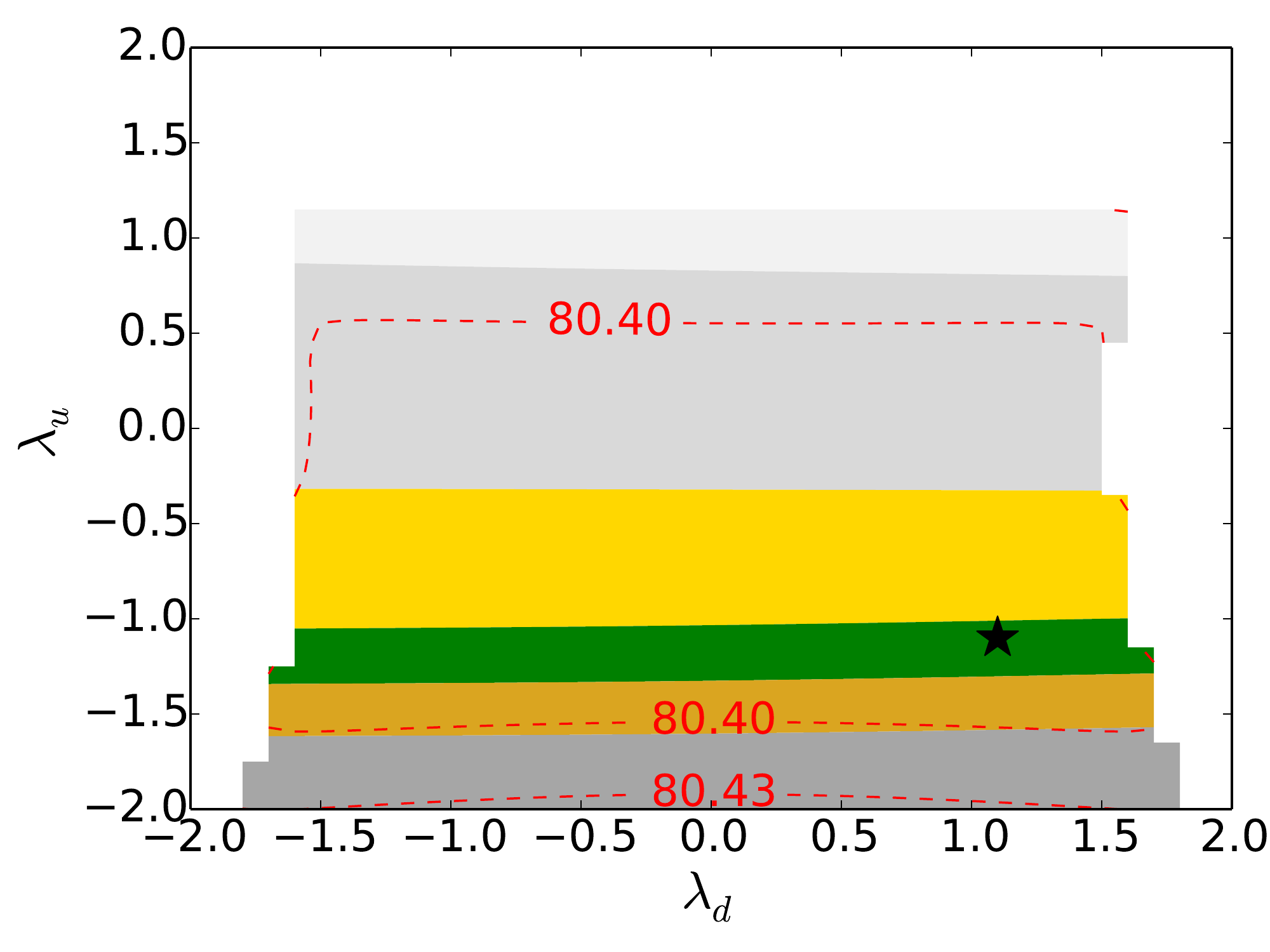}\\
\includegraphics[width=\textwidth]{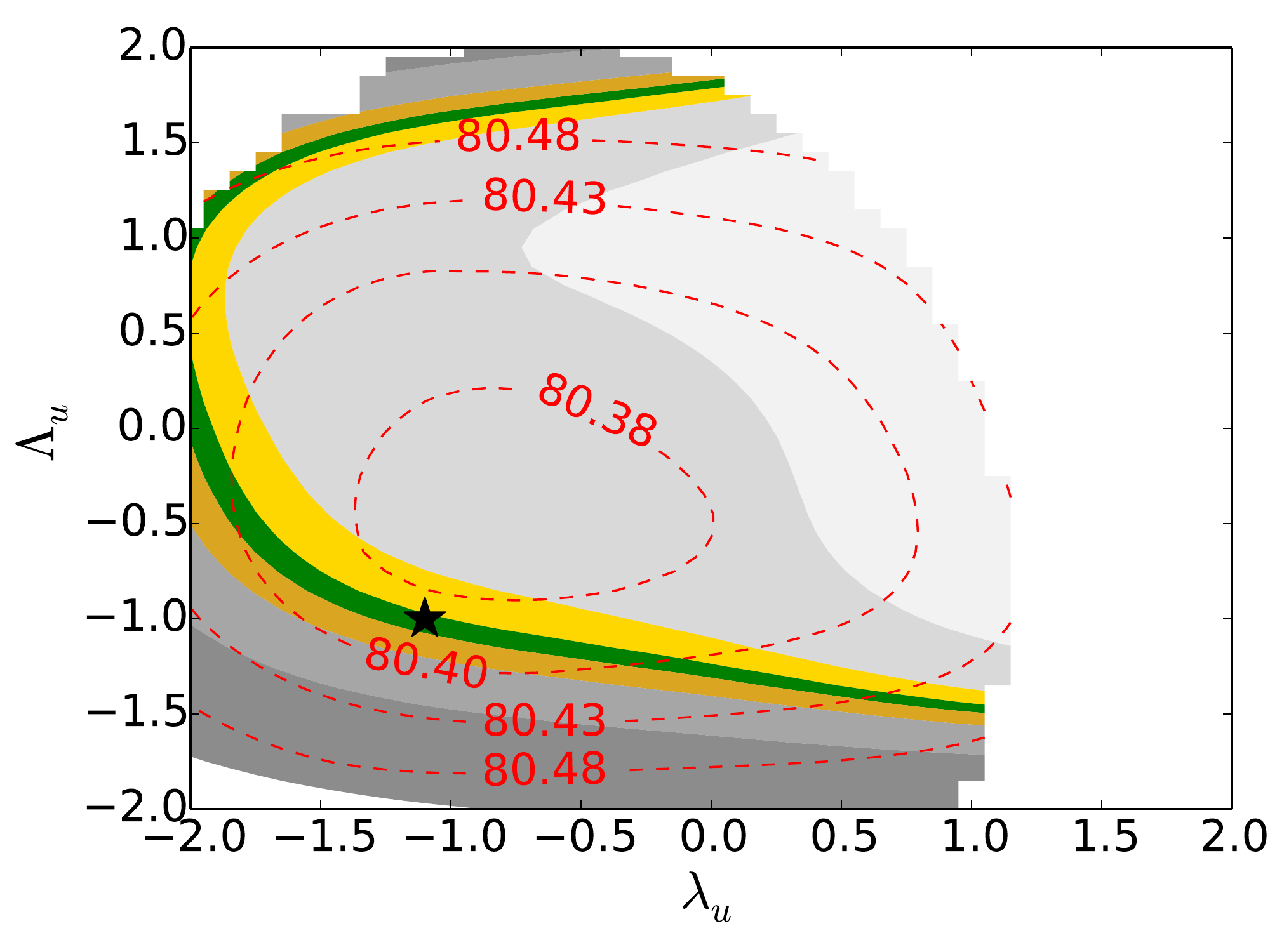}
\end{minipage}
\begin{minipage}{0.3\textwidth}
\includegraphics[width=\textwidth]{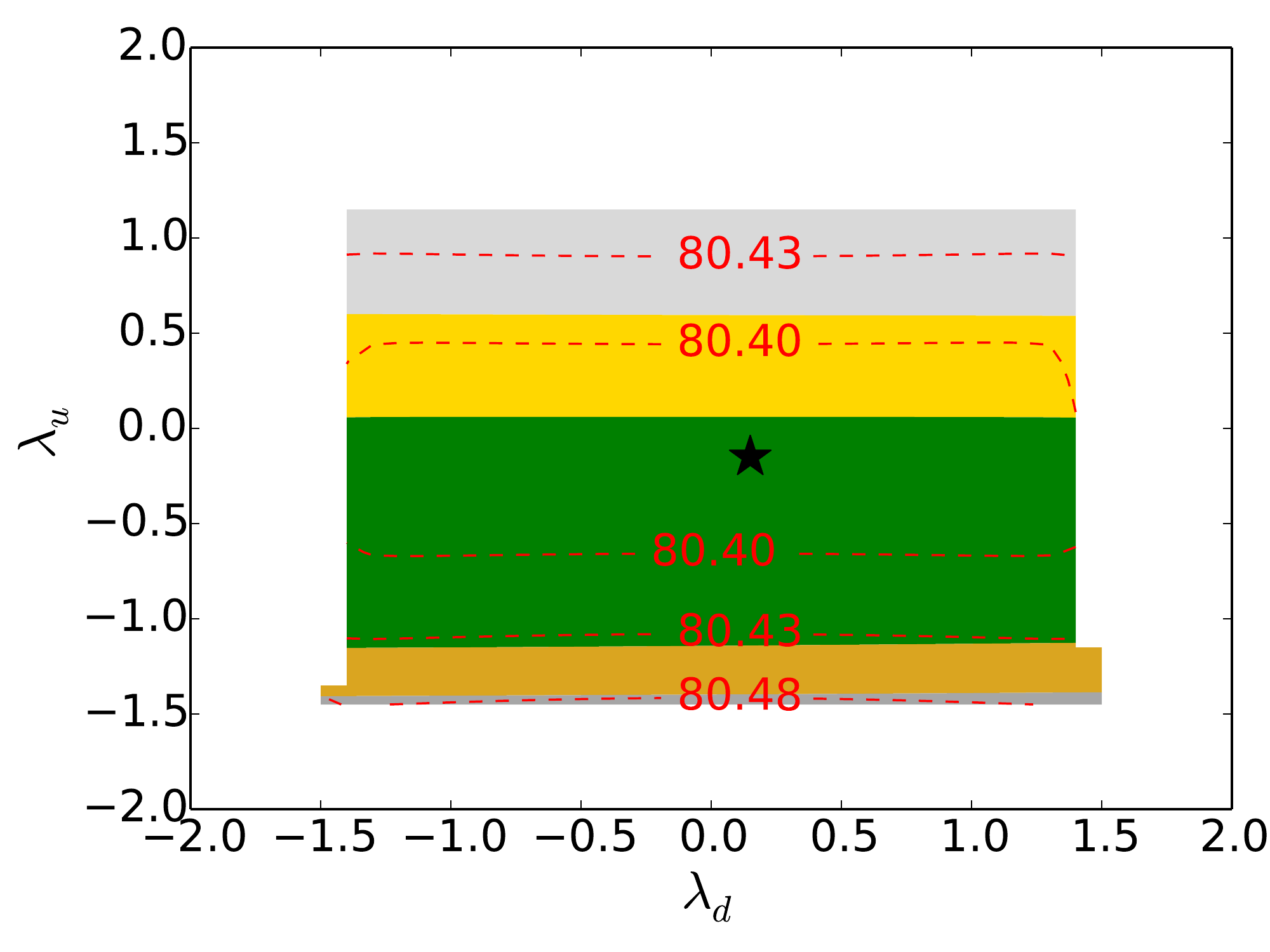}\\
\includegraphics[width=\textwidth]{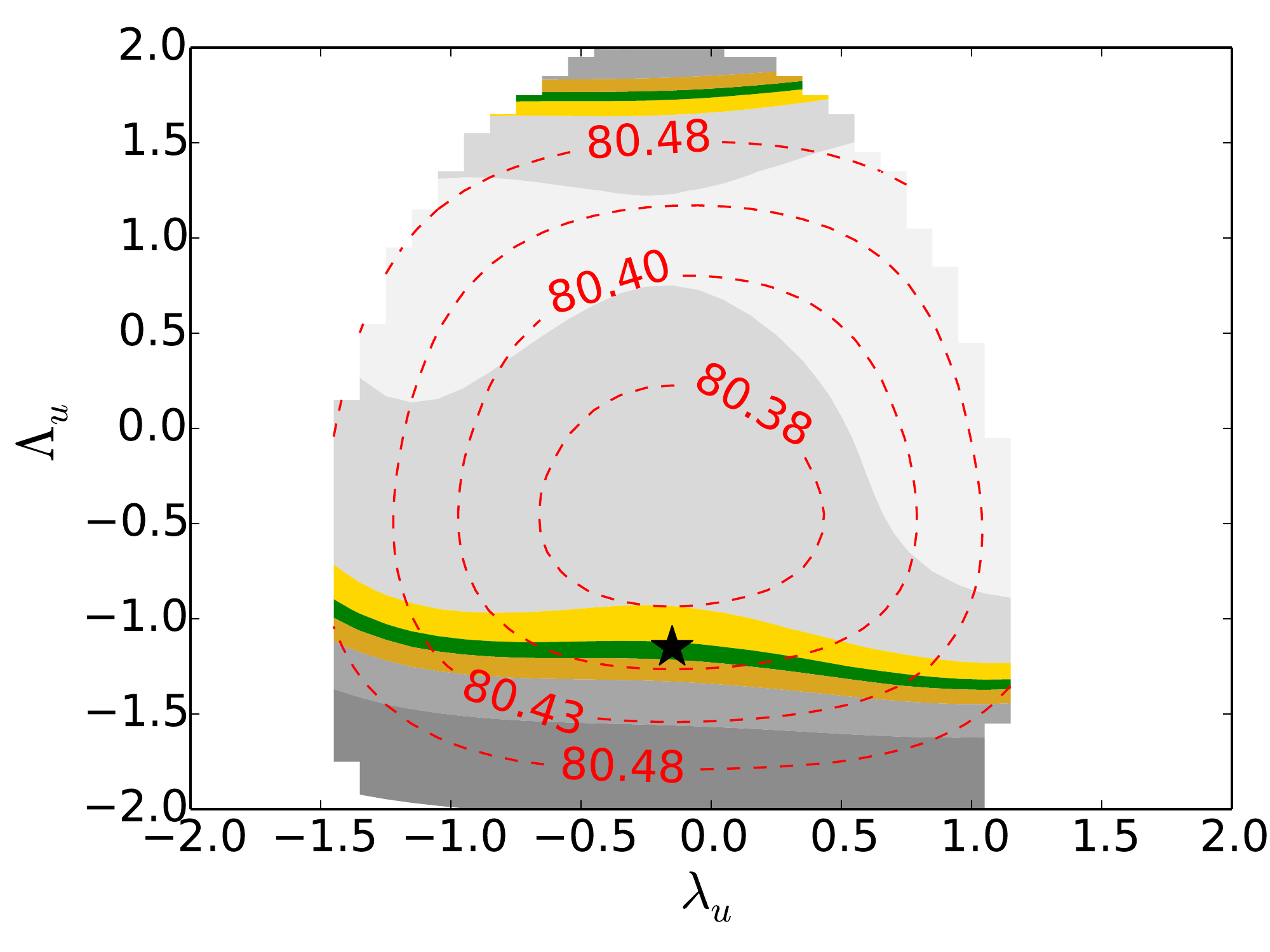}
\end{minipage}
\begin{minipage}{0.05\textwidth}
\includegraphics[width=\textwidth]{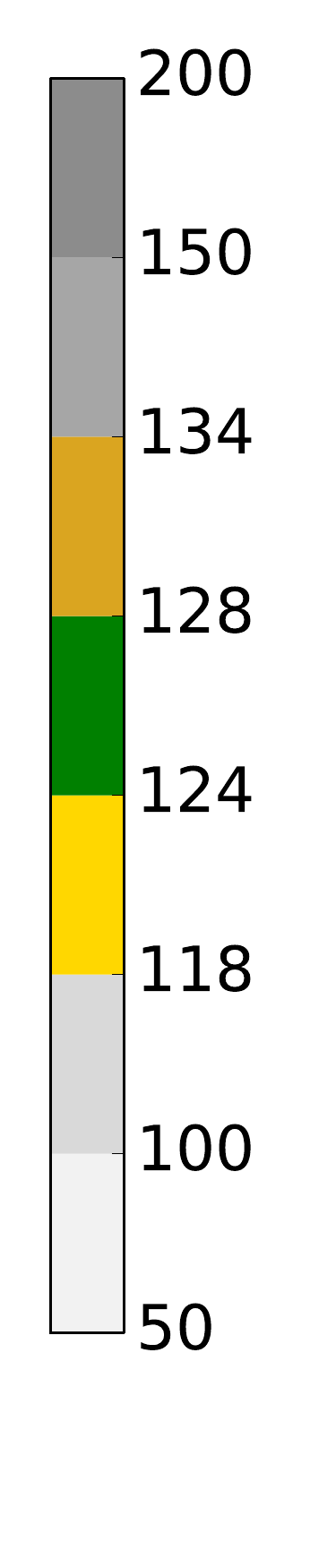}\\
\includegraphics[width=\textwidth]{img/mhmw_colorbar.pdf}
\end{minipage}
\begin{minipage}{0.3\textwidth}
\includegraphics[width=\textwidth]{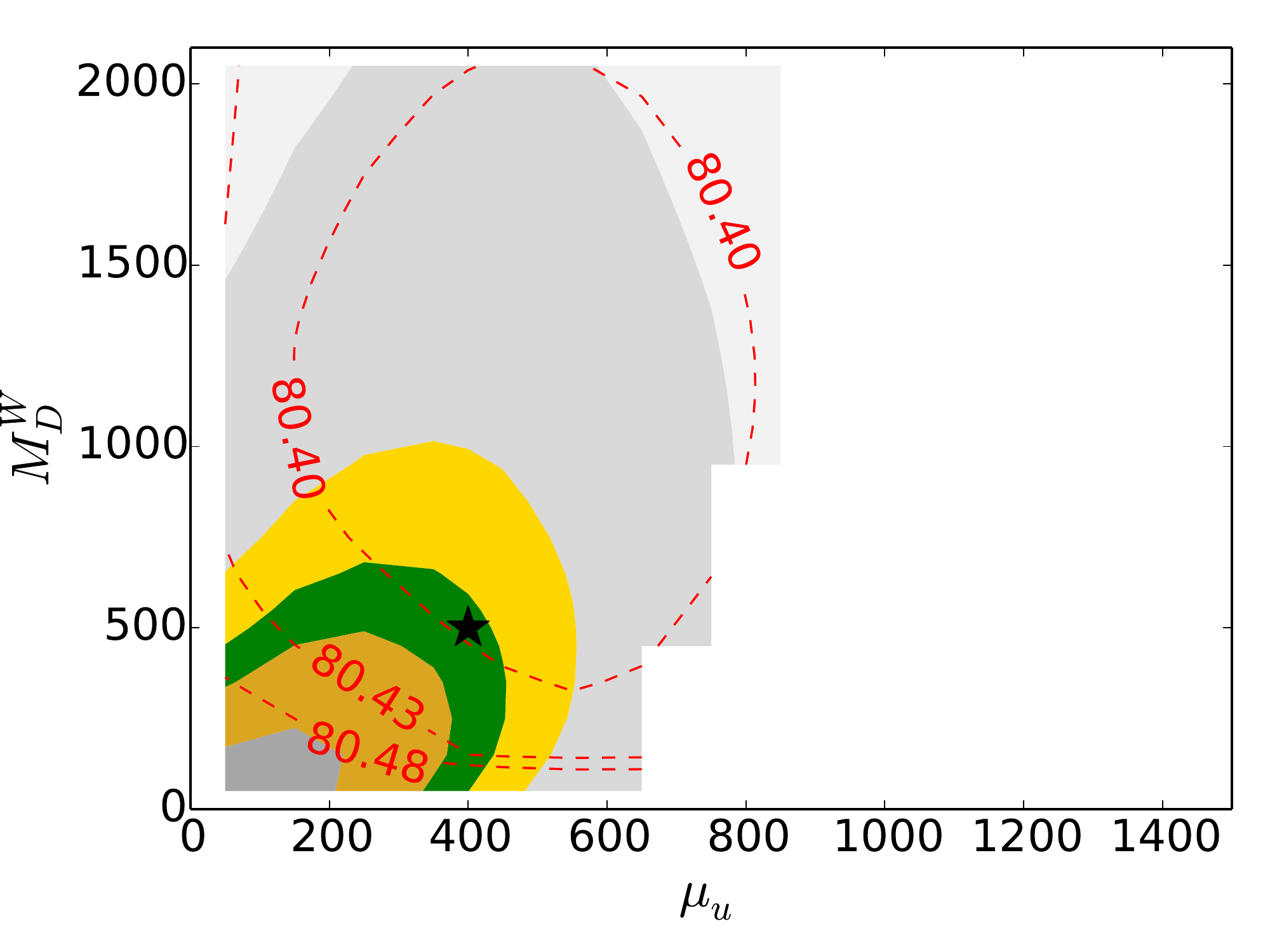}\\
\includegraphics[width=\textwidth]{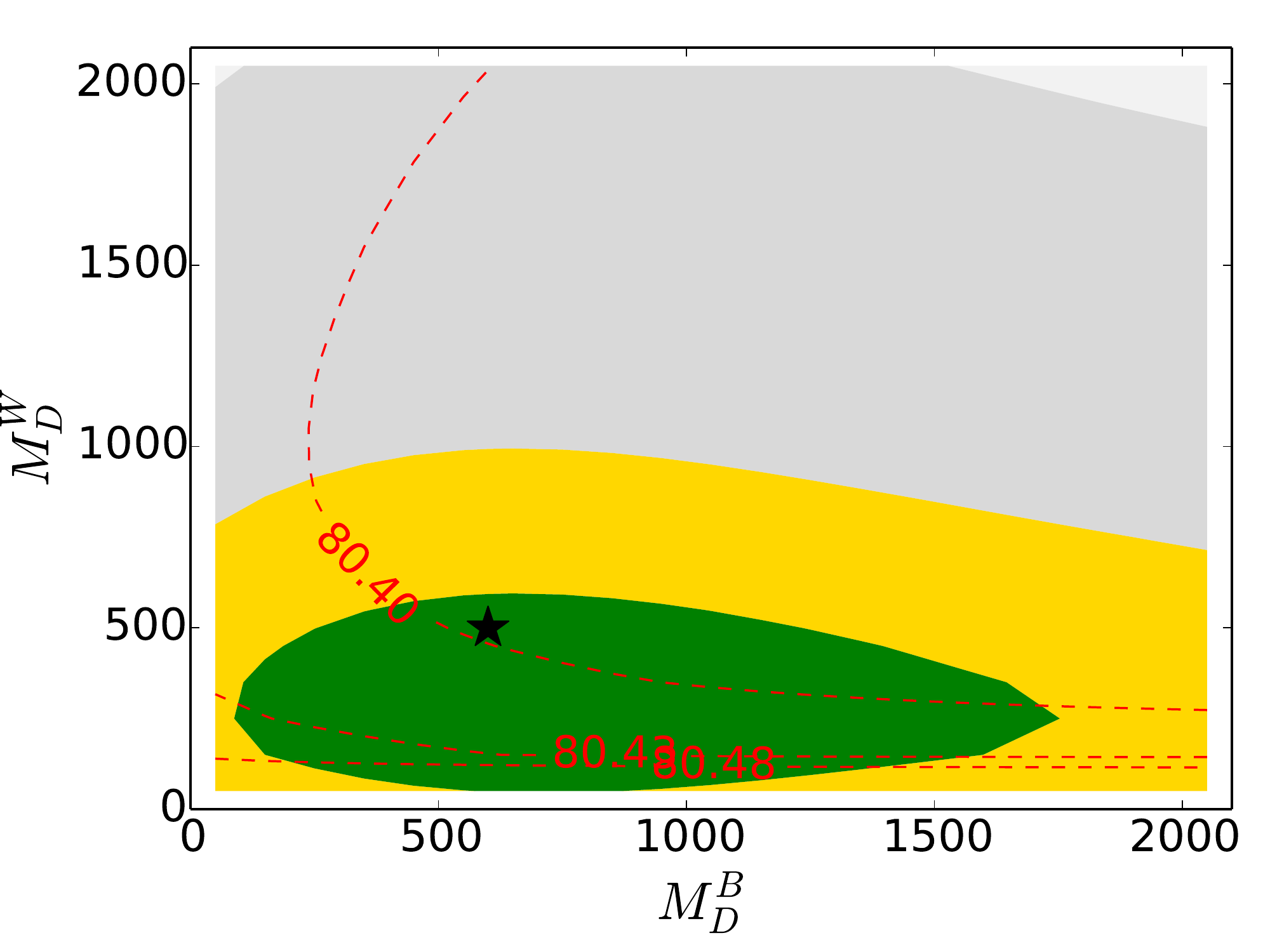}
\end{minipage}
\begin{minipage}{0.3\textwidth}
\includegraphics[width=\textwidth]{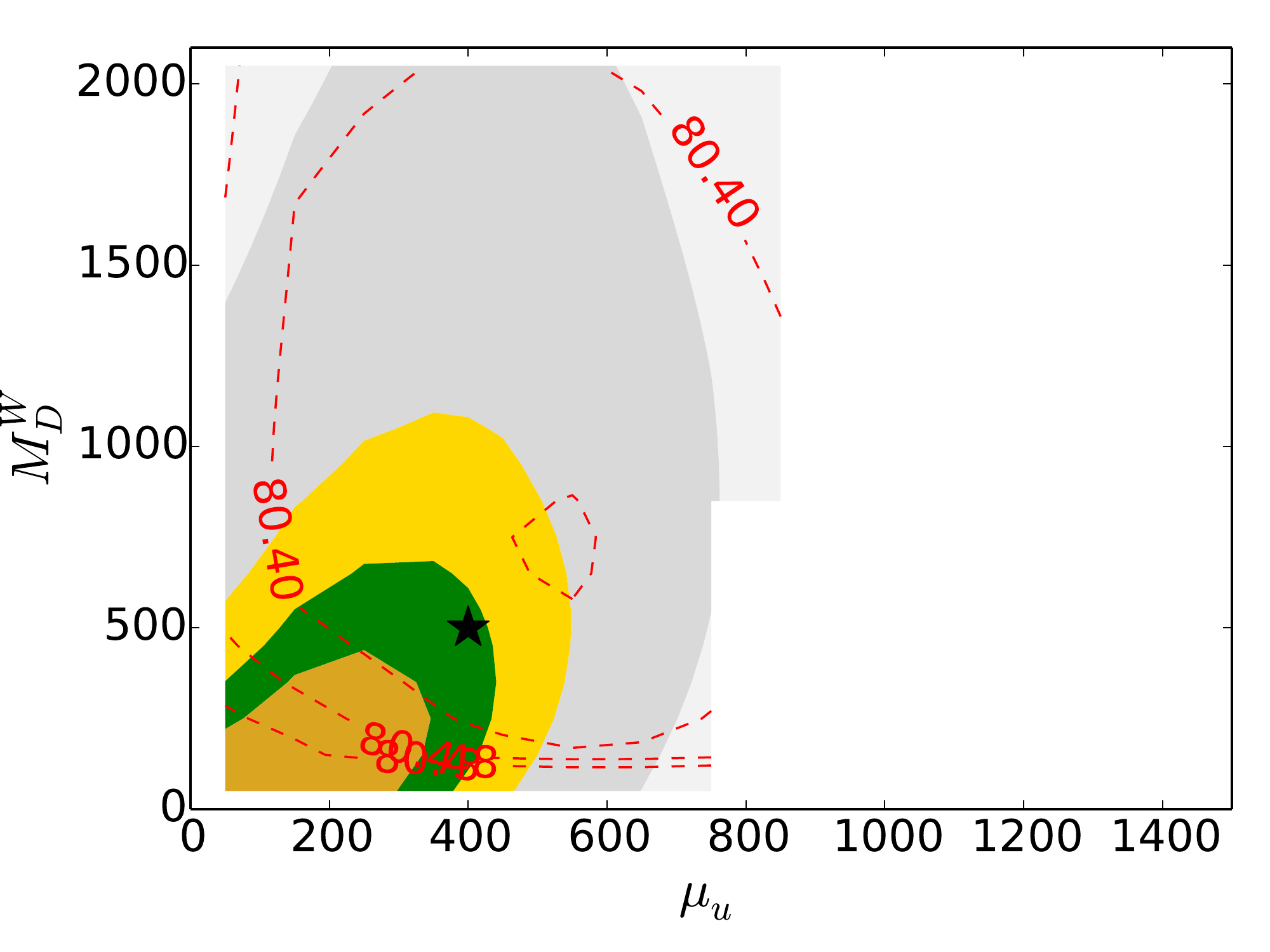}\\
\includegraphics[width=\textwidth]{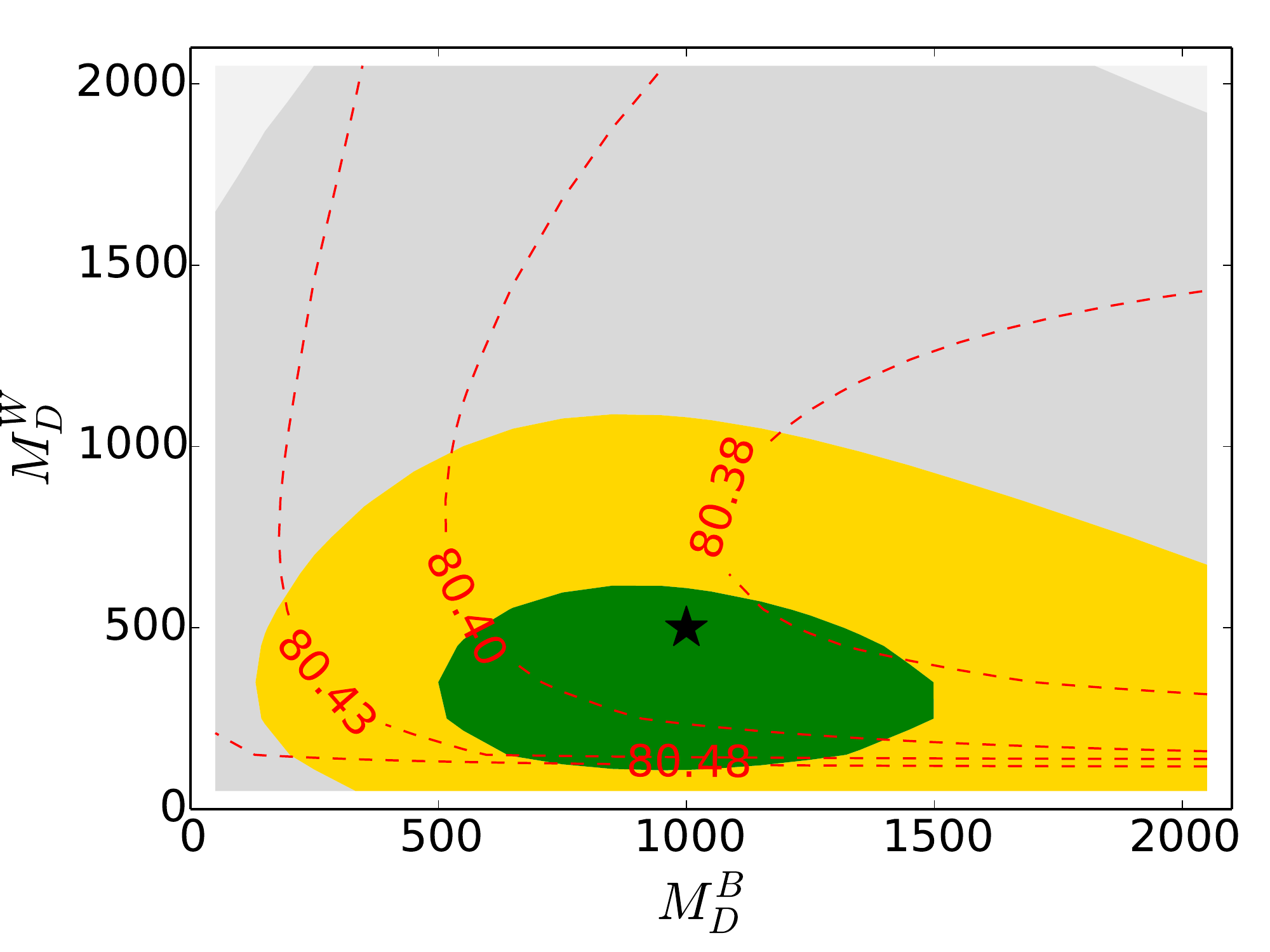}
\end{minipage}
\begin{minipage}{0.3\textwidth}
\includegraphics[width=\textwidth]{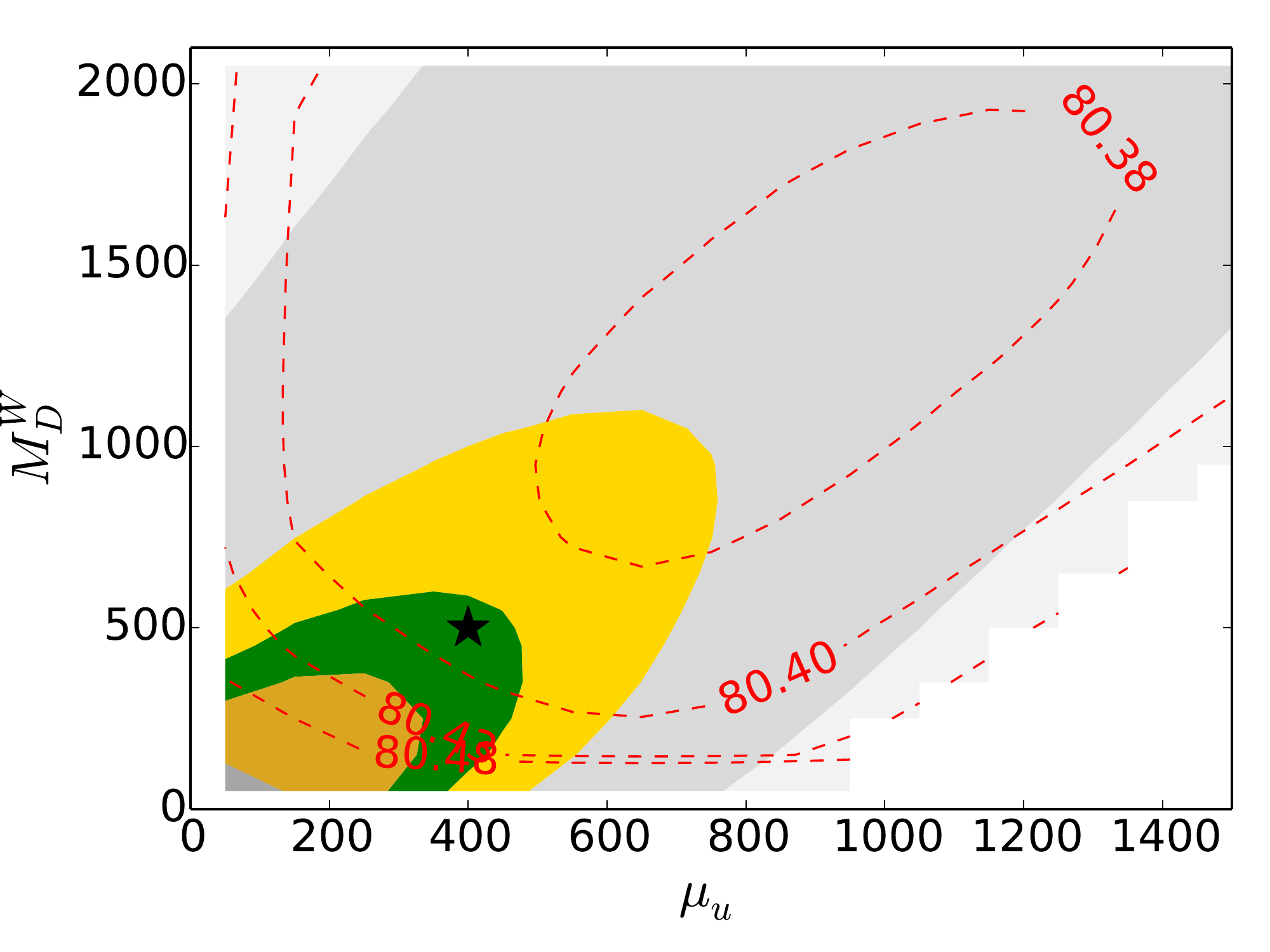}\\
\includegraphics[width=\textwidth]{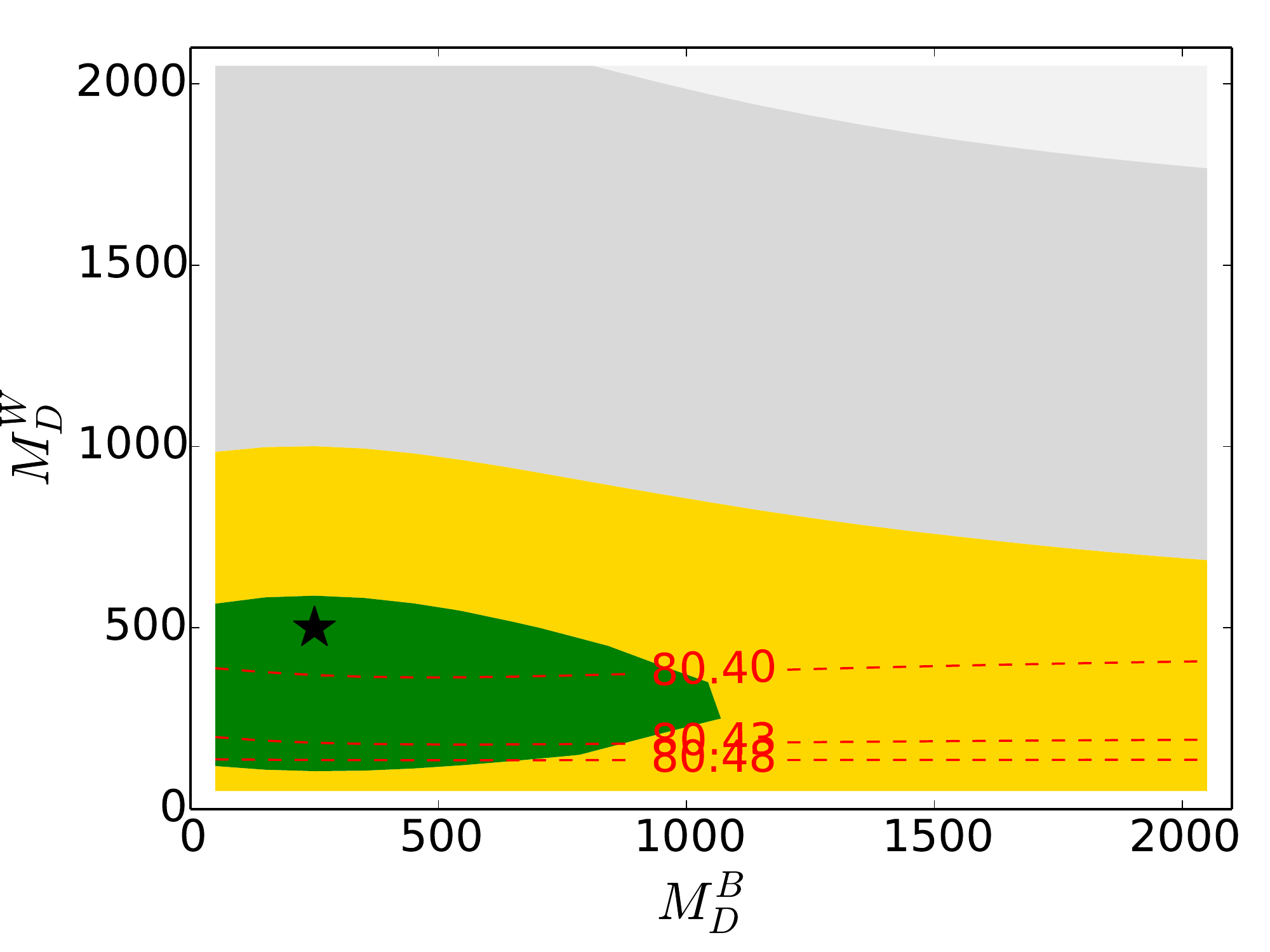}
\end{minipage}
\begin{minipage}{0.05\textwidth}
\includegraphics[width=\textwidth]{img/mhmw_colorbar.pdf}\\
\includegraphics[width=\textwidth]{img/mhmw_colorbar.pdf}
\end{minipage}
\caption{Contour plots showing the behavior of $m_{H_1}$ given by the color map and $m_W$ given by
the red contour lines. 
The plots are ordered horizontally by benchmark points with
rising $\tan\beta$, while vertically different combination of model parameters are varied.
The values for the benchmark point of each plot are marked by a star.}
\label{img:mhmw1}
\end{figure}

\begin{figure}[th]
\centering
\begin{minipage}{0.3\textwidth}
\includegraphics[width=\textwidth]{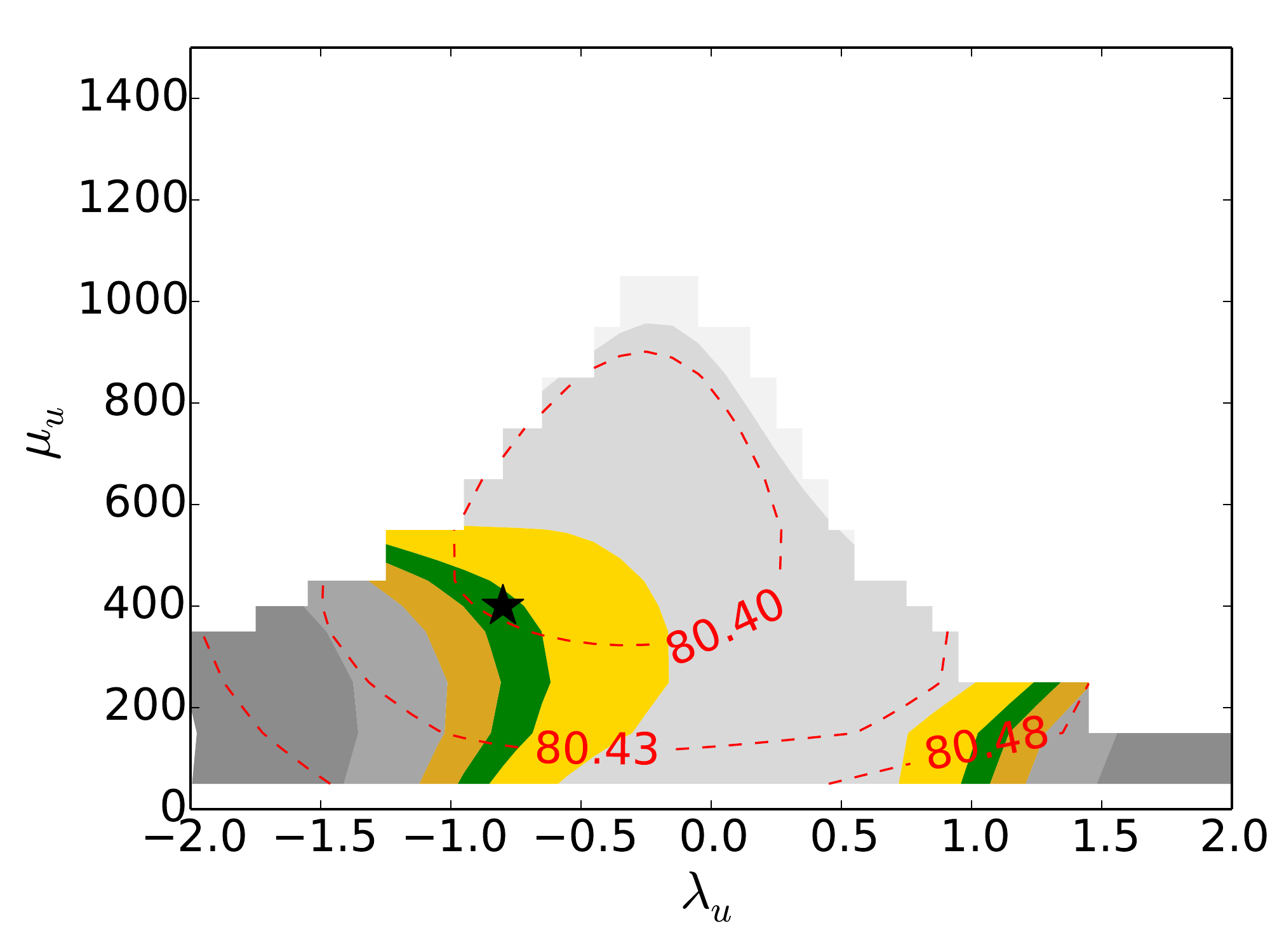}\\
\includegraphics[width=\textwidth]{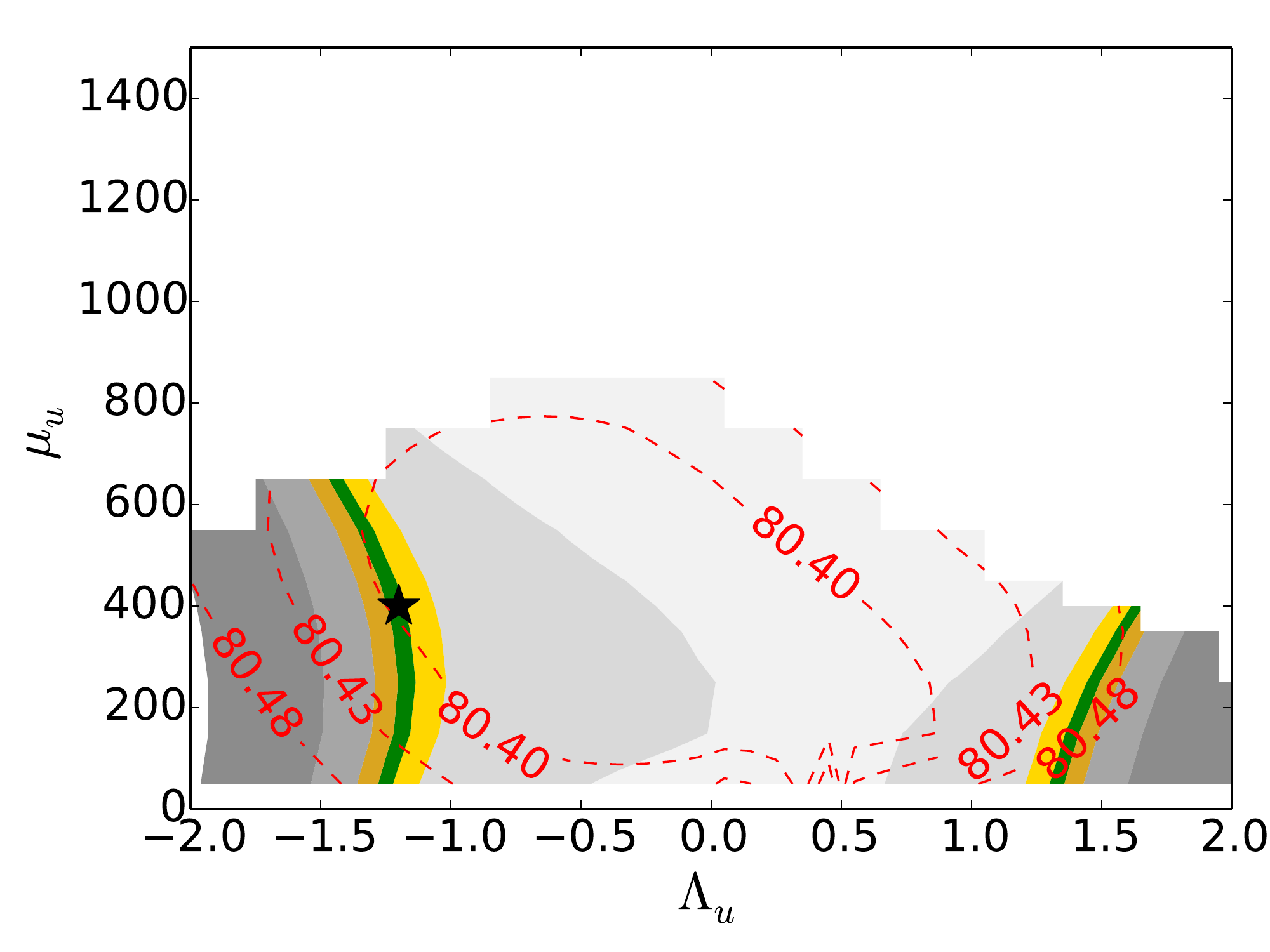}\\
\includegraphics[width=\textwidth]{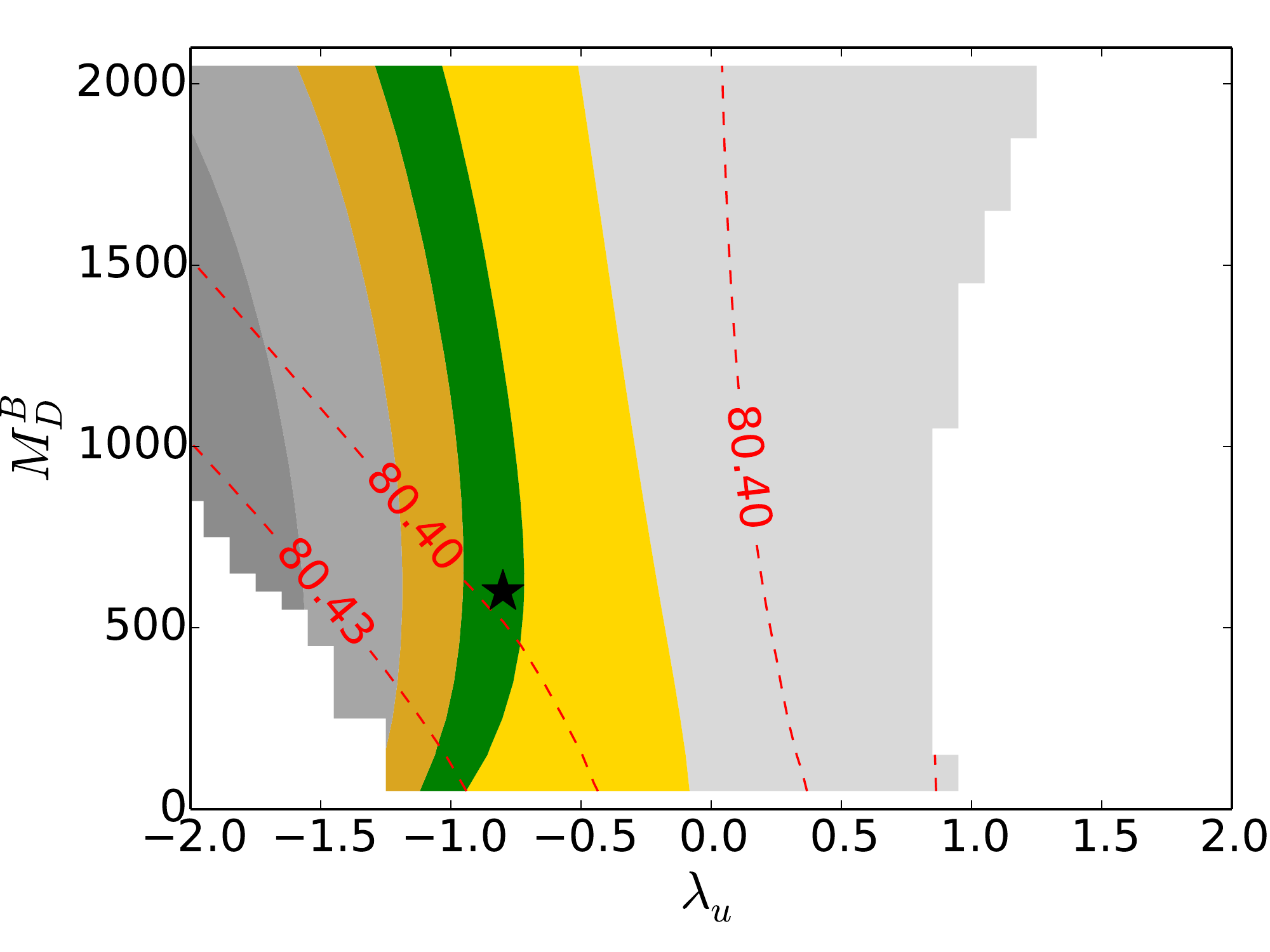}
\end{minipage}
\begin{minipage}{0.3\textwidth}
\includegraphics[width=\textwidth]{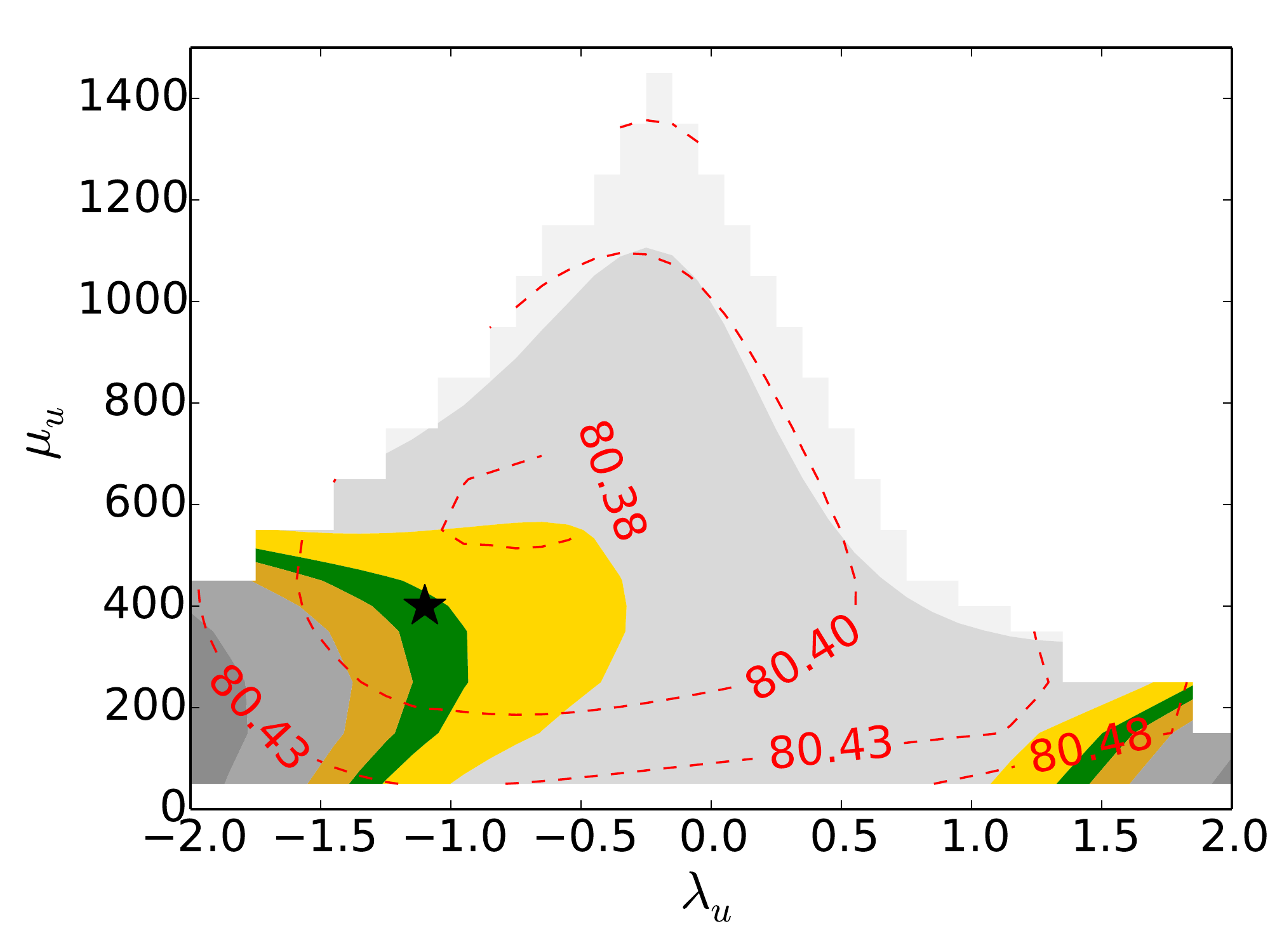}\\
\includegraphics[width=\textwidth]{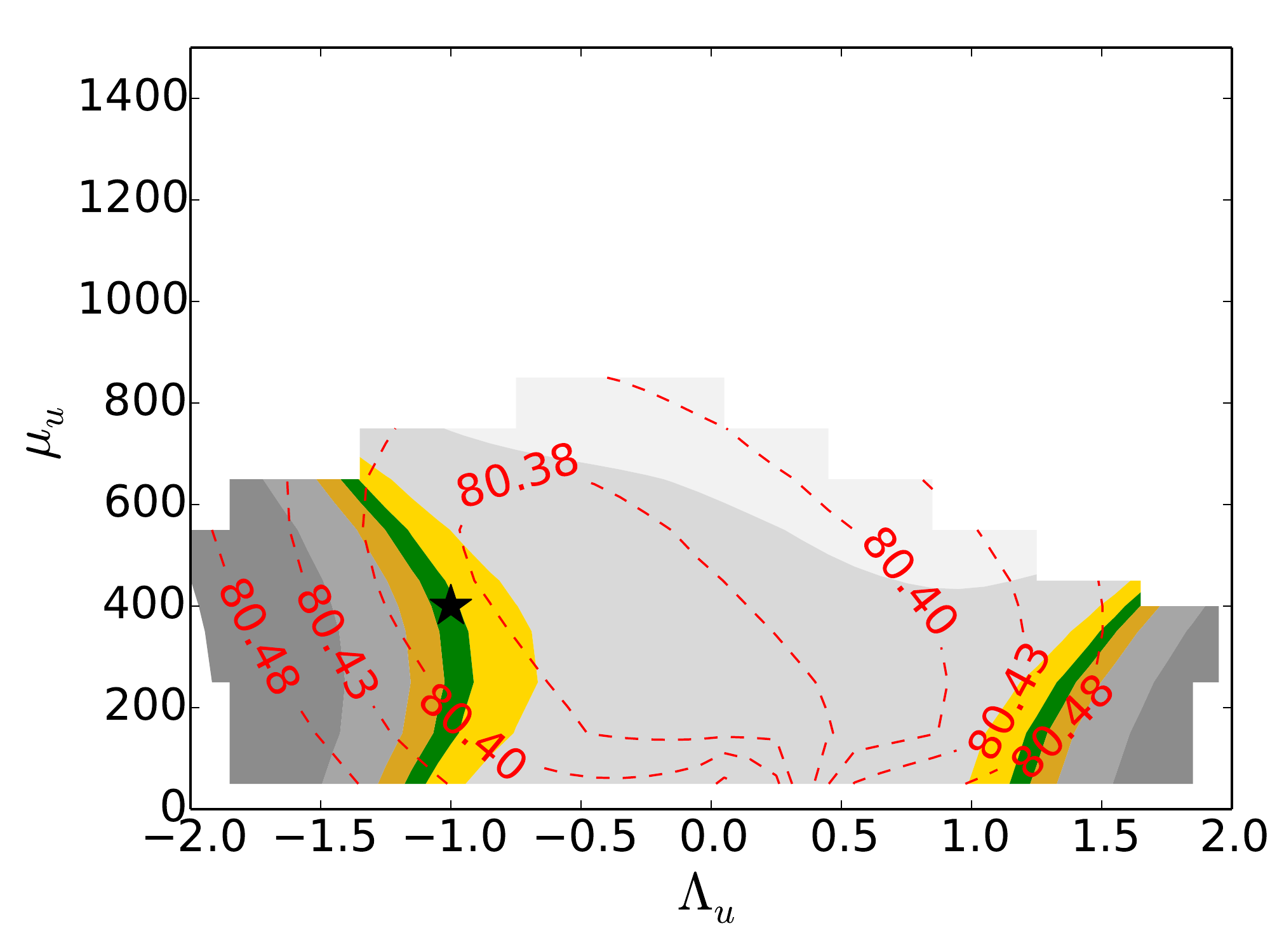}\\
\includegraphics[width=\textwidth]{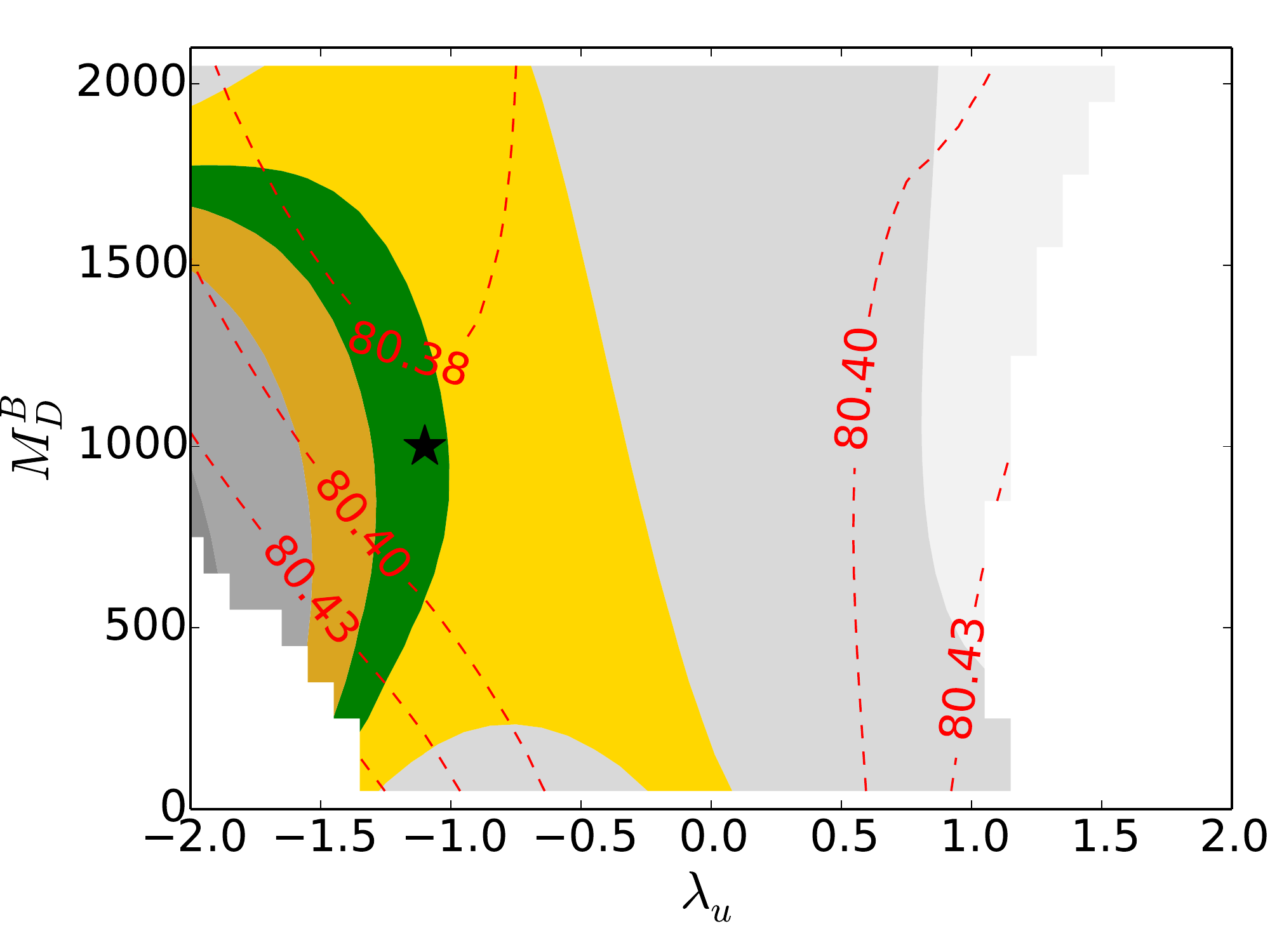}
\end{minipage}
\begin{minipage}{0.3\textwidth}
\includegraphics[width=\textwidth]{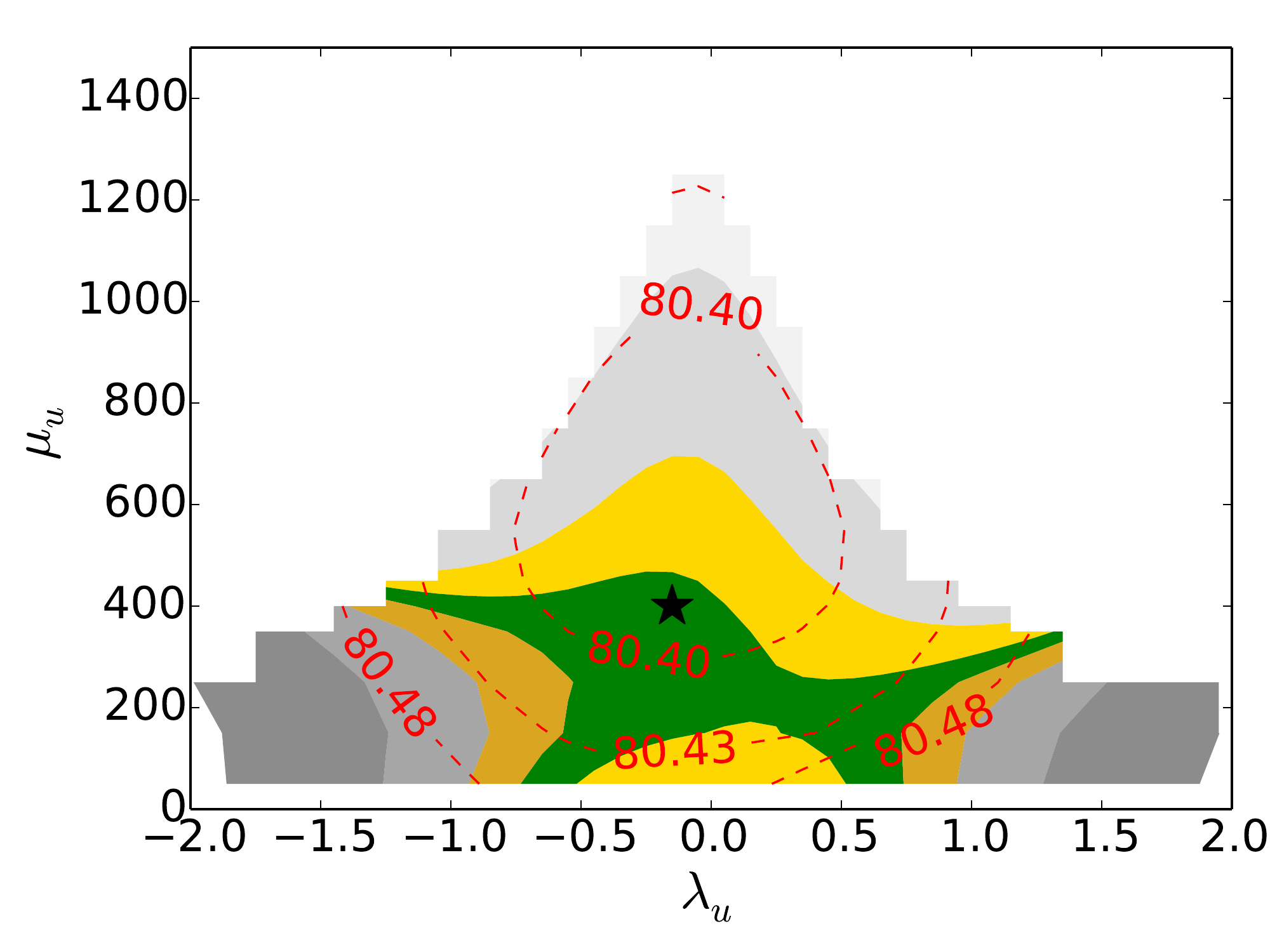}\\
\includegraphics[width=\textwidth]{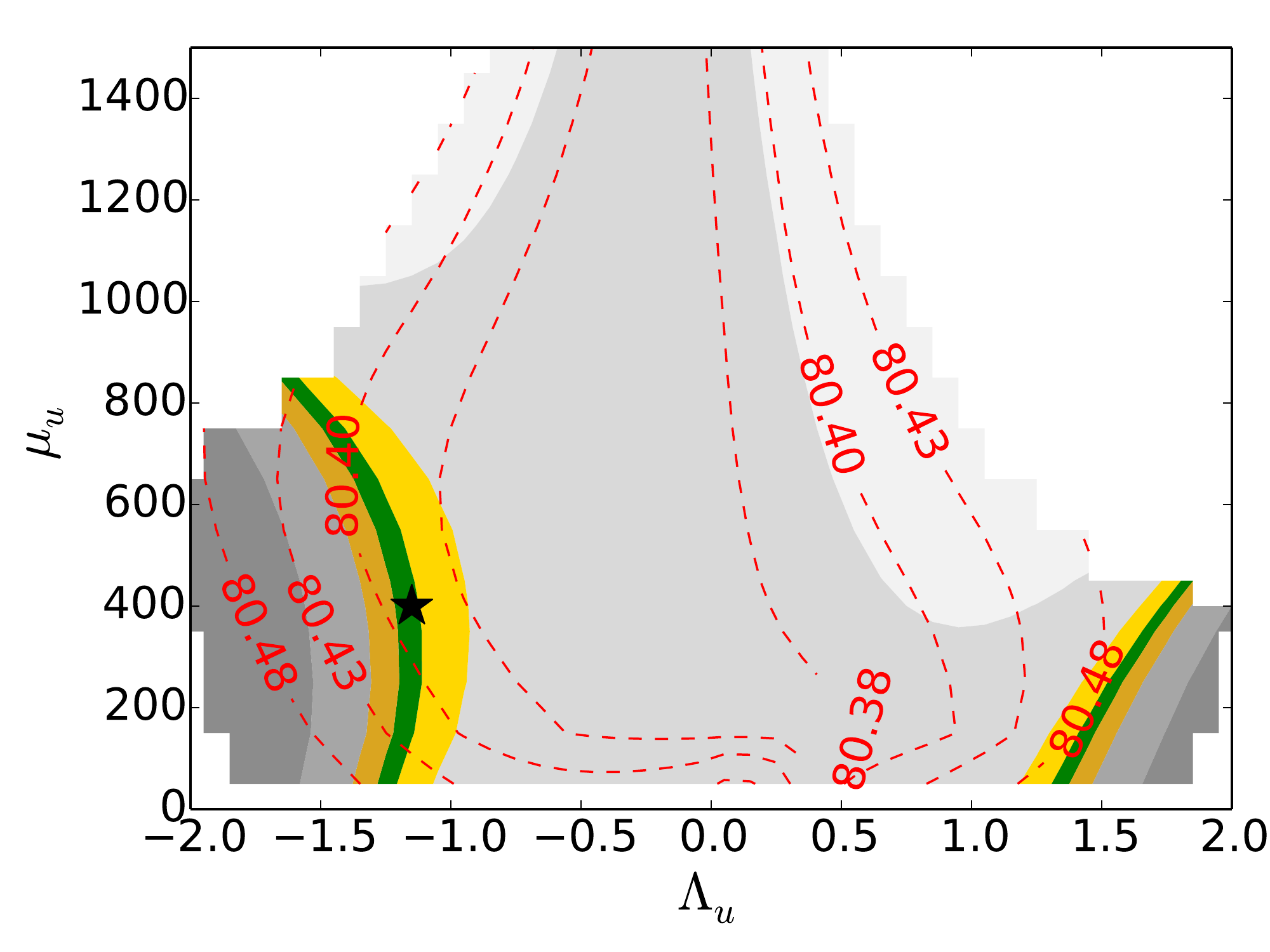}\\
\includegraphics[width=\textwidth]{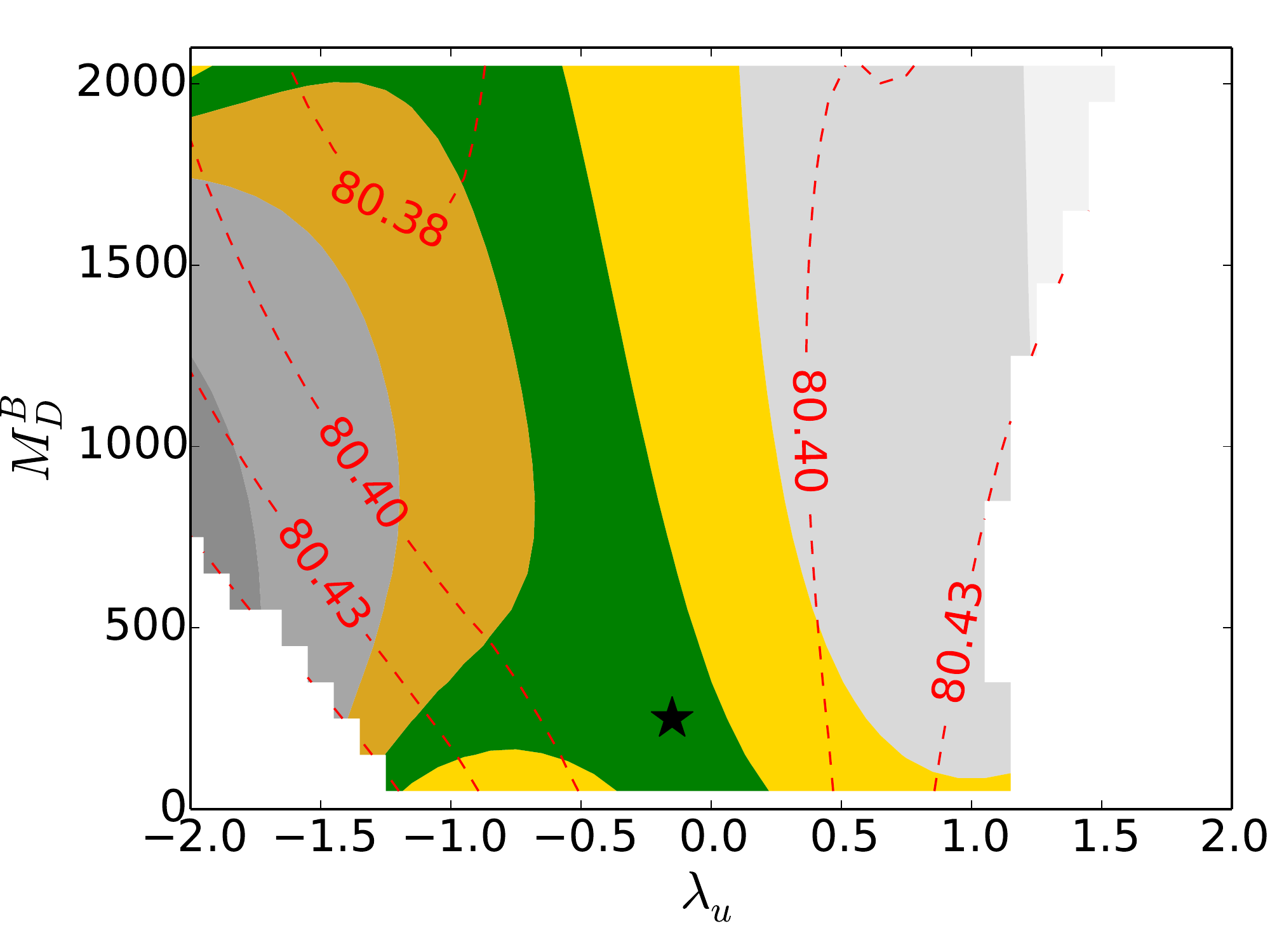}
\end{minipage}
\begin{minipage}{0.05\textwidth}
\includegraphics[width=\textwidth]{img/mhmw_colorbar.pdf}\\
\includegraphics[width=\textwidth]{img/mhmw_colorbar.pdf}\\
\includegraphics[width=\textwidth]{img/mhmw_colorbar.pdf}
\end{minipage}
\caption{Contour plots with labeling and ordering as for figure \ref{img:mhmw1}.}
\label{img:mhmw2}
\end{figure}

It is interesting to explore the parameter space  around our benchmark points 
to show the range of parameters satisfying 
both the Higgs boson mass and $m_W$ constraints. Figures  \ref{img:mhmw1} and 
\ref{img:mhmw2} show scans as functions of two chosen parameters with 
other parameters fixed and the corresponding benchmark point is marked in 
each plot.  The plots are ordered from left to right by benchmark points 
with rising $\tan\beta$.
The green (yellow) band shows the region of parameter space, where the mass of the 
lightest Higgs boson is $m_{H_1} = 126 \pm 2 \; (\pm 8)$ GeV. The red contour lines 
give the mass of the
W boson. The white regions denote regions in which the mass spectra
contain tachyonic states.

We first recall that for both the Higgs and W boson masses $m_{H_1}$ and $m_W$, large scalar
masses (compared to chargino/neutralino masses) are favorable, and the
benchmark points are chosen accordingly.
The large mass ratio enhances 
the loop contribution to $m_{H_1}$ and suppresses the contribution of the 
new scalars to $m_W$,  see the 
approximations in eqs.~\eqref{eq:effpothiggs} and \eqref{eq:drhorh} and comments there.
The exact values of the scalar masses are thus not very important, and therefore in all plots 
they will always remain fixed to their benchmark values.

The plots can be understood by noting that the new contributions to
the $m_W$ prediction are dominated by the neutralino/chargino  
contributions to $T$. The parameter dependence of the lightest Higgs boson mass $m_{H_1}$ is described by the
tree-level reduction, eq.~\eqref{eq:approx_treehiggs} and loop
corrections,  eq.~\eqref{eq:effpothiggs}. 

The top row in figure \ref{img:mhmw1} gives the behavior of  the
parameter combination $\lambda_u$, $\lambda_d$. For this combination
the change in $\tan\beta$ with the benchmark points has the strongest
influence.  As is to be expected, the down-type parameter $\lambda_d$
has a particularly noteworthy effect for small
$\tan\beta$. Nevertheless, even for $\tan\beta=3$, the dependence of
both $m_{H_1}$ and $m_W$ on the up-type parameter $\lambda_u$ is much
stronger. Therefore, in all the following plots the value of
$\lambda_d$ is kept fixed to the values of the benchmark points.

The second row in figure \ref{img:mhmw1} shows the $\Lambda_u^4$ and 
$\lambda_u^4$ behavior for $m_W$, as expected  from the approximate 
result  in eq.~\eqref{chaneumainlimit}. The Higgs boson mass dependence is
also compatible with the previous estimates and
figure \ref{img:effpotvs1l}, although some contours 
in the plots are cut off due to tachyonic states. 
Both the $\Lambda^4,\lambda^4$-behavior from eq.~\eqref{eq:effpothiggs}
and the asymmetry in $\Lambda_u$ and $\lambda_u$ from the tree-level reduction 
in eq.~\eqref{eq:approx_treehiggs} are clearly visible.
Owing to the sensitivity of both observables $m_{W,H_1}$ to both
parameters, the region where both experimental constraints are
satisfied is quite narrow, but agreement with 
experiment is possible.

In the third row the scan over the two Dirac mass parameters $\mu_u$ and $M_D^W$ is shown. The correct 
Higgs boson mass can be found in a region where the tree-level reduction, eq.~\eqref{eq:approx_treehiggs}, 
is sufficiently small. The tree-level reduction stays small for
sufficiently small Dirac masses (compared to the scalar masses
$m_{T,S}$), resulting in an elliptic shape of the contours of correct
Higgs boson mass (the axes are controlled by the ratio of $g_2$ and
$\Lambda_u$).
The behaviour of $m_W$ can be described using the vector-like limit in
eq.~\eqref{vectorlikelimit}. The Higgs boson mass prefers smaller values
of the Dirac masses, while the W boson mass prefers larger values, but
there is a significant overlap.

The lower row in figure \ref{img:mhmw1} shows the behavior with respect the two Dirac gaugino masses $M_D^B$, $M_D^W$.
As in the previous plots, too large Dirac masses reduce the loop contributions 
to $m_{H_1}$ significantly and contribute strongly to the tree-level reduction, so
that it is not possible to reach the correct Higgs boson
mass. This is the main driving force for relatively light gauginos in all benchmark points. Note however that too small Dirac masses provide too large contributions
to $m_W$, as can be understood from the $T$-parameter approximation in
the vector-like limit. It should be added that the impact of the bino mass parameter 
$M_D^B$ is far
less important than the one of the wino mass parameter $M_D^W$.

Figure \ref{img:mhmw2} shows the parameter planes $(\mu_u,\lambda_u)$,
$(\mu_u,\Lambda_u)$, and $(M_D^B,\lambda_u)$. It confirms the fact
that the $\lambda/\Lambda$ couplings are most important. The
$\Lambda_u^4$ and $\lambda_u^4$ behavior is again visible; the
$\tan\beta=40$ plot in the top row shows a rather flat behavior of
$m_{H_1}$ for a range of $\lambda_u$, which is equivalent to the
$\lambda_u$-dependence in figure \ref{img:effpotvs1l}. 

Generally, the plots in figure \ref{img:mhmw2} explicitly show that
changes in the Dirac masses can be compensated by changes in
$\lambda_u/\Lambda_u$ such that $m_{W,H_1}$ agree with experiment. The
required values of $\lambda_u$ or$\Lambda_u$ are typically close to $-1$;
their absolute values are thus similar to the top Yukawa coupling.
In particular, the plots again show that the higgsino Dirac mass
$\mu_u$ has larger influence than the bino one, $M_D^B$.  In the first two rows of figure \ref{img:mhmw2} one of the couplings $\Lambda_u/\lambda_u$ is varied against 
$\mu_u$. The
dependence of $m_{H_1}$ on $\mu_u$ shows a structure which again can be understood by the reduction formula, eq.~\eqref{eq:approx_treehiggs}, while the vector-like 
approximation of $T$ \eqref{vectorlikelimit} for $m_W$. 
Here, the minimum of $T$ is given by $\mu_u=M_D^W$ and the contributions are increasing for a larger mass splitting.
The last row of figure \ref{img:mhmw2} shows the variation of $\lambda_u$ and $M_D^B$ and makes it
clear that the bino Dirac mass only has a mild influence on the mass predictions, also justifying
that the minimum of $T$ as a function of $\mu_u$ in the plots before is given by $M_D^W$ instead.

All these plots show that a suitable range for the input parameters exists to meet 
the correct value for $m_W$ and $m_{H_1}$ but that it is non-trivial to find regions
where both constraints are fulfilled.
These 
regions are characterized by at least one large $\lambda/\Lambda$ coupling, 
high scalar masses to reduce the tree-level reduction of $m_{H_1}$ and minimize the $T$ parameter, 
as well as relatively low higgsino and gaugino Dirac masses to minimize tree-level reduction. 
 
 %%%%
\subsection{Experimental constraints}
\begin{table}
\begin{center}
\begin{tabular}{l|ccc}
  & BMP1 & BMP2 & BMP3 \\
  \hline
$m_{H_1}$ & 125.3 GeV & 125.1 GeV & 125.1 GeV \\
$m_W$ &80.399 GeV & 80.385 GeV & 80.393 GeV\\
\texttt{HiggsBounds}'s \texttt{obsratio} & $0.61$ & $0.61$ & $0.63$\\
 \texttt{HiggsSignals}'s p-value & 0.42 & 0.40 & 0.40 \\
$S$  & 0.0097 &0.0092 & 0.0032\\
$T$  & 0.090 & 0.091 & 0.085 \\
$U$  & 0.00067 &0.00065 & 0.0010 \\
\texttt{Vevacious} & $\checkmark$ & $\checkmark$ &$\checkmark$ \\
selected $b$ physics observables & $\checkmark$ &$\checkmark$ & $\checkmark$
\end{tabular}
\caption{Collection of different predictions for the benchmark points
defined in \ref{tab:BMP}. For details see the full text. \label{tab:exp-constraints} }
\end{center}
\end{table}

Table \ref{tab:exp-constraints} shows a compilation of different predictions for our
benchmark points. As argued in the introduction, the MRSSM can accommodate proper 
Higgs boson mass of around $125$ GeV with relatively light stops of around 1 TeV, without any 
left-right mixing which is absent in the MRSSM.
The W boson mass is also found in agreement with the experimental value from combined LEP 
and Tevatron measurements $m_W^{\text{exp}} = 80.385 \pm 0.015 \text{ GeV}$ \cite{PDG}.
This can be compared with the SM theory prediction 
$m^{\text{theory},\text{SM}}_W = 80.362 \pm 0.010 \text{ GeV}$ 
\cite{w-mass-sm-prediction,{Ferroglia:2012ir}}.
The Higgs sector of the benchmark points was checked against existing experimental data using 
\texttt{HiggsBounds}~\cite{HiggsBounds1,HiggsBounds2,HiggsBounds3,HiggsBounds4} \textit{v4.1.3} and 
\texttt{HiggsSignals}~\cite{Bechtle:2013xfa,Stal:2013hwa} \textit{v1.2.0}. The corresponding
results for the statistical  significance of exclusion  
are given in table \ref{tab:exp-constraints} and show
that all benchmark points are not excluded by the experimental data included in those codes. 

In section \ref{ch:mW} we described that the main contribution to $m_W$ from the MRSSM 
can be effectively described using the $S$, $T$, $U$ parameters. The values of the three 
parameters for our BMPs given in table \ref{tab:exp-constraints} fulfill the expectation that
the $T$ parameters gives the largest contribution in eq.~\eqref{eq:mW-STU},
while the next-to-largest is always given by $S$ and an order of magnitude smaller.
All values are allowed by the fits to electroweak precision observables.

The Higgs potential of the MRSSM  was checked for possible presence of deeper minima in the 
$\{v_d, v_u, v_S, v_T \}$ space using \texttt{Vevacious}~\cite{Camargo-Molina:2013qva} 
\textit{v1.1.00}. 
Within the validity of approximations used by \texttt{Vevacious}, none were found. 
Searches for deeper charge- and/or color-breaking minima were not performed, as we expect none of them due to the 
absence of A-terms in the MRSSM~\cite{Casas:1995pd}.
 
We checked also selected low energy $b-$physics observables, namely
 $B \to X_s \gamma$, $B_{s/d} \to \mu^+ \mu^-$ and $\Delta M_{B_{s/d}}$ using \texttt{SARAH}'s 
\texttt{FlavorKit} interface and found agreement with measurements,  as expected from 
\cite{Kribs:2007ac}. Finally the effective couplings  of the lightest Higgs boson to 
quarks, leptons and gauge bosons, normalized to the SM value, are listed in table  \ref {tab:higgs_coupling_strength}. 
The enhanced couplings to the down-type fermions result from the admixture of $\phi_d$ to the lightest Higgs boson.
\begin{table}
\centering
\begin{tabular}{l|ccccccc}
& $\gamma \gamma$ & gg & $W^\pm W^{\mp*}$ & $Z Z^*$ & $\tau^+ \tau^-$ & $c \bar{c}$ & $b \bar{b}$ \\
\hline
BMP1 & 0.909 & 1.003 & 0.999 & 0.999 & 1.212 & 0.980 & 1.212 \\
BMP2 & 0.917 & 1.023 & 0.999 & 0.999 & 1.192 & 0.998  & 1.192 \\
BMP3 & 0.906 & 1.027 & 1     & 1     & 1.130 & 1  & 1.130  
\end{tabular}
\caption{\label{tab:higgs_coupling_strength} Normalized (squared) effective couplings of the lightest  Higgs particle to gauge boson and fermion pairs at leading order.}
\end{table}

\section{Conclusions}
Models with a continuous R-symmetry contain the complete set of
possible types of symmetries in a relativistic quantum field
theory. They are phenomenologically appealing for a variety of reasons, e.g. 
they solve the SUSY flavor problem and relax direct search limits
from the LHC. The minimal model, the MRSSM, contains many new states,
in particular R-Higgs doublet and singlet/triplet superfields, whose
fermionic components partner with the usual higgsinos and
SU(2)$\times$U(1) gauginos to give Dirac-type gaugino and higgsino
masses.

In this paper we have shown how the MRSSM can accommodate the measured values
of the Higgs boson and W boson masses. We have computed both
observables via the full one-loop calculations of Higgs boson self
energies and muon decay. The most important new model
parameters are the superpotential couplings $\lambda_{u,d}$,
$\Lambda_{u,d}$. They enter in a way similar to the top/bottom Yukawa
couplings, with the role of the quark doublets and singlets played by
the R-Higgs doublets and the singlet/triplet.

For the case of the Higgs boson mass, the
$\lambda/\Lambda$-contributions have essentially the same structure as
the top/stop contribution. If the $\lambda/\Lambda$'s are of the order
of the top Yukawa couplings, the Higgs boson mass can easily be as
large as 125 GeV or even larger for a broad range of $\tan\beta$. Stop masses beyond 
the TeV are not required.

For the case of the W boson mass, there exists a limit (for vanishing
Dirac higgsino/gaugino masses) in which the
$\lambda/\Lambda$-contributions have the same structure as the ones
from the top/bottom sector. However, in the more likely case of
non-negligible Dirac mass parameters (which have no counterpart in the
quark sector), the behavior is different. The interplay between the
$\lambda/\Lambda$'s and the Dirac mass parameters has a strong impact
on the W boson mass. 

To summarize: the experimental values of $m_W$ and $m_{H_1}$ impose stringent and non-trivial constraints on the model parameters. 
Nevertheless it is easy to identify regions in the parameter space which
accommodate the observed values. These regions are characterized by at least one large $\lambda/\Lambda$ 
coupling and high scalar masses to reduce the tree level reduction of $m_{H_1}$ and minimize the $T$ parameter, 
as well as relatively low higgsino and gaugino Dirac masses to minimize tree level reduction. 

The proposed benchmark points contain many of the new states within the  
kinematical reach at the LHC, in particular supersymmetric fermions.  
Given the Dirac nature of gauginos, together with R-charge conservation, distinctly different signatures 
in comparison to the MSSM are expected. 
This motivates further dedicated studies of phenomenological implications of
our benchmark points for LHC physics.

\acknowledgments

P.D. and W.K. would like to thank Florian Staub for continuous help on \texttt{SARAH} and Alex Voigt for help in cross-checking results with \texttt{FlexibleSUSY}. P.D. thanks Gregor Hellwig for providing his STU package for use. We also thank Tim Stefaniak for help concerning \texttt{HiggsBounds} and \texttt{HiggsSignals}, and Tania Robens for useful discussions and comments.  

This work was supported in part by the Polish National Science Centre
grants under OPUS-2012/05/B/ST2/03306, DEC-2012/05/B/ST2/02597,
DEC-2011/01/M/ST2/02466,  the European Commission through the contract PITN-GA-2012-316704 (HIGGSTOOLS), the German DAAD PPP Poland Project
56269947 "Dark Matter at Colliders", the German DFG Research
Training Group 1504 and the DFG grant STO 876/4-1.

%\paragraph{Note added.} This is also a good position for notes added
%after the paper has been written.

\appendix
% Please always give a title also for appendices.

\section{Feynman rules}
In this Appendix we present Feynman rules for vertices needed in Appendix B, which are peculiar for the MRSSM model due to the Dirac nature of neutralinos and different composition of charginos.

\noindent
\begin{longtable}{ll}
\raisebox{-0.475\height}{
\begin{fmffile}{chi0chi+W-} 
\fmfframe(20,20)(20,20){ 
\begin{fmfgraph*}(45,45) 
\fmfleft{l1}
\fmfright{r1,r2}
\fmf{fermion}{v1,l1}
\fmf{fermion}{r1,v1}
\fmf{boson}{r2,v1}
\fmflabel{$\overline{{\chi}_{{i}}}$}{l1}
\fmflabel{${\chi}^+_{{j}}$}{r1}
\fmflabel{$W^-_{{\mu}}$}{r2}
\end{fmfgraph*}} 
\end{fmffile}
}
&
%  \chi^0_i \chi^+_j W^-_\mu:\;\;
$\frac{\imath}{2} g_2 \gamma_\mu \left [\left(2 V^{1*}_{j 1} N_{{i 2}}^{1}  - \sqrt{2} V^{1*}_{j 2} N_{{i 3}}^{1} \right)\mathbb{P}_L
  +  \left (2 N^{2*}_{i 2} U_{{j 1}}^{1}  + \sqrt{2} N^{2*}_{i 3} U_{{j 2}}^{1} \right )\mathbb{P}_R \right] $\\[3mm]
\raisebox{-0.475\height}{
\begin{fmffile}{chi-chi0W+} 
\fmfframe(20,20)(20,20){ 
\begin{fmfgraph*}(45,45) 
\fmfleft{l1}
\fmfright{r1,r2}
\fmf{fermion}{v1,l1}
\fmf{fermion}{r1,v1}
\fmf{boson}{v1,r2}
\fmflabel{$\overline{{\chi}^+_{{i}}}$}{l1}
\fmflabel{${\chi}_{{j}}$}{r1}
\fmflabel{$W^+_{{\mu}}$}{r2}
\end{fmfgraph*}} 
\end{fmffile} 
}
&
%&&\chi^-_i\chi^0_jW^+_\mu:\;\; 
$ \frac{\imath}{2} g_2 \gamma_\mu \left [ \left (2 N^{1*}_{j 2} V_{{i 1}}^{1}  - \sqrt{2} N^{1,*}_{j 3} V_{{i 2}}^{1} \right ) \mathbb{P}_L  
  + \, \left (2 U^{1*}_{i 1} N_{{j 2}}^{2}  + \sqrt{2} U^{1*}_{i 2} N_{{j 3}}^{2} \right ) \mathbb{P}_R \right ]$\\[3mm]
\raisebox{-0.475\height}{
\begin{fmffile}{chi0rho-W+} 
\fmfframe(20,20)(20,20){ 
\begin{fmfgraph*}(45,45) 
\fmfleft{l1}
\fmfright{r1,r2}
\fmf{fermion}{v1,l1}
\fmf{fermion}{r1,v1}
\fmf{boson}{v1,r2}
\fmflabel{$\overline{{{\chi}}_{{i}}}$}{l1}
\fmflabel{${\rho}^-_{{j}}$}{r1}
\fmflabel{$W^+_{{\mu}}$}{r2}
\end{fmfgraph*}} 
\end{fmffile}   
}
&
 %\chi^0_i\rho^-_jW^+_\mu:\;\; 
$-\frac{\imath}{2} g_2 \gamma_\mu \left [\left(2 U^{2*}_{j 1} N_{{i 2}}^{1}  + \sqrt{2} U^{2*}_{j 2} N_{{i 4}}^{1} \right)  \mathbb{P}_L 
   -\left({\sqrt{2}}  N^{2*}_{i 4} V_{{j 2}}^{2}  -2 N^{2*}_{i 2} V_{{j 1}}^{2} \right)  \mathbb{P}_R \right ] $\\[3mm]
\raisebox{-0.475\height}{
\begin{fmffile}{rho+chi0W-} 
\fmfframe(20,20)(20,20){ 
\begin{fmfgraph*}(45,45) 
\fmfleft{l1}
\fmfright{r1,r2}
\fmf{fermion}{v1,l1}
\fmf{fermion}{r1,v1}
\fmf{boson}{r2,v1}
\fmflabel{$\overline{{\rho}^-_{{i}}}$}{l1}
\fmflabel{${\chi}_{{j}}$}{r1}
\fmflabel{$W^-_{{\mu}}$}{r2}
\end{fmfgraph*}} 
\end{fmffile} 
}
&
%&&\bar{\rho^-_i}\chi^0_jW^-_\mu:\;\;  
$-\frac{\imath}{2} g_2 \gamma_\mu \left [ \left (2 N^{1*}_{j 2} U_{{i 1}}^{2}  + \sqrt{2} N^{1*}_{j 4} U_{{i 2}}^{2} \right )  \mathbb{P}_L - \,\left({\sqrt{2}}  V^{2*}_{i 2} N_{{j 4}}^{2}  - V^{2*}_{i 1} N_{{j 2}}^{2} )  \mathbb{P}_R \right)\right]$\\[3mm]
\raisebox{-0.475\height}{
\begin{fmffile}{chi+ellsnu} %FeynDia276
\fmfframe(20,20)(20,20){ 
\begin{fmfgraph*}(45,45) 
\fmfleft{l1}
\fmfright{r1,r2}
\fmf{fermion}{v1,l1}
\fmf{fermion}{r1,v1}
\fmf{scalar}{v1,r2}
\fmflabel{$\overline{{\chi}^-_{{i}}}$}{l1}
\fmflabel{$\ell$}{r1}
\fmflabel{$\tilde{\nu}^*_{{\ell}}$}{r2}
\end{fmfgraph*}} 
\end{fmffile} 
}
&  
%&&\chi^+_i e_j \tilde{\nu}^*_k:\;\;
$-\imath g_2 V^{1*}_{i 1}    \mathbb{P}_L  
   + \,i Y^*_{\ell}   U_{{i 2}}^{1}  \mathbb{P}_R $\\[3mm]
% 
%\raisebox{-0.475\height}{
%\begin{fmffile}{Diagrams/chi0nusnu} %FeynDia291 
%\fmfframe(20,20)(20,20){ 
%\begin{fmfgraph*}(45,45) 
%\fmfleft{l1}
%\fmfright{r1,r2}
%\fmf{fermion}{l1,v1}
%\fmf{fermion}{r1,v1}
%\fmf{scalar}{v1,r2}
%\fmflabel{${\chi}^0_{{i}}$}{l1}
%\fmflabel{$\nu_{{\ell}}$}{r1}
%\fmflabel{$\tilde{\nu}^*_{{\ell}}$}{r2}
%\end{fmfgraph*}} 
%\end{fmffile}
% below is h.c. of above figure 
\raisebox{-0.475\height}{
\begin{fmffile}{chi0nusnu} %FeynDia291 
\fmfframe(20,20)(20,20){ 
\begin{fmfgraph*}(45,45) 
\fmfleft{l1}
\fmfright{r1,r2}
\fmf{fermion}{v1,l1}
\fmf{fermion}{r1,v1}
\fmf{scalar}{v1,r2}
\fmflabel{$\overline{{\chi}_{{i}}^c}$}{l1}
\fmflabel{${\nu_{{\ell}}}$}{r1}
\fmflabel{$\tilde{\nu}^*_{{\ell}}$}{r2}
\end{fmfgraph*}} 
\end{fmffile} 
}
%\chi^0_i\nu_j\tilde{\nu}^*_k:\;\;
&$
\frac{\imath}{\sqrt{2}} \left (g_1 N^{1^*}_{i 1}  - g_2 N^{1*}_{i 2} \right ) \mathbb{P}_L $
\\[3mm] 
\raisebox{-0.475\height}{
\begin{fmffile}{rho-nusel} %FeynDia283
\fmfframe(20,20)(20,20){ 
\begin{fmfgraph*}(45,45) 
\fmfleft{l1}
\fmfright{r1,r2}
\fmf{fermion}{v1,l1}
\fmf{fermion}{r1,v1}
\fmf{scalar}{v1,r2}
\fmflabel{$\overline{{\rho}^+_{{i}}}$}{l1}
\fmflabel{$\nu_{{\ell}}$}{r1}
\fmflabel{$\tilde{\ell}^*_{{L}}$}{r2}
\end{fmfgraph*}} 
\end{fmffile} 
}
& 
%\rho^-_i\nu_j\tilde{e}^*_k:\;\;
$ -\imath g_2 U^{2*}_{i 1}    \mathbb{P}_L $
\\[3mm]
\raisebox{-0.475\height}{
\begin{fmffile}{chi0esel} % FeynDia289
\fmfframe(20,20)(20,20){ 
\begin{fmfgraph*}(45,45) 
\fmfleft{l1}
\fmfright{r1,r2}
\fmf{fermion}{v1,l1}
\fmf{fermion}{r1,v1}
\fmf{scalar}{v1,r2}
\fmflabel{$\overline{{\chi}_{{i}}^c}$}{l1}
\fmflabel{${\ell}$}{r1}
\fmflabel{$\tilde{\ell}^*_{{L}}$}{r2}
\end{fmfgraph*}} 
\end{fmffile} 
}
& %&\chi^0_ie_j\tilde{e}^*_k:\;\;
$   \frac{\imath}{\sqrt{2}} \left(g_1 N^{1*}_{i 1}  + g_2 N^{1*}_{i 2} \right)
        \mathbb{P}_L $ 
    $ -\imath Y_{\ell}^* N_{{i 3}}^{2}  \mathbb{P}_R $     
\\[3mm]   
%% %
%% \raisebox{-0.475\height}{
%% \begin{fmffile}{Diagrams/nuchi+sel} %FeynDia280
%% \fmfframe(20,20)(20,20){ 
%% \begin{fmfgraph*}(45,45) 
%% \fmfleft{l1}
%% \fmfright{r1,r2}
%% \fmf{fermion}{v1,l1}
%% \fmf{fermion}{r1,v1}
%% \fmf{scalar}{r2,v1}
%% \fmflabel{$\bar{\nu}_{{\ell}}$}{l1}
%% \fmflabel{${\chi}^+_{{j}}$}{r1}
%% \fmflabel{$\tilde{\ell}_{{R}}$}{r2}
%% \end{fmfgraph*}} 
%% \end{fmffile} 
%% }
%% %&&\bar{\nu}_i\chi^+_j\tilde{e}_k:\;\;
%% &
%% $\,\imath Y^*_{\ell}   U_{{j 2}}^{1} \mathbb{P}_R$
%% \\[3mm]       
%
%\raisebox{-0.475\height}{
%\begin{fmffile}{Diagrams/echi0sel} %FeynDia295 
%\fmfframe(20,20)(20,20){ 
%\begin{fmfgraph*}(45,45) 
%\fmfleft{l1}
%\fmfright{r1,r2}
%\fmf{fermion}{v1,l1}
%\fmf{fermion}{r1,v1}
%\fmf{scalar}{r2,v1}
%\fmflabel{$\bar{\ell}$}{l1}
%\fmflabel{$\tilde{\chi}^0_{{j}}$}{r1}
%\fmflabel{$\tilde{\ell}_{{R}}$}{r2}
%\end{fmfgraph*}} 
%\end{fmffile} 
%}
%%
%&%&\bar{e}_i\chi^0_j\tilde{e}_k:\;\;
%$\bar{e}_i\chi^0_j\tilde{e}_k:\;\; -i \sqrt{2} g_1 N^{1,*}_{j 1}   \mathbb{P}_L  
%    \,-i Y^*_{\ell}   N_{{j 3}}^{2}  \mathbb{P}_R $
%\\[3mm]
%%
\raisebox{-0.475\height}{
\begin{fmffile}{chi0esel2} %FeynDia305
\fmfframe(20,20)(20,20){ 
\begin{fmfgraph*}(45,45) 
\fmfleft{l1}
\fmfright{r1,r2}
\fmf{fermion}{v1,l1}
\fmf{fermion}{r1,v1}
\fmf{scalar}{v1,r2}
\fmflabel{$\overline{{{\chi}}_{{i}}}$}{l1}
\fmflabel{$\ell$}{r1}
\fmflabel{$\tilde{\ell}^*_{{R}}$}{r2}
\end{fmfgraph*}} 
\end{fmffile} 
}
&%&\bar{\chi}^0_i e_j\tilde{e}^*_k:\;\;
$   -i N^{2*}_{i 3} Y_{\ell}     \mathbb{P}_L  
    \,-i \sqrt{2} g_1   N_{{i 1}}^{1}  \mathbb{P}_R $
\\[3mm]
\raisebox{-0.535\height}{
\begin{fmffile}{chi-nusel} %FeynDia314
\fmfframe(20,20)(20,20){ 
\begin{fmfgraph*}(45,45) 
\fmfleft{l1}
\fmfright{r1,r2}
\fmf{fermion}{v1,l1}
\fmf{fermion}{r1,v1}
\fmf{scalar}{v1,r2}
\fmflabel{$\overline{{\chi}^+_{{i}}}$}{l1}
\fmflabel{$\nu_{{\ell}}$}{r1}
\fmflabel{$\tilde{\ell}^*_{{R}}$}{r2}
\end{fmfgraph*}} 
\end{fmffile} 
}
&%&\chi^-_i\nu_j\tilde{e}^*_k:\;\;
$  \imath \, U^{1*}_{i 2} Y_{\ell}\,\mathbb{P}_L $
\\[3mm]
\end{longtable}

\noindent where $ \mathbb{P}_L =(1-\gamma_5)/2$ and $ \mathbb{P}_R =(1+\gamma_5)/2$.  The couplings to quarks/squarks can easily be reproduced by simple substitutions. For simplicity the above rules assume no inter-generation mixing of sfermions.

\section{Calculation of vertex and box contributions to muon decay \label{sec:deltaVB}}
\begin{figure}[t]
\centering
\includegraphics[width=0.34\textwidth]{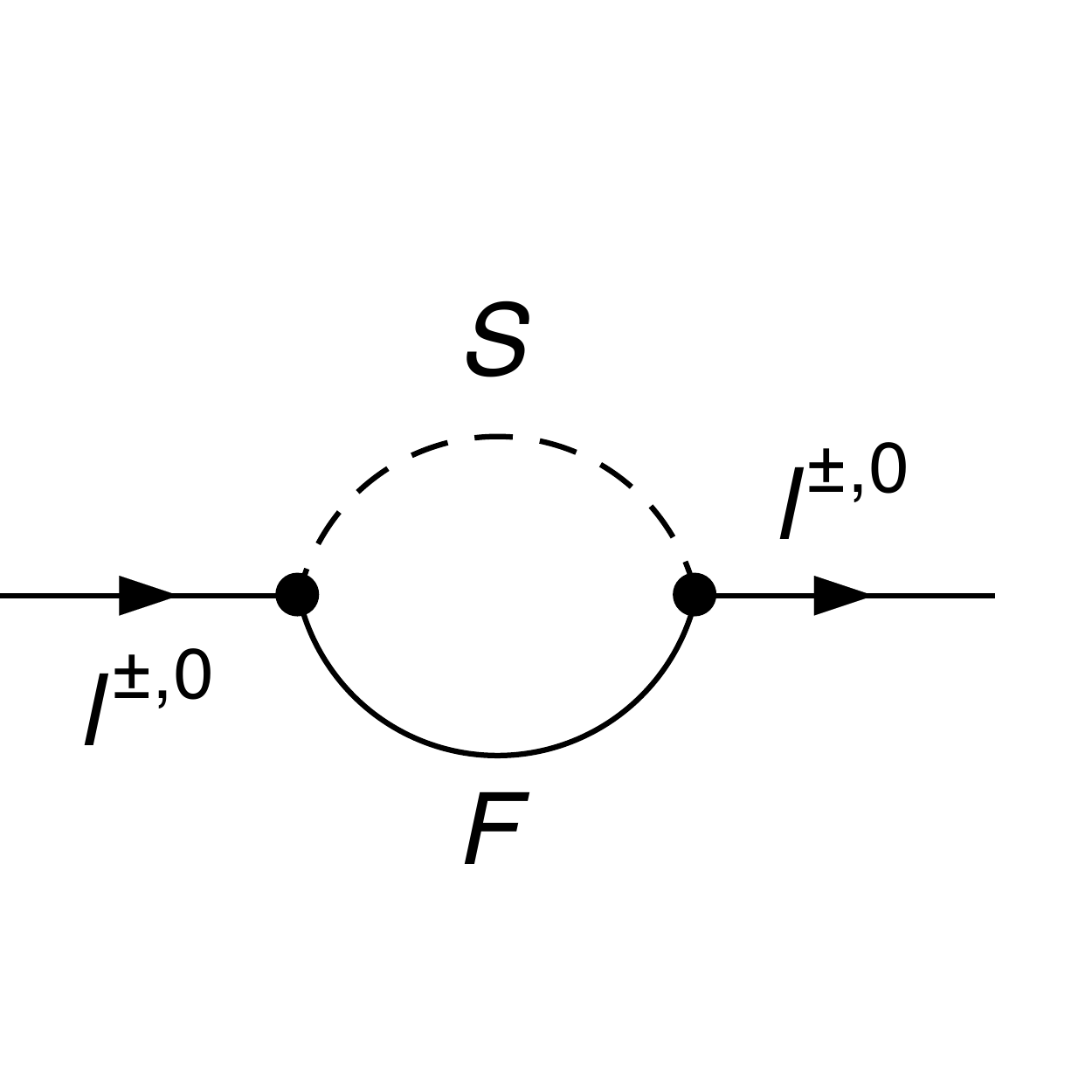}
\caption{Generic form of a non SM-like contribution to the external wave function renormalization.}
\label{external-wave-function}
\end{figure}

Here we give an explicit formula for the MRSSM-specific contributions to $\delta_{VB}$\begin{equation}
  \delta_{VB} = 2\cdot\frac{\sqrt{2}}{g_2} \delta V + \frac{2 m_W^2}{g_2^2}  B + \frac{1}{2}\delta Z^e_L + \frac{1}{2} \delta Z^\mu_L + \frac{1}{2}\delta Z^{\nu_e}_L + \frac{1}{2} \delta Z^{\nu_\mu}_L,
\label{eq:deltaVB}
\end{equation}
where $\delta V$ and $B$ are vertex- and box-corrections for muon decay and $\delta Z^i_L$ stands for the external wave function renormalization of the lepton $i$.\\[2mm]
%Symbolic calculimgation setup follows the one from the section 3.
%
\noindent $\bullet$ External wave functions renormalization\\
A generic diagram contributing to the renormalization of the external wave functions is shown in figure \ref{external-wave-function}. Decomposing the general fermion self-energy $-\imath \Sigma_{ff}$ as
\begin{equation}
\Sigma_{ff} = \Sigma^L_{ff} \slashed{p} P_L + \Sigma^R_{ff} \slashed{p} P_R + \Sigma^M_{ff} M_f,
\end{equation}
one finds, in the flavor conserving limit and assuming that leptons are massless, for the left-handed projector
\begin{eqnarray}
\Sigma_{\mu \mu}^{L,\text{MRSSM-SM}} &=& \frac{g_2^2}{16 \pi^2} \sum_{i=1}^2 \left | V^{1}_{i1} \right|^2 B_1 \left ( m_{\chi^\pm_i}^2, m_{\tilde{\nu}_\mu}^2 \right )  \\
&& + \frac{1}{16 \pi^2} \sum_{i=1}^4 \left | \frac{g_2 N^1_{i2} + g_1 N^1_{i1} }{\sqrt{2}} \right |^2 B_1 \left ( m_{\chi_i}^2, m_{\tilde{\mu}_L}^2 \right ), \nonumber \\
\Sigma_{\nu_\mu \nu_\mu}^{L,\text{MRSSM-SM}} &=& \frac{g_2^2}{16 \pi^2} \sum_{i=1}^2 \left | U^{2}_{i1} \right|^2 B_1 \left (m_{\rho^\pm_i}^2, m_{{\tilde{\mu}_L}}^2 \right )  \\
&& + \frac{1}{16 \pi^2} \sum_{i=1}^4 \left | \frac{g_2 N^1_{i2} - g_1 N^1_{i1} }{\sqrt{2}} \right |^2 B_1 \left (m_{\chi_i}^2, m_{\tilde{\nu}_\mu}^2 \right ) \nonumber,
\end{eqnarray}
where  $B_1$ is defined according to the \texttt{LoopTools} \cite{Hahn:1998yk} convention.
Because we neglected masses of leptons, the same equations hold for leptons of the first generation. External wave functions are renormalized in the \textit{on-shell} scheme, i.e.
%  on - shell field strength renormalization
%  d/dp (i Z - i Sigma) = 0 (term proportional to psibar p P_L psi
%  sign difference with 19.24 of Matthew Schwarz because we define self-energies with a minus
\begin{equation}
 \delta Z^L_i = \Sigma_{i i}^{L,\text{MRSSM}} .
\end{equation}

\begin{figure}[t]
\centering
\includegraphics[width=0.8\textwidth]{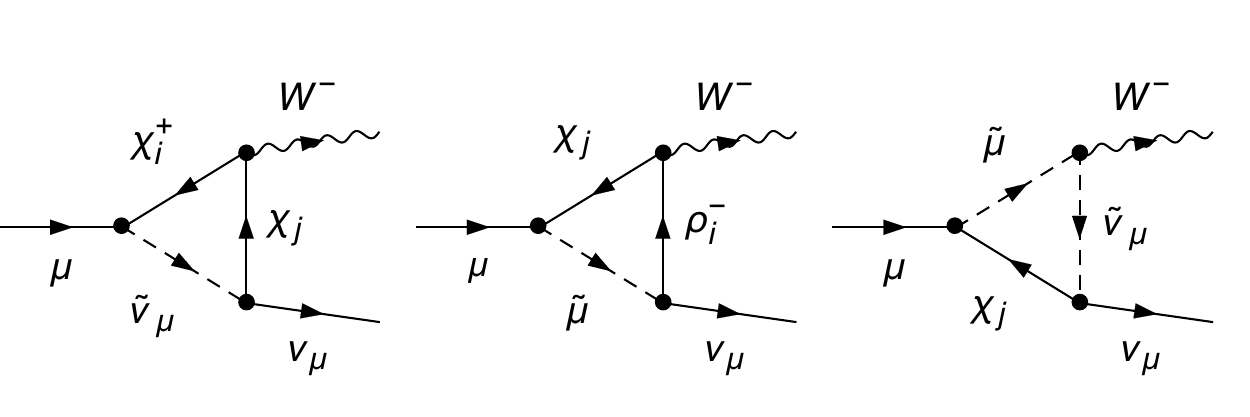}
\caption{Non-SM corrections to $\mu^- \to W^- \nu_\mu$ decay vertex. Diagrams proportional to the muon Yukawa are not shown. \label{fig:vertex-diagrams}}
\end{figure}

%  Vertex corrections
\noindent $\bullet$ {Vertex correction}\\
Figure \ref{fig:vertex-diagrams} contains non-SM corrections to the muon decay vertex. The corresponding analytic expression for the amplitude reads
%
%  It's not easy to check if this formula is correct or not numerically due to the running of parameters which can be read of SPheno.spc.MRSSM at \mu = 1000 but should be used at \mu = mZ in DeltaVB subroutine. 
%  tanb = 40 |SPheno| = 1.92, 16 pi^2 |formula|= 2.39
\begin{eqnarray}
% first diagram
-\imath \delta V &=& 
-\frac{\imath g_2}{16 \sqrt{2} \pi^2} \sum_{i=1}^2 \sum_{j=1}^4 g_2 V^{1*}_{i1} \left ( \frac{g_2 N^1_{j2} - g_1 N^1_{j1} }{\sqrt{2}} \right ) \left \{ \left ( \sqrt{2} N^{1*}_{j2} V^1_{i1} - N^{1*}_{j3} V^1_{i2} \right )\right .\\
&&  \left [m_{\tilde{\nu}_\mu}^2 C_{0} \left ( m_{\chi^\pm_i}^2, m_{\chi_j}^2, m_{\tilde{\nu}_\mu}^2 \right ) + B_0 \left (m_{\chi^\pm_i}^2, m_{\chi_j}^2 \right ) - 2 C_{00} \left ( m_{\chi^\pm_i}^2, m_{\chi_j}^2, m_{\tilde{\nu}_\mu}^2 \right ) \right ] \nonumber \\ 
 && -  \left . m_{\chi^\pm_i}^2  m_{\chi_j}^2 \left (\sqrt{2} N^2_{j2} U^{1*}_{i1} + N^2_{j3} U^{1*}_{i2} \right ) C_{0} \left ( m_{\chi^\pm_i}^2, m_{\chi_j}^2, m_{\tilde{\nu}_\mu}^2 \right ) \right \}  \nonumber\\ 
% second diagram
&& - \frac{\imath g_2}{16 \sqrt{2} \pi^2} \sum_{i=1}^2 \sum_{j=1}^4 g_2 U^{2}_{i1} \left ( \frac{g_2 N^{1*}_{j2} + g_1 N^{1*}_{j1}}{\sqrt{2}} \right ) \left \{ \left ( \sqrt{2} N^1_{j2} U^{2*}_{i1} + N^1_{j3} U^{2*}_{i2} \right )\right . \nonumber\\
&&  \left [m_{\tilde{\mu}}^2 C_{0} \left ( m_{\rho^\pm_i}^2, m_{\chi_j}^2, m_{\tilde{\mu}}^2 \right ) + B_0 \left (m_{\rho^\pm_i}^2, m_{\chi_j}^2 \right ) - 2 C_{00} \left ( m_{\rho^\pm_i}^2, m_{\chi_j}^2, m_{\tilde{\mu}}^2 \right ) \right ] 
   \nonumber\\ 
 && - \left . \left . m_{\rho^\pm_i}^2  m_{\chi_j}^2 \left (\sqrt{2} N^2_{j2} U^{1*}_{i1} - N^{2*}_{j4} V^2_{i2} \right )\right ] C_{0} \left ( m_{\rho^\pm_i}^2, m_{\chi_j}^2, m_{\tilde{\mu}}^2 \right ) \right \}  \nonumber\\
% third diagram
 && + \frac{\imath g_2}{8 \sqrt{2} \pi^2} \sum_{i=1}^4 \left ( \frac{g_2 N^1_{j2} - g_1 N^1_{j1} }{\sqrt{2}} \right ) \left ( \frac{g_2 N^{1*}_{j2} + g_1 N^{1*}_{j1} }{\sqrt{2}} \right ) C_{00} \left ( m_{\chi_i}^2, m_{\tilde{\mu}}^2, m_{\tilde{\nu}_\mu}^2 \right ) \nonumber ,
\end{eqnarray}
where all Passarino-Veltman's functions follow, as before,  the \texttt{LoopTools}  convention.
Although diagrams from figure \ref{fig:vertex-diagrams} have the same analytic expression 
as in the MSSM, due to the absorption of $\tilde{W}^\pm$ into $\chi^+$ and $\rho^-$, 
respectively,  diagram 2 is in principle independent in magnitude from diagram 1, 
as it depends on the  mass of a different particle.\\[2mm]
\begin{figure}[b]
\centering
\includegraphics[width=0.525\textwidth]{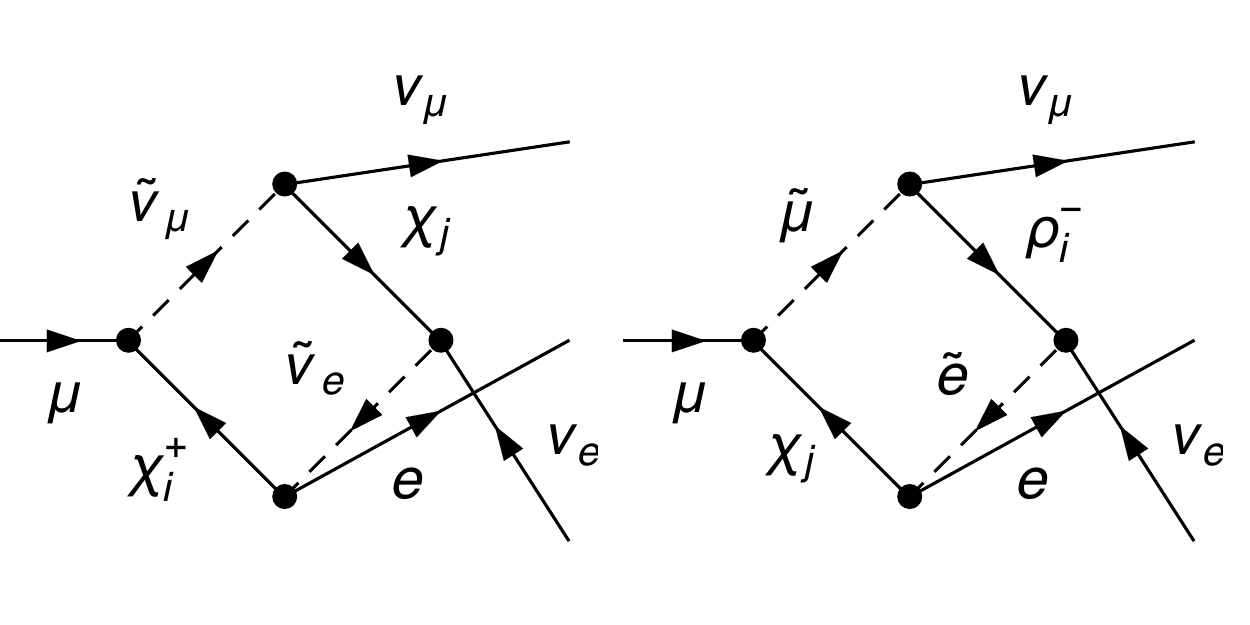}
\caption{Non-SM box contributions to $\mu^- \to \nu_\mu e^- \bar{\nu}_e$ in the MRSSM. Diagrams proportional to lepton's Yukawa or vanishing in the flavor-conserving limit are not shown. \label{box-diagrams}}
\end{figure}
\noindent $\bullet$ {Box correction} \\
Figure \ref{box-diagrams} contains the most relevant box-type contributions to muon decay in the MRSSM. Those diagrams are both UV and IR finite. The expression for them, after factorizing out the spinor structure (cf. eq. (\ref{eq:deltaVB})), reads 
%
%  Overal factor in agreement with P. H. Chankowski, A. Dabelstein, W. Hollik, W. M. Mosle, S. Pokorski, others, “Delta R in the MSSM,” Nucl.Phys., vol. 417, no. 1, pp. 101–129, 1994.
%  In DRED formcalc multiplies it by a factor of -2 which is probably wrong!
%
\begin{multline}
-\imath B  =  -\frac{\imath}{16 \pi^2} \sum_{i=1}^2 \sum_{j=1}^4  \left |g_2^2 V^1_{i1} \left (\frac{g_1 N^1_{j1} - g_2 N^1_{j2}}{\sqrt{2}} \right )\right |^2 D_{00} \left ( m_{\chi^\pm_i}^2, m_{\chi_j}^2, m_{\tilde{\nu}_\mu}^2, m_{\tilde{\nu}_e}^2 \right ) \\
 - \frac{\imath}{16 \pi^2} \sum_{i=1}^2 \sum_{j=1}^4  \left |g_2^2 U^2_{i1} \left (\frac{g_1 N^1_{j1} + g_2 N^1_{j2}}{\sqrt{2}} \right )\right |^2 D_{00} \left ( m_{\rho^\pm_i}^2, m_{\chi_j}^2, m_{\tilde{\mu}}^2, m_{\tilde{e}}^2 \right ) ,
\end{multline}
where $D_{00}$ is defined according to the \texttt{LoopTools} convention.
The structure of this correction is different from the one in the MSSM due to the Dirac nature of MRSSM's neutralinos and the fact that $\tilde{W}^+$ and $\tilde{W}^-$ are parts of two different types of charginos. This forbids the existence of two additional MSSM-like diagrams with mass-term induced $\tilde{W}^+-\tilde{W}^-$. Although distinct from the MSSM, for benchmark points under consideration, the contribution from the box-correction is below 1 MeV.

In total, for the benchmark points considered, the $\delta_{VB}$ accounts for about $150 - 175$ MeV downward shift on $m_W$, which comprises $-350$ MeV pure SM correction and $+175$ MeV MRSSM part. It should be recalled that $\delta_{VB}$ is not finite nor gauge-independent and as such these numbers should be interpreted with care. Also, the parameter dependence of $\delta_{VB}$ is mild and it generates mostly a constant shift in $m_W$.

\end{document}